\documentclass{aa}  
\usepackage{graphicx}
\usepackage{hyperref}
\usepackage{xcolor}
\usepackage{lscape}

\usepackage{txfonts}
\newcommand\micron{$\mu$m}
\definecolor{orange}{rgb}{1.0, 0.5, 0.3}

\begin{document}

   \title{Chemical exploration of Galactic cold cores}

   \author{Chenlin Zhou \inst{1}
          \and
          Charlotte Vastel \inst{1}
          \and 
          Julien Montillaud \inst{2}
          \and 
          Cecilia Ceccarelli \inst{3}
          \and 
          Karine Demyk \inst{1}
          \and
          Jorma Harju \inst{4,5}
          \and
          Mika Juvela \inst{5}
          \and
          Isabelle Ristorcelli \inst{1}
          \and
          Tie, Liu \inst{6}
          }

   \institute{IRAP, Universit\'e de Toulouse, CNRS, UPS, CNES, 31400 Toulouse, France \\
         \email{cvastel@irap.omp.eu}
         \and
             Institut UTINAM - UMR 6213 - CNRS - Univ. Bourgogne Franche Comt\'e, OSU THETA, 41bis avenue de l'Observatoire, 25000 Besan\c con, France
         \and
         Universit\'e Grenoble Alpes, CNRS, IPAG, F-38000 Grenoble, France
         \and
         Max-Planck-Institut f\"ur extraterrestrische Physik, Gie{\ss}enbachstra{\ss}e 1, D-85748 Garching, Germany
         \and
         Department of Physics, P.O. BOX 64, FI-00014 University of Helsinki, Finland 
         \and
         Department of Astronomy, Peking University, 100871 Beijing, China        
             }

   \date{Received October 11th, 2021; accepted January 11th, 2022}

 
  \abstract
 {A solar-type system starts from an initial molecular core that acquires organic complexity as it evolves. The so-called prestellar cores that can be studied are rare, which has hampered our understanding of how organic chemistry sets in and grows.  }
{We selected the best prestellar core targets from the cold core catalogue (based on Planck and Herschel observations) that represent a diversity in terms of their environment to explore their chemical complexity: 1390 (in the compressed shell of Lambda Ori), 869 (in the MBM12 cloud), and 4149 (in the California nebula). }
{We obtained a spectral survey with the IRAM 30 m telescope in order to explore the molecular complexity of the cores.  We carried out a radiative transfer analysis of the detected transitions in order to place some constraints on the physical conditions of the cores and on the molecular column densities.  We also used the molecular ions in the survey to estimate the cosmic-ray ionisation rate and the S/H initial elemental abundance using a gas-phase chemical model to reproduce their abundances. }
{We  found large differences in the molecular complexity (deuteration, complex organic molecules, sulphur, carbon chains, and ions) and compared their chemical properties with a cold core and two prestellar cores. The chemical diversity we found in the three cores seems to be correlated with their chemical evolution: two of them are prestellar (1390 and 4149), and one is in an earlier stage (869). }
{The influence of the environment is likely limited because cold cores are strongly shielded from their surroundings. The high extinction prevents interstellar UV radiation from penetrating deeply into the cores. Higher spatial resolution observations of the cores are therefore needed to constrain the physical structure of the cores, as well as a larger-scale distribution of molecular ions to understand the influence of the environment on their molecular complexity.}


 \maketitle


\section{Introduction}
\label{sec: Intro}

Observations of molecular complexity in cold cores have long been limited by the sensitivity of the instruments. This is particularly true for heavy interstellar complex organic molecules (iCOMs) whose many transitions are split and whose intensities are therefore weaker. These iCOMS have been defined by astronomers as species containing at least six atoms and the element carbon \citep{Herbst2009}. To date, it is still not possible to explore the molecular complexity in interstellar ices, which might only be possible when the James Webb Space Telescope (JWST) becomes operational, but it is possible to detect the gas-phase emission of iCOMs in the earliest phases of star-forming regions. These observations are crucial for understanding when and where molecular complexity is initiated  in the interstellar medium (ISM) prior to delivery to a planetary system such as our own Solar System. 

The iCOMs were first detected in the hot cores of massive star-forming regions \citep{Cummins1986,Blake1987}.
Later, they were detected in the warm central regions of low-mass protostars, the so-called hot corinos \citep{Cazaux2003,Bottinelli2004}. More studies reported the detection of iCOMs in colder regions, for example in cold clouds toward the Galactic centre \citep{Requena-Torres2006}, and cold envelopes of solar-type protostars \citep{Oberg2010,Cernicharo2012ApJ,Jaber2014}.  The first evidence of the efficient production of iCOMs in cold environments was reported by \citet{Oberg2010} towards Barnard~1, a dense core hosting a low-mass protostar. The detected iCOMs (methanol: CH$_3$OH; methyl formate: HCOOCH$_3$; and acetaldehyde: CH$_3$CHO) seem to trace the quiescent core and not the outflow driven by the protostar. An active chemistry is then likely to proceed at low temperatures in the ice and in the gas phase. However, the presence of the outflows in this source could affect the chemistry, as discussed by \citet{Fuente2016}. More recently, 
\citet{Cernicharo2012ApJ} observed the methoxy radical (CH$_3$O) towards Barnard~1 as well as methyl mercaptan (CH$_3$SH), propynal (HCCCHO), acetaldehyde (CH$_3$CHO), and dimethyl ether (CH$_3$OCH$_3$). These observations suggested that these species are formed on the surface of dust grains and are ejected to the gas phase through non-thermal desorption processes, most likely cosmic rays, which are expected to penetrate deep into the cloud and react with the abundant gaseous H$_{2}$, resulting in the emission of UV photons. Following these observations, many spectral surveys have been performed towards either cold cores or towards the more evolved prestellar cores. 


All these detections reveal that the chemical processes at work are not yet understood, and understanding the processes that lead to the production of these large iCOMs under very cold conditions is still a mystery.\\

Prestellar cores represent a subclass of starless cold cores that is more centrally concentrated and gravitationally bound. They are typically detected in (sub)millimeter dust continuum emission and dense molecular gas tracers such as NH$_3$ or N$_2$H$^+$.  They show high degrees of CO (among other molecules) depletion and a high deuterium fractionation. They are rare, as their life times are short \citep{Konyves2015}. All prestellar cores are starless, but only a fraction of the starless cores will eventually evolve into prestellar cores. The best-studied cores are in the solar neighbourhood, at distances below 200 pc. Their sizes are about 10$^4$ au, with centrally concentrated structures, reaching central visual extinctions higher than 50 magnitudes and volume densities higher than 10$^5$ cm$^{-3}$ \citep{Crapsi2005ApJ, Keto2008ApJ}, with a clear drop in temperature  \citep[$\sim$ 7\,K:][]{Crapsi2007}.\\
L1544 is one of the most frequently studied of these cores. An unbiased spectral survey at the IRAM 30m telescope explored its chemical composition and led to the detection of a very rich chemistry. For instance, more than 20 sulphur-bearing species have been detected \citep{Vastel2018a, Cernicharo2018}, and a high sulphur depletion in the gas phase during the earliest phase was determined with a gas and grain surface model. Many nitrogen-bearing species have also been detected, such as the cyanomethyl radical CH$_2$CN, including its hyperfine structure \citep{Vastel2015}, HNC$_3$, HCCNC and HC$_3$N \citep{Vastel2018b,Hily-Blant2018}, HC$_3$NH$^+$ and HCNH$^{+}$ \citep{Quenard2017}, CNCN, NCCNH$^+$, C$_3$N, CH$_3$CN, C$_2$H$_3$CN, and H$_2$CN \citep[][]{Vastel2019}. Many iCOMs have been detected \citep{Bizzocchi2014,Vastel2014,Jimenez-Serra2016,Chacon2019}, such as methanol (CH$_3$OH and its deuterated forms), acetaldehyde (CH$_3$CHO), and dimethyl ether (CH$_3$OCH$_3$), although with integrated intensities of just a few mK~km/s. From these detections, \citet{Quenard2017,Vastel2018a,Vastel2018b,Vastel2019} built a detailed chemical network to reproduce the molecular abundances and found that most emission occurs in the external envelope of the prestellar core \citep{Vastel2014,Vasyunin2017}, while the deuterated species (e.g. the deuterated forms of H$_3^+$, methanol and formaldehyde) trace the internal structure where CO is depleted {\citep[e.g.][]{Sipila2018}}. The very presence of methanol in the gas phase in such a cold environment has been difficult to explain, as methanol is thought to be a grain surface product \citep{Watanabe2002ApJ,Rimola2014}, the temperature being too cold for thermal desorption to occur. Non-thermal desorption in a layer at least 8000 au from the centre where the temperature is $\sim$ 10 K and the H$_2$ volume density is $\sim$ 10$^4$ cm$^{-3}$ accounts for the detection of iCOMs in L1544. In this layer, secondary UV photons produced by the interaction between cosmic rays and H$_2$ molecules might lead to the desorption of water and ethene as well as to photo-fragments such as CH$_{3}$O in the gas phase \citep[see e.g.][]{Bertin2016}. Other chemical processes that have been proposed to explain the presence of iCOMs in the cold ISM include chemical desorption \citep{Vasyunin2017}, cosmic-ray induced chemistry \citep{Shingledecker2018}, non-diffusive reactions \citep{Chang2016}, chemical explosions \citep{Rawlings2013,Ivlev2015}, non-thermal cosmic-ray induced diffusion of radicals on ices \citep{Reboussin2014}, and sputtering by cosmic rays \citep{Dartois2019,Wakelam2021}. However, all these processes are still poorly constrained and need to be tested against observations. More detections of iCOMs in the cold regions of the ISM, especially prestellar cores, are therefore needed so that they can help us understand where the emissions come from in cold cores, and to better determine their abundances in different environments. 

We started a pilot program to extend the observations of the molecular complexity in the earliest phases of star-forming regions towards a sample of three prestellar core candidates selected from the catalogue in \citet{Montillaud2015}. These nearby cores are located in diverse Galactic environments and are presented in Sec.~\ref{sec: Select}. The survey, performed at the IRAM 30m, is presented in Sec.~\ref{sec: Obs} together with the data reduction method. The results are presented in Sec.~\ref{sec: Result}, along with the method, including line identification. This led to the determination of the molecular complexity in all three cores, as well as the local thermodynamic equilibrium (LTE) vs non-LTE analysis we used to determine the column densities for each species. Sec.~\ref{sec:discuss} finally presents the evolutionary stages of the cores, a comparison with the prototypical cores observed previously, and references in the literature. In the last section we present our conclusions as well as what must be done in the near future to better characterise the core to understand the molecular complexity presented in this manuscript.

\section{Source selection}
\label{sec: Select}

The \emph{Planck} space telescope \citep{Tauber2010} has mapped the whole sky at several submillimetre wavelengths with high sensitivity and a beam size of a few arcminutes, providing data for an all-sky inventory of the coldest structures of the interstellar medium. 
The cold ($T_\mathrm{dust}<14$ K) and compact sources were  listed  in  a  catalogue  containing  more  than 10000 objects. 
This Cold Clump Catalogue of Planck Objects (C3PO, \citealt{C3PO}) contains clumps that may host prestellar and starless cores at sub-parsec scales.
It was further developed and led to the Planck Catalogue of Galactic Cold Clumps (PGCC, \citealt{PGCC}), containing 13188 Galactic sources that are thought to be good candidates for star-forming regions in early evolutionary stages.
The \emph{Herschel} open-time  key  project  Galactic cold  core (GCC) has mapped selected \emph{Planck} C3PO objects with the \emph{Herschel} PACS (at 100 and 160 $\mu m$) and SPIRE (at 250, 350, and 500 $\mu m$) instruments with sub-arcminute angular resolution to reveal the sub-structure of the \emph{Planck} sources and to derive their physical properties \citep{Juvela2010}.
\citet{Montillaud2015} generated dust temperature and N$_{\mathrm{H}_2}$ maps using modified black-body SED fitting of SPIRE data. A catalogue of $\sim$ 4000 cold and compact sources was provided, called the Galactic Cold Cores (GCC) catalogue. The physical properties were derived from SED fitting results. 

\citet{Montillaud2015} separated the sources into protostellar and starless sources based on two methods. 
One method was based on submillimetre dust temperature profiles. The sources with a clear increase or decrease in temperature from the edge to the centre were selected, and the temperature profiles were fitted as Gaussians. The source was classified as protostellar if $\Delta T > 2$ K and |$\Delta T / \delta (\Delta T)| > 5$. Here $\Delta T$ is the profile amplitude from Gaussian fitting, and $\delta (\Delta T)$ is the uncertainty on $\Delta T$. Similarly,  the source was classified as starless if $\Delta T < -2$ K and |$\Delta T / \delta (\Delta T)| > 5$. Based on this method, the authors found 195 protostellar cores and 44 starless cores.

In their second method, \citet{Montillaud2015} used WISE PSC data to identify Class I and Class II YSOs adopting the colour criteria from \citet{Koenig2012}. They also adopted AKARI PSC to identify Class 0 objects based on the criteria as sources that had significant fluxes in all the bands of WISE and in the AKARI 65 \micron \ band.
Sources identified as Class 0, Class I, or Class II were considered to be protostellar. Sources without a counterpart at wavelengths shorter than or equal to 65 \micron \ were considered to be starless. Based on this method, the authors  found 449 protostellar cores and 1384 starless cores.

Taking advantage of these  two  methods  by  combining  the  two  diagnostics, \citet{Montillaud2015} reported 14 starless cores with maximum reliability and 1383 good candidates of starless cores. These cores are good candidates for searching and studying prestellar cores.

\begin{figure*}
    \centering
    \includegraphics[width=0.45\textwidth]{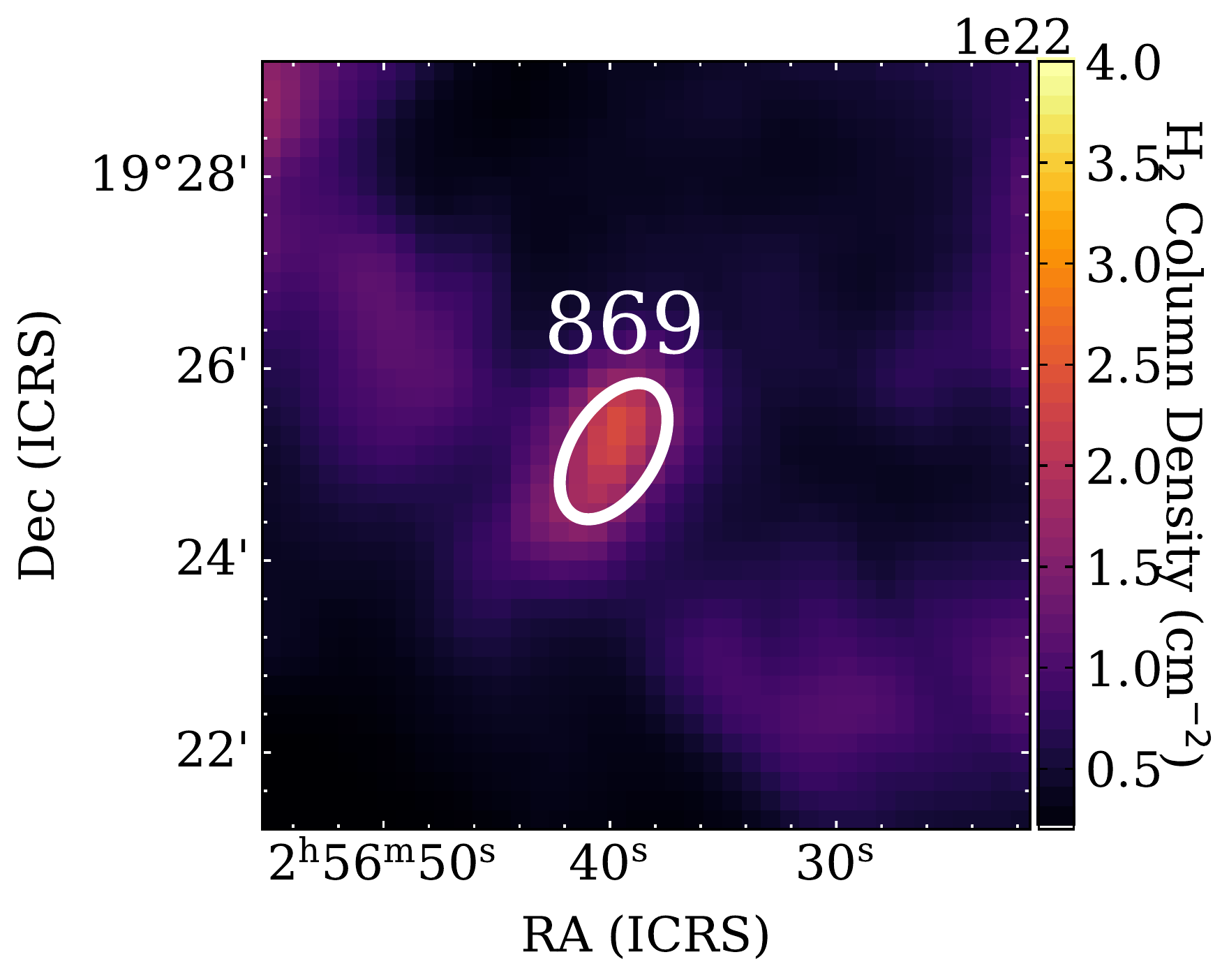} \ 
    \includegraphics[width=0.45\textwidth]{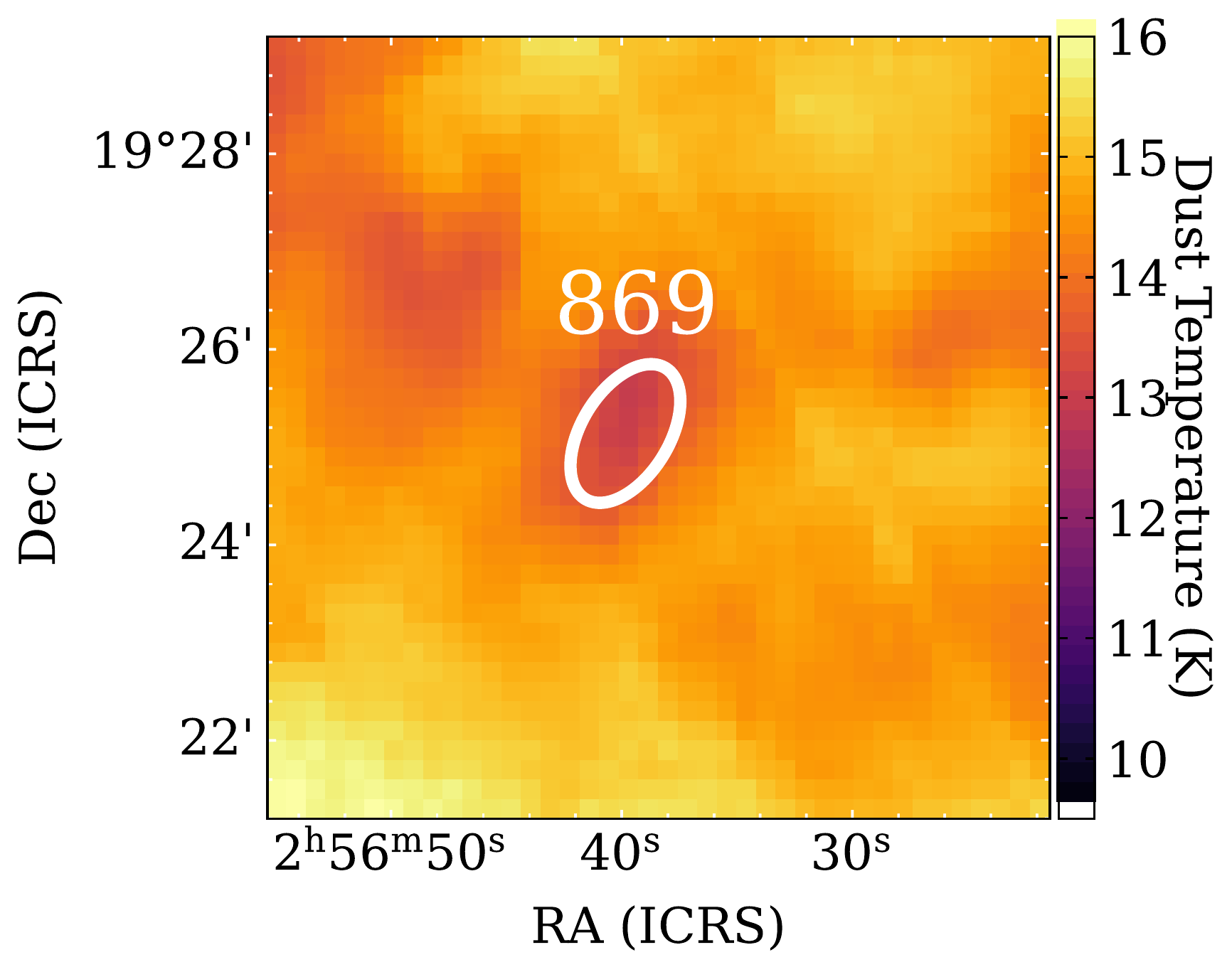} \\
    \includegraphics[width=0.45\textwidth]{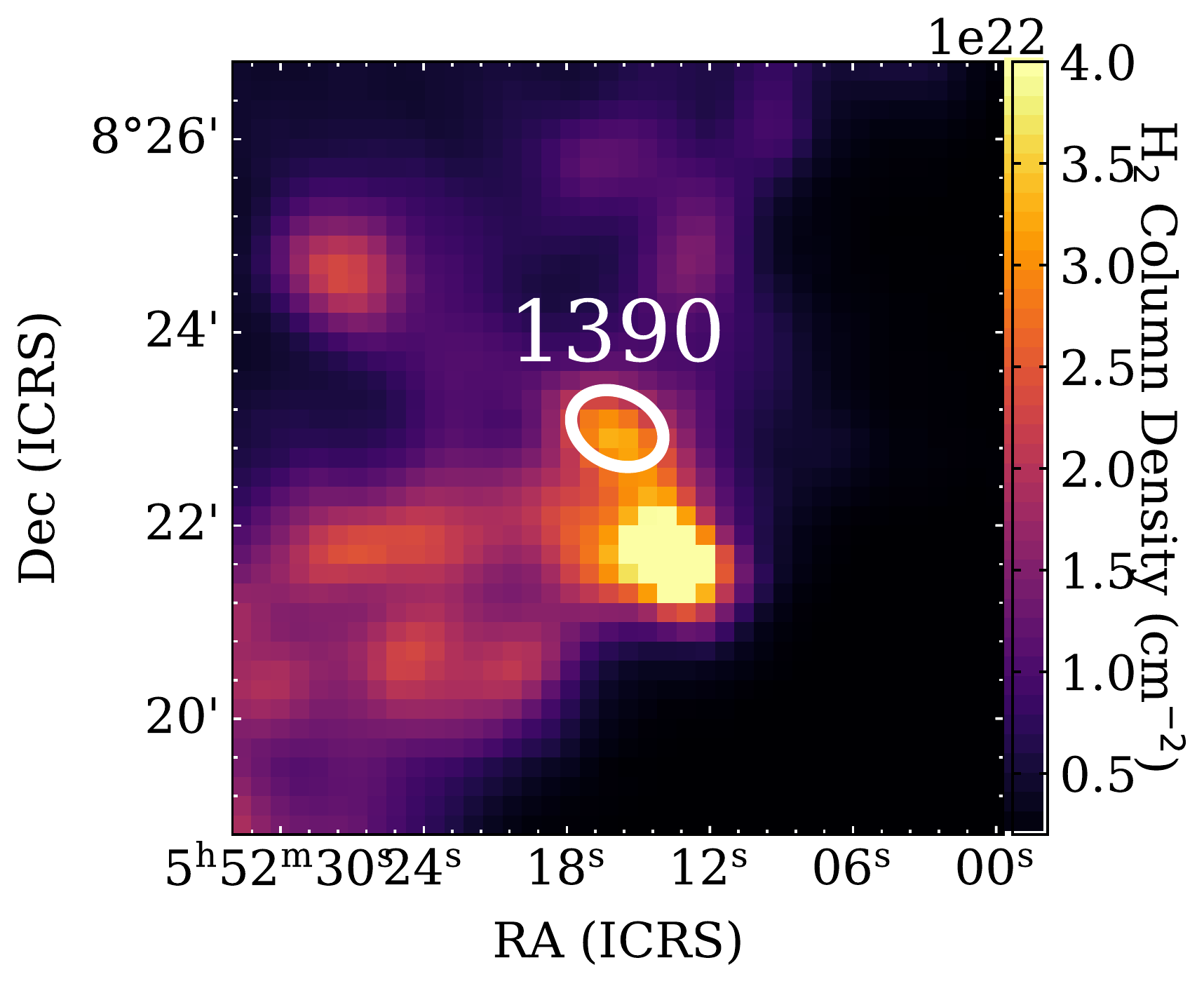} \ 
    \includegraphics[width=0.45\textwidth]{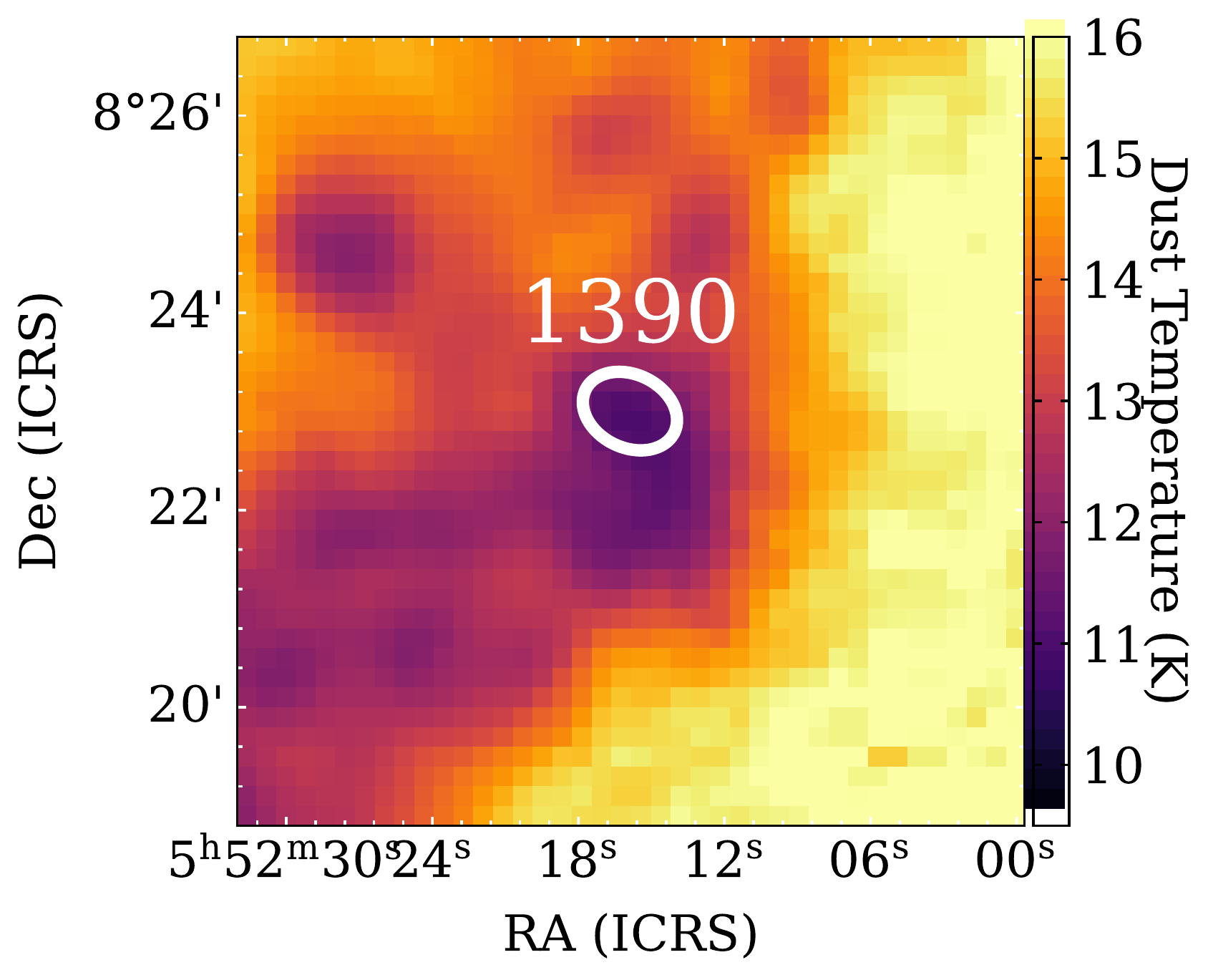}\\
    \includegraphics[width=0.45\textwidth]{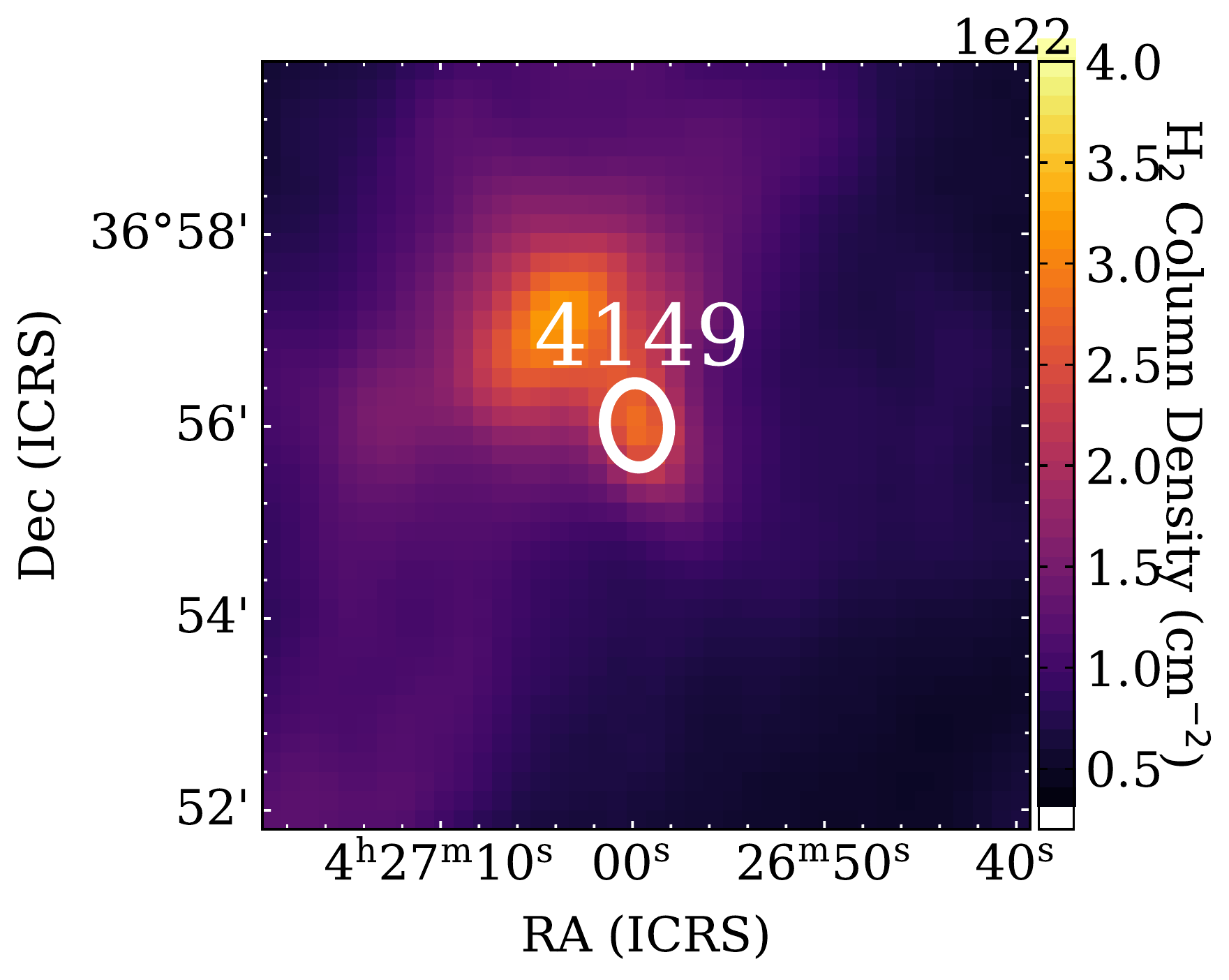} \ 
    \includegraphics[width=0.45\textwidth]{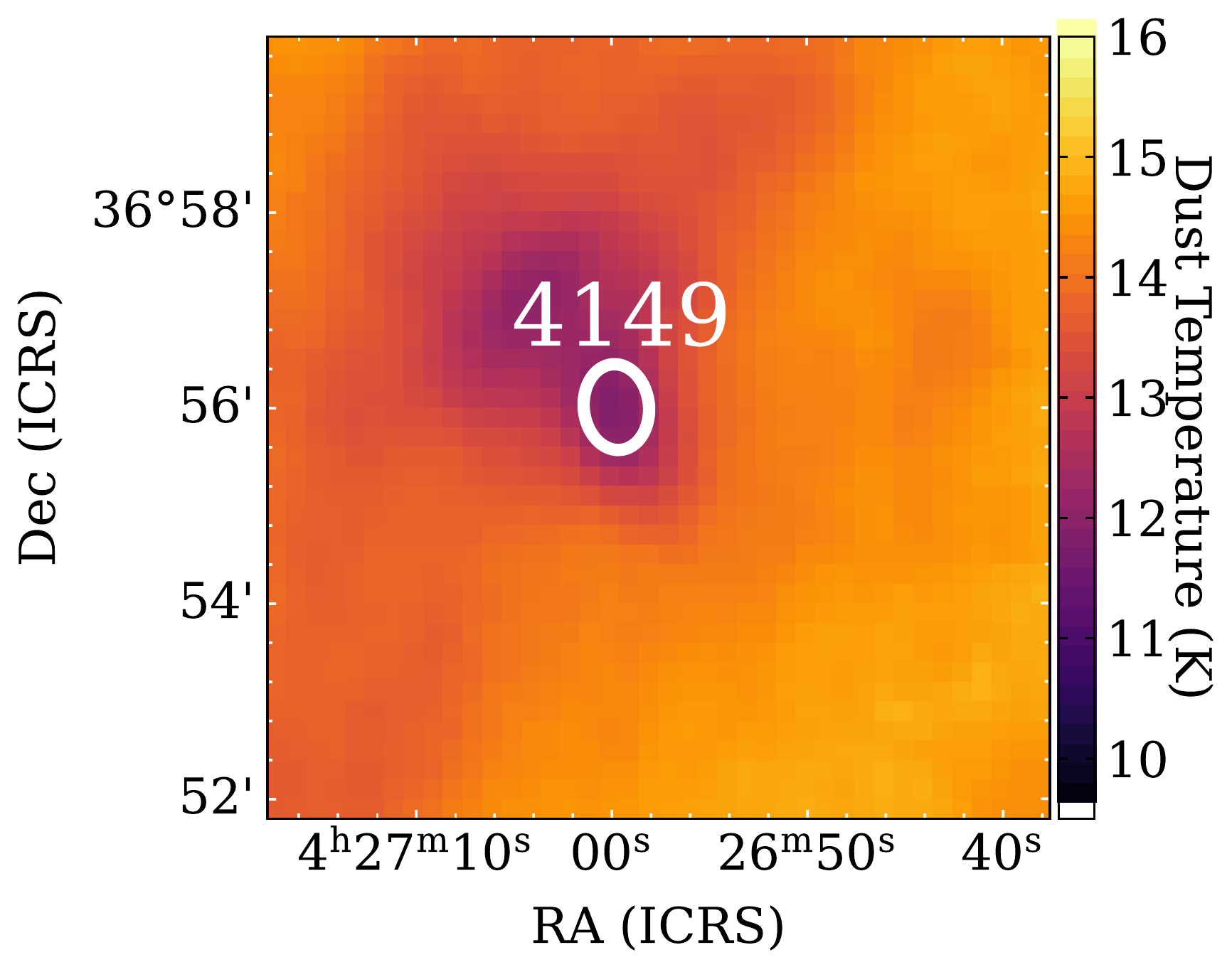}
    \caption{ {\it Herschel} H$_2$ column density (left) and dust temperature (right) maps of the three cores. The maps are generated by SED fitting of {\it Herschel} data \citep{Montillaud2015}. The map sizes are $8 \arcmin \times 8 \arcmin$. The white ellipses mark the size and the position of the cores extracted from the GCC catalogue, as described in Sec. \ref{sec: Select}.}
    \label{fig: T&N}
\end{figure*}

\begin{figure}
    \centering
    \includegraphics[width=0.47\textwidth,trim=100 150 80 250 clip]{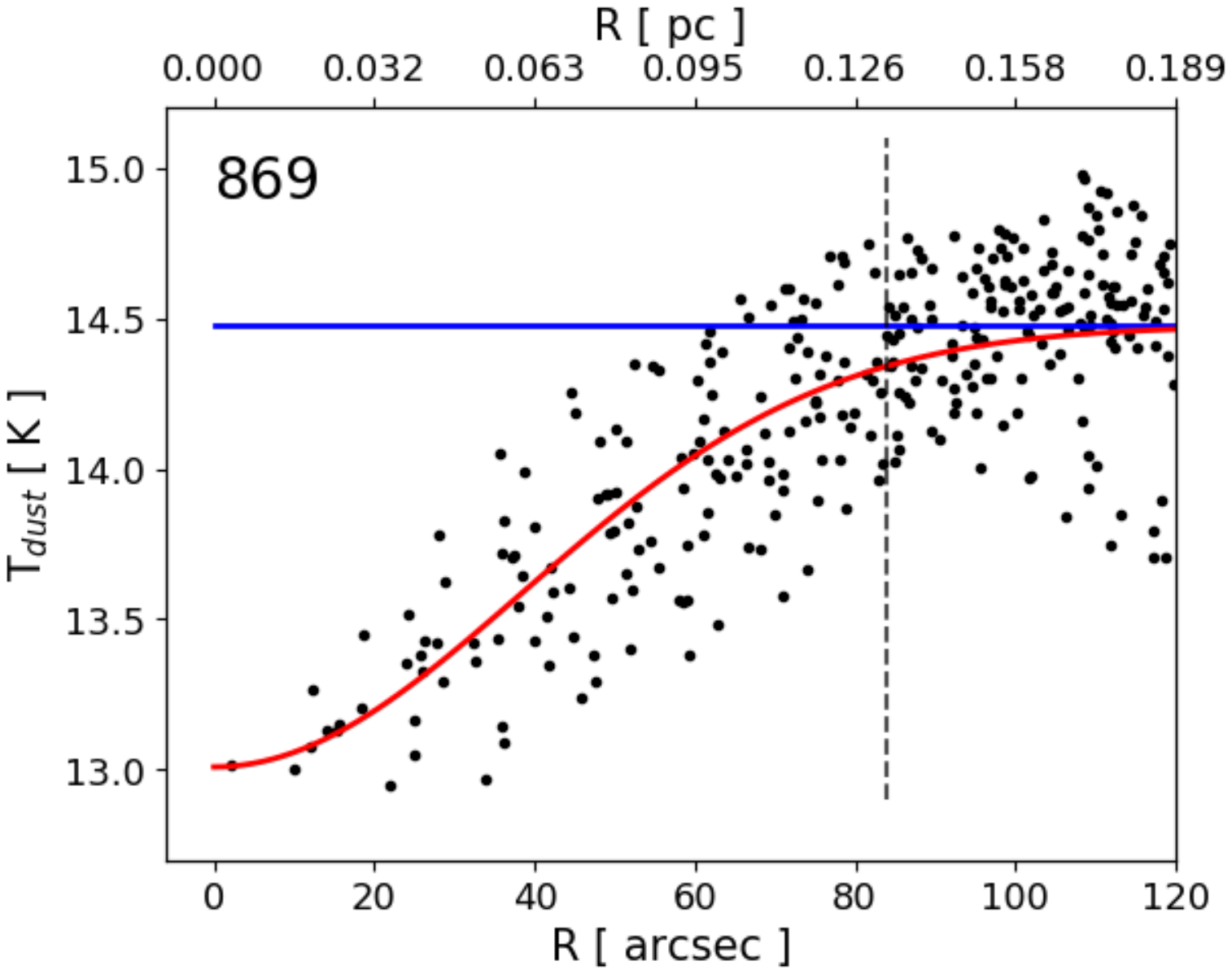} \ \
    \includegraphics[width=0.47\textwidth,trim=100 130 80 330 clip]{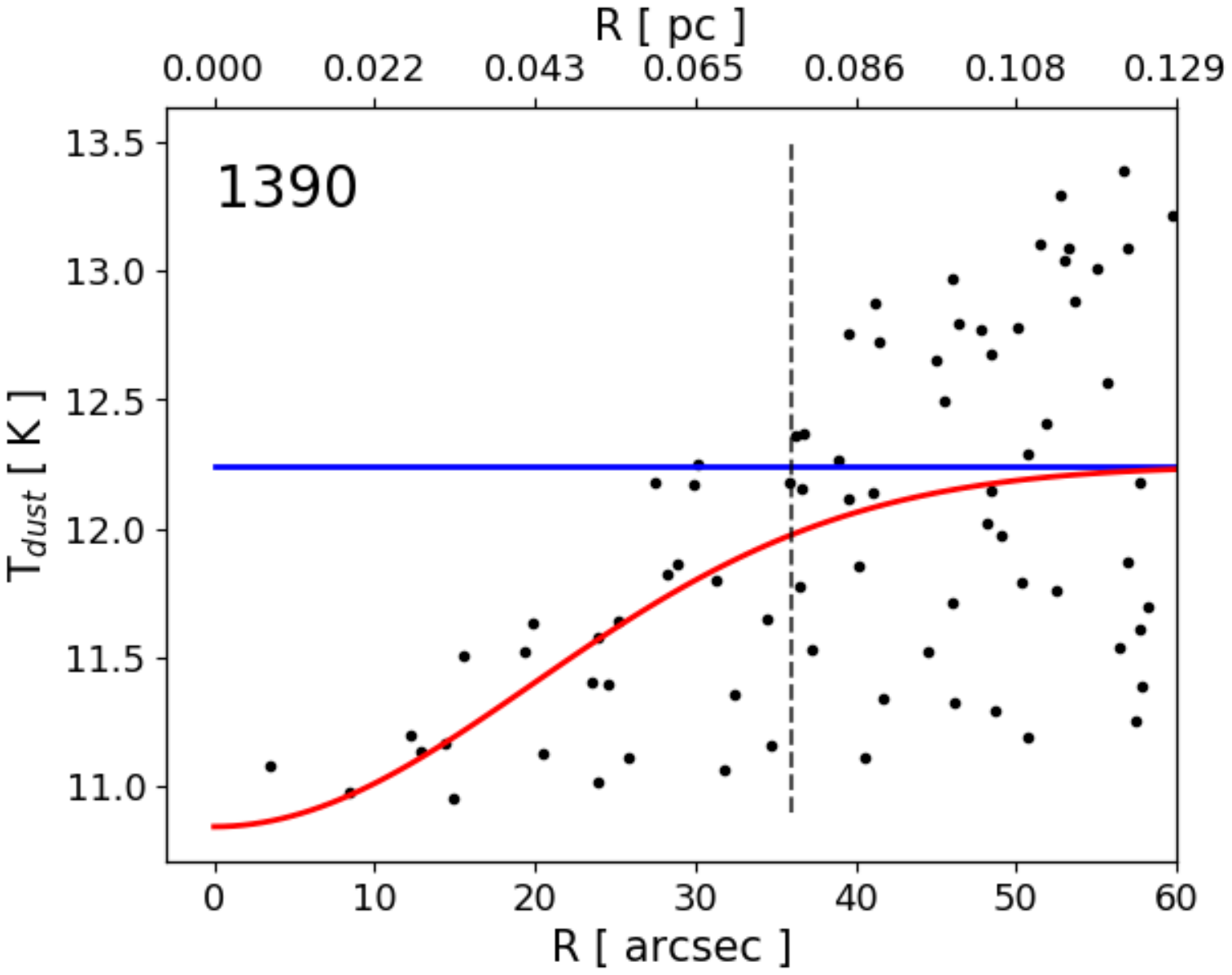} \ 
    \includegraphics[width=0.47\textwidth,trim=100 220 80 350 clip]{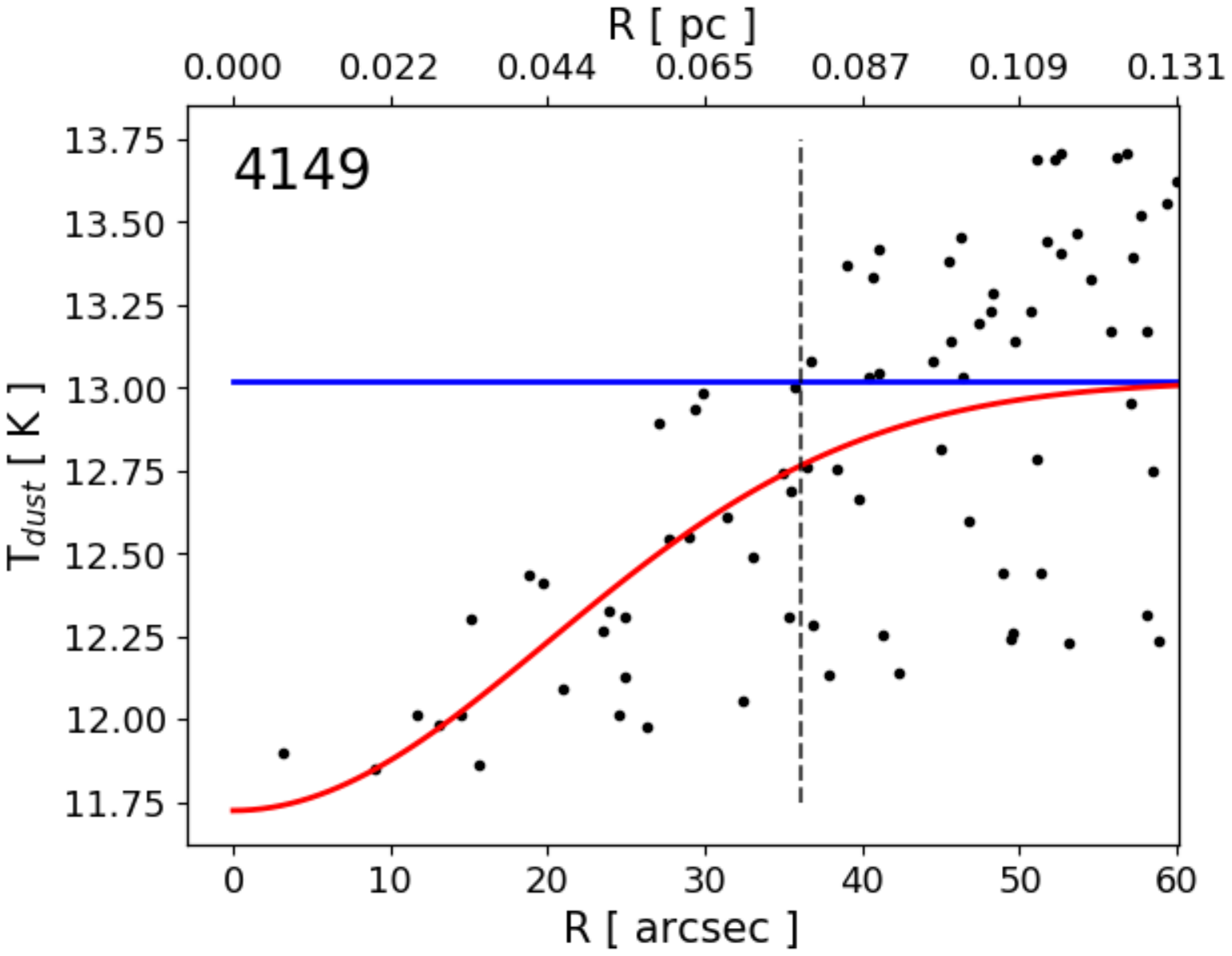}
    \caption{Dust temperature profiles of the three targeted sources (869, 1390, and 4149 from top to bottom). The black points are individual pixels in the dust temperature maps derived from Herschel data by \citet{Montillaud2015}, the blue lines show the average dust temperature of the points beyond the vertical dashed line, and the red lines are Gaussian fits.}
    \label{fig: T_profile}
\end{figure}

We have selected three of the best cold prestellar core candidates in the GCC catalogue of \citet{Montillaud2015}, namely sources 869, 1390, and 4149. They are shown in Fig.~\ref{fig: T&N} in white ellipses whose semi-major axes, semi-minor axes, position angles, and centre positions are adopted from \citet{Montillaud2015}. The column density and dust temperature maps derived from Herschel data are presented in Fig.~\ref{fig: T_profile}.
The selection criteria were as follows: To maximise the chances that the sources are genuine starless cores, we first examined the dust temperature profiles shown in Fig.~\ref{fig: T_profile}. In this figure, each pixel of the dust temperature map in the neighbourhood of the source is represented according to its distance to the source centre as given by \citet{Montillaud2015}. The surrounding temperature is provided by averaging $T_{\rm dust}$ outside the vertical dashed line, and the profile is fitted by a Gaussian function. The dip in temperature (defined by the difference between the surrounding temperature and the central temperature provided by the Gaussian fit) was required to be greater than 1\,K.
Using the colour criterion by \citet{Koenig2012} for WISE data, we also verified that no mid-IR sources classified as YSOs were  found in the area of the sources. Another selection criterion was the distance, which was chosen to be within 600 pc to avoid dilution in the IRAM 30 m beam, considering a typical core size of 0.1 pc. Finally, the three selected sources lie in different environments, as explained below.

The properties of the selected sources are listed in the Table~\ref{tab: cores}. We list the host cloud names, the coordinates, the distances, the dust temperatures, and the H$_2$ column densities of the three cores. 
The dust temperatures and H$_2$ column densities in the table are derived from an SED fitting of background subtracted \emph{Herschel} data. 
The dust temperatures and H$_2$ column densities are averaged over the whole source and therefore represent a lower limit for 
$N_{\mathrm{H}_2}$ and an upper limit for the dust temperature \citep{Montillaud2015}.

The source 869 is part of the MBM\,12 molecular cloud. The MBM\,12  molecular  cloud  was  catalogued  as  L1453-L1454, L1457, and L1458 in Lynds' Dark Cloud Catalog \citep{Lynds1962} and is one of the nearby high-latitude molecular clouds \citep{Magnani1985}.  

The source 1390 is part of the $\lambda$ Ori cloud, a roughly spherical shell known as the head of the Orion molecular complex, with an OB association in the centre of an HII region. Stars in this association are unbound because molecular gas is rapidly removed by a supernova explosion that occurred about 1 Myr ago \citep{Dolan2001}. 
The molecular clumps in the $\lambda$ Ori cloud show clear velocity and temperature gradients \citep{Liu2012,Liu2016,Goldsmith2016}, indicating external compression by the HII region. This suggests that the gas located in $\lambda$ Ori is greatly affected by stellar feedback. 
Our source 1390 is located at the inner side of the compressed shell, where star formation is probably triggered.

The source 4149 is located in the California molecular cloud (CMC). The CMC is so called due to its physical association with the well-known California nebula located at the cloud's southern border. The cloud is in an extended ($\sim80$ pc) and massive ($\sim10^5$ $M_\sun$) filamentary structure. The shape and mass of the CMC is similar to that of the Orion A molecular cloud (OMC). Compared to the OMC, the CMC displays significantly less star formation activity. The number of YSOs is more than an order of magnitude smaller in the CMC than in the OMC, suggesting that the star formation rate in the CMC might correspondingly be lower by an order of magnitude or more than in the OMC. The CMC is in an earlier evolutionary stage \citep{Lada2009}.

\begin{table*}
{\tiny
\caption{Parameters of the targeted sources. }
\label{tab: cores}
\centering
\begin{tabular}{c c c c c c c c c c c}
\hline \hline
         ID  &  Cloud  & Distance & RA  & DEC  & $T_\mathrm{dust}$ & $N_{\mathrm{H}_2}$ & Angular Size & Physical Size & Mass & $\alpha_{vir}$ \\
              &      &     (pc)     &      (2000)             &      (2000)                &            (K)                 & (cm$^{-2}$) & (arcsec) & (pc) & M$_\sun$  &\\
\hline
       869 & MBM12 & 252 $\pm$ 12 & 02:56:39.35 & 19:25:07.6 & 12.0 & $1.57 \times 10^{22}$ & 71.2 & 0.086 & 2.9 & 1.4 $\pm$ 0.2\\
       1390 & $\lambda$ Ori& 402 $\pm$ 20 & 05:52:15.88 & 08:23:00.3&  8.8 & $3.2 \times 10^{22}$ & 51.8 & 0.102 & 5.9 & 1.0 $\pm$ 0.1\\
       4149 & California & 470 $\pm$ 24 & 04:26:59.47 & 36:55:52.9& 7.5 & $1.02 \times 10^{22}$ & 46.0 & 0.104 & 2.4 & 0.7 $\pm$ 0.1\\
\hline
\end{tabular}
}
\tablefoot{
The sources are based on the Galactic cold core catalogue \citep{Montillaud2015}. The temperature and the H$_2$ column density in the table are averaged over the whole core. The angular size and physical size here have an equivalent diameter: $2 \times \sqrt{ab}$, where a and b are the semi-major axis and semi-minor axis. The core mass is estimated using Herschel data. The adopted distances are from \citet{Zucker2019} based on GAIA observations.
}
\end{table*}

Finally, we estimated the virial parameter \citep{Bertoldi1992},
\begin{equation}
\alpha_{\rm vir} = 5\sigma^{2}_{\rm tot} R/(GM_{\rm core})
,\end{equation}
where $G$ is the gravitational constant. The total velocity dispersion $\sigma_{\rm tot}$ includes both the thermal velocity dispersion ($\sigma_{\rm th}$) and the non-thermal velocity dispersion ($\sigma_{\rm NT}$), given by
\begin{equation}
        \sigma_{\rm tot}^2 = \sigma_{\rm th}^2 + \sigma_{\rm NT}^2, \\
              \sigma_{\rm th} = \sqrt{\frac{k_{\rm b} T_{\rm k}}{ \mu m_{\rm H}}},
\end{equation}
where $k_{\rm b}$ is the Boltzmann constant, $T_{\rm k}$ is the kinetic temperature, $\mu=2.33$ is the mean weight of the particle, and $m_{\rm H}$ is hydrogen mass. The non-thermal velocity dispersion is calculated following
\begin{equation}
        \sigma_{\rm NT}^2 = \sigma_{\rm species}^2 - \frac{k_{\rm b} T_{\rm k}}{\mu_{\rm species} m_{\rm H}},
\end{equation}
where $\sigma_{\rm species}$ is the velocity dispersion of an observed species, and $\mu_{\rm species}$ is its molecular weight.
We used methanol to determine the non-thermal velocity dispersion, so that the molecular weight  $\mu_{\rm species}=32$.
The derived $\alpha_{\rm vir}$ are 1.4,  1.0, and 0.7 for 869, 1390, and 4149, which are listed in Table~\ref{tab: cores} along with the core masses, which were estimated using Herschel data. The virial parameter is below the critical value of two for all three cores. It can be used to gauge whether a cloud fragment is subcritical or supercritical.  Supercritical fragments are bound or marginally gravitationally bound and can undergo collapse when perturbed. They can be characterised by a virial parameter lower than a critical value equal to 2, although magnetisation or external pressure can lead to different critical values \citep{Kauffmann2013,Evans2021}. Our methanol $\alpha_{\rm vir}$ values indicate that all three cores are gravitationally bound and might be collapsing (the gravitational binding energy is higher than the kinetic energy) if no additional support, such as from magnetic field, is present. The situation is not so clear if we take a factor of two of uncertainty on the mass of the cores.  

\section{Observations}
\label{sec: Obs}

We launched a pilot study focused on these three cores with the IRAM-30m telescope to explore their molecular content and abundances in cores with different evolutionary stages and different environments.
The pointed observations were performed in ten days from June 2018 to October 2018, and in four days in May 2019, using the 3 mm Eight Mixer Receivers (EMIR, 16 GHz of total instantaneous bandwidth per polarisation), and the fast Fourier transform  spectrometers (FTS)  with  a  spectral  resolution  of 50  kHz (0.18 km\,s$^{-1}$ at 83 GHz and 0.15 km\,s$^{-1}$ at 99 GHz), allowing  for  the  observation  of  the  inner  1.82  GHz of each of the four sub-bands. We used two setups to cover the gaps between the sub-bands. The  final frequency range is 79 to 86 GHz and 95 to 102 GHz.  

In order to maximise the integration time, we used the frequency-switching mode with a frequency throw of 7.14 MHz.
The total integration time of each source was between 4.7 and 5.5 hours. The weather was good, resulting in stable system temperatures ($\sim$90 K at the lower frequencies and $\sim$140 K at the higher frequencies). 

The  data  were  reduced using  the  IRAM  CLASS/GILDAS package\footnote{\href{http://www.iram.fr/IRAMFR/GILDAS}{http://www.iram.fr/IRAMFR/GILDAS}}. 
The telescope and receiver parameters (main-beam  efficiency,  half -power  beam  width,  and forward  efficiency) were adopted from the IRAM website\footnote{\href{http://www.iram.fr}{http://www.iram.fr}}: at 86 GHz, $\eta_{mb}$=81\%, $\eta_{ff}$=95\%, $\theta$=28.6$^{\prime\prime}$, and at 115 GHz, $\eta_{mb}$=78\%, $\eta_{ff}$=94\%, $\theta$=21.4$^{\prime\prime}$.

\begin{figure*}
    \centering
    \includegraphics[width=0.4\textwidth,trim=100 10 0 20 clip]{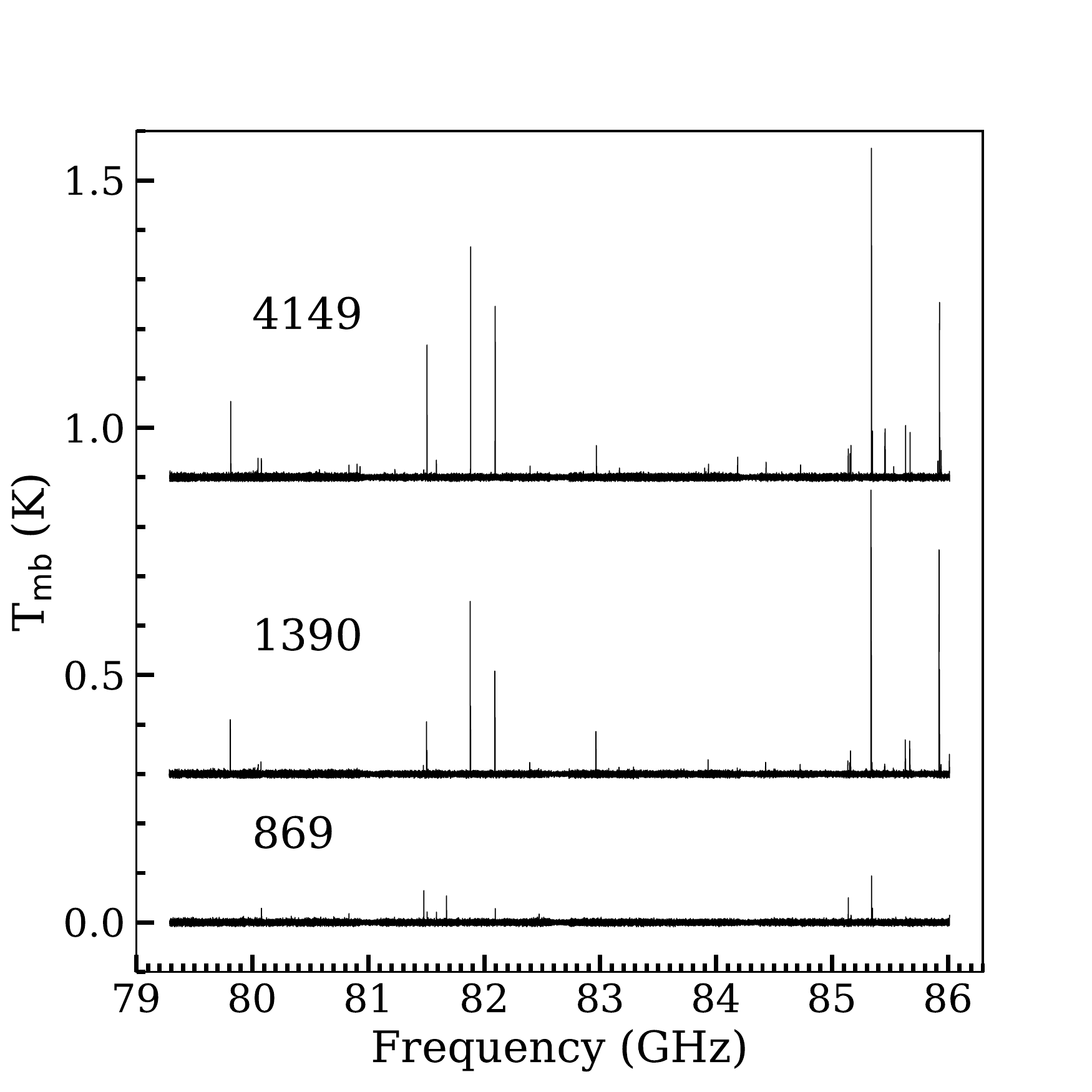} \ 
    \includegraphics[width=0.4\textwidth,trim=0 10 100 20 clip]{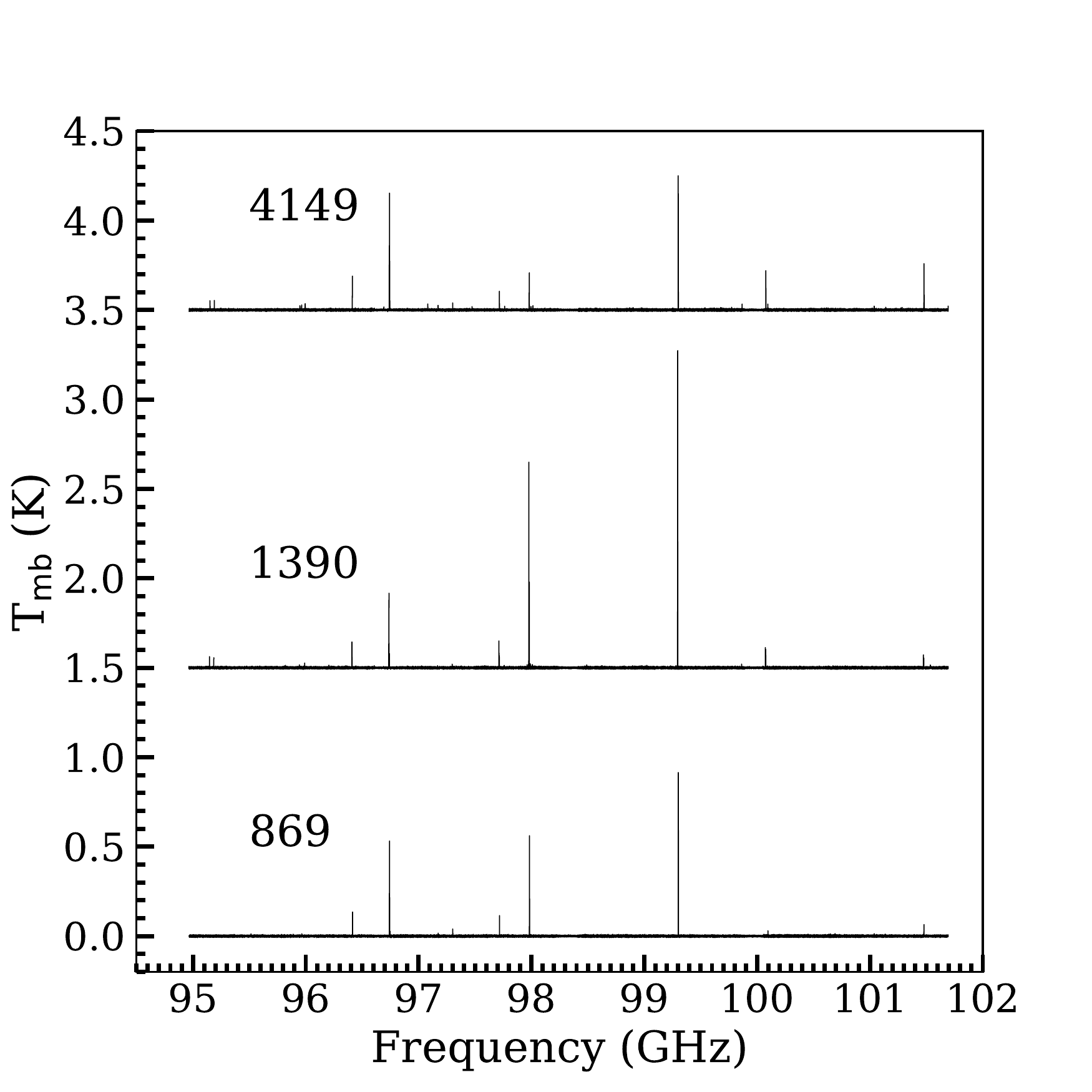} \ 
    \caption{Final baseline-subtracted spectrum in main-beam temperature of each core after the frequency-switching absorptions are removed (see Sec. \ref{sec: Obs}). Left: 869, 1390+0.3K, and 4149+0.9K. Right: 860, 1390+1.5K, and 4149+3.5K. }
    \label{fig: spec}
\end{figure*}

The frequency-switching technique was used to optimise the telescope time, but unfortunately, it results in poor baselines over the sub-bands. In order to obtain flat baselines, we first cropped the spectrum into spectra with a smaller frequency range ($\sim 5$\,MHz) adapted to avoid molecular transitions located at the edges. Then, we made masks for the identified lines where the intensities are brighter than four times the rms and fitted the baseline with polynomials of order between 3 and 5. The baseline-subtracted spectra were checked again to avoid missing some weak lines  in baseline subtraction. We then replaced the symmetric absorption artefacts with the noise interpolated on either side. Finally, we stitched all the spectra into a single spectrum for each core (shown in Fig.~\ref{fig: spec}). These spectra were used for line identification and further analysis. 
The noise was computed around each line within 20 km\,s$^{-1}$ and estimated to be 3 mK in the [79--86] GHz range and 4 mK in the [95--102] GHz range for each source.

\section{Results}
\label{sec: Result}

\subsection{Line identification}
\label{subsec: LineID}

\begin{figure}
    \centering
    \includegraphics[width = 0.45\textwidth]{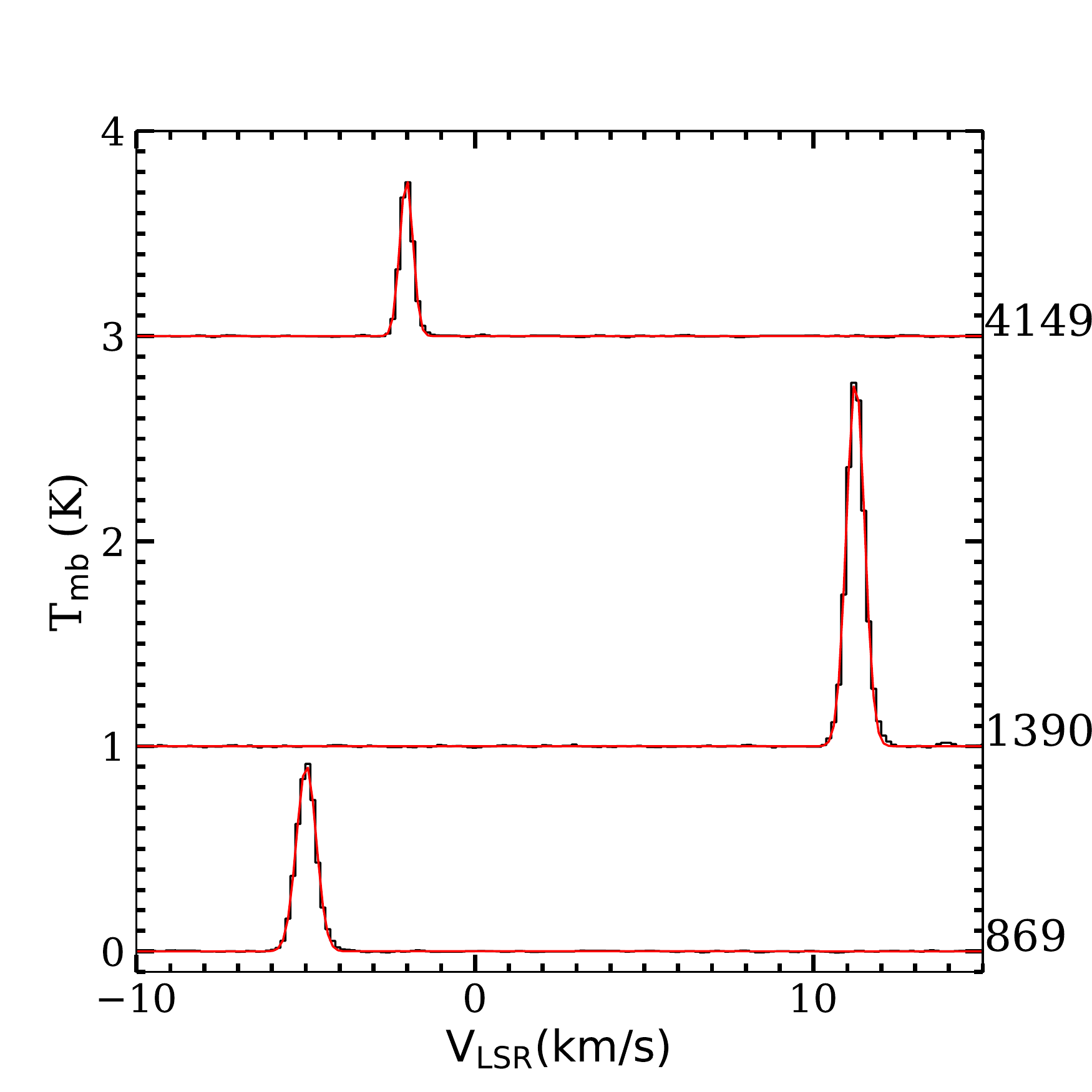}
    \caption{SO $2_3$ -- $1_2$ transition at 99.3\,GHz of each core. The observation line (black) is overlaid by Gaussian fitting (red): $\rm V_{LSR}$(869) = -4.98 km~s$^{-1}$, $\rm V_{LSR}$(1390) = 11.24 km~s$^{-1}$, $\rm V_{LSR}$(4149) = -2.03 km~s$^{-1}$ (uncertainty of $\sim$ 0.02 km~s$^{-1}$). }
    \label{fig:SO}
\end{figure}

The line identification was performed using the  CASSIS\footnote{\href{http://cassis.irap.omp.eu }{http://cassis.irap.omp.eu }} software, which connects to the JPL and CDMS databases. This is a standalone software, written in Java, and is freely delivered to the community to help with visualising, analysing, and modelling observations from ground- or space-based observatories. It has been developed at IRAP since 2005 and is part of the OVGSO\footnote{\href{https://ov-gso.irap.omp.eu/}{https://ov-gso.irap.omp.eu/}} data centre, which aims at promoting the Virtual Observatory technology. CASSIS displays any spectra (ASCII, FITS or GILDAS/CLASS format or the result from the query to any SSAP VO or EPN-TAP VO service from IVOA registries) and identifies atomic and molecular species through its link to the databases, such as CDMS \citep{MULLER2005} or JPL \citep{PICKETT1998}, via a SQLite database, or via a direct access to VAMDC\footnote{\href{http://www.vamdc.org/}{http://www.vamdc.org/}} for any available spectroscopic database. CASSIS also provides ortho, para, A and E forms for some species (provided on the CASSIS website, Catalogs section), which is necessary to discern when the cold interstellar medium is traced. 
The first step towards line identification is to measure the radial velocity in the local standard of rest (LSR) for the three cores. To do this, we used the methanol triplet at $\sim$ 96.7 GHz as well as the strong SO line at 99.3 GHz. We present the latter in Fig.~\ref{fig:SO}, where the observation line (black) is overlaid with the best-fit adjustment (red). The $V_\mathrm{LSR}$ is measured from the results of a single-Gaussian fit using the Levenberg-Marquardt fitting within CASSIS and many iterations:  -5.0 km~s$^{-1}$ for 869, 11.2 km~s$^{-1}$ for 1390, and -2.0 km~s$^{-1}$ for 4149. 

The spectral surveys obtained for our three cores are broad enough to cover many transitions for the same species, and non-detections are crucial to confirm the line identification through LTE modelling in the temperature range deduced from molecular species for which we have the collision coefficients (e.g. methanol). Because the cores are cold, we first searched for transitions in which the upper energy levels are lower than 50 K. Some species present a very large number of transitions. CH$_{3}$CHO, for example, presents 36 transitions in the frequency range of our observations. Another selection, using the Einstein coefficient values, was used to reduce the number of potential transitions, using $\rm A_{ij}$ larger than 10$^{-6}$ s$^{-1}$. The detection criterion was fixed at 4 $\sigma$, and the detection was confirmed using a simple LTE model assuming a temperature in the range of 10-15 K to verify whether the model also agrees with the non-detected transitions. For stronger transitions, we verified whether negative lines were caused by frequency-switching observation on either side of the lines at 7.14 MHz from the line centre, symmetric and representative of the flux of the line in emission.

With this strategy, we also detected a ghost line in 869 at 81.70 GHz that is produced by a strong line in the image sideband, for which the 10 dB rejection is not enough. In our case, this ghost line stems from the strong SO $2_{3} -1_{2}$ transition at 99.29987 GHz (2$\nu_\mathrm{LO} - 81.70$ GHz, $\nu_\mathrm{LO}=90.5$ GHz). The strength of the line (which only exists in the H polarisation sub-band) is about 54 mK, reaching less than 5\% of the strength of the original line. 

We present the detected species and transitions in all three cores in Table~\ref{tab: lines} and their observation parameters in Table~\ref{tab: lines_para}. The noise was measured over a range of 20 km\,s$^{-1}$ around the central velocity of the line (the zone excluding the negative lines due to the frequency throw), and the line widths and central velocities were derived from Gaussian fitting. The FWHMs of the measured transitions are compatible with what is expected for a cold medium ([0.3-0.6] km~s$^{-1}$), taking the spectral resolution of [0.15-0.18] km~s$^{-1}$ in the observed frequency range into account. 


The overall error on the line flux is the combination of the statistical and calibration error. The calibration uncertainty is estimated to be $10\%$ given by the IRAM website. The uncertainty ($\Delta$W) of the integrated area is therefore computed through the following formula: 
\begin{equation}
\rm \Delta W = \sqrt{(\rm \alpha \times W)^2 + (rms \sqrt{2 \times \rm FWHM \times \delta V})^2}
\label{eq: error}
,\end{equation}
where $\alpha$ is the calibration value (0.10), W is the integrated area (in K~km~s$^{-1}$), rms is the noise around the selected species (in K), FWHM is the full width at half maximum (km~s$^{-1}$), and $\delta$V is the bin size (in km~s$^{-1}$). We assumed that the number of channels in the line is 2 $\times$ FWHM/$\delta$V. \\
For undetected transitions, we estimated the upper limit of 3$\sigma$,
\begin{equation}
    \rm W \ (\mathrm{K~km~s^{-1}}) \le  3(rms \times FWHM) \times \sqrt{(\rm 2\times\alpha)^2 + (2\times\delta V/FWHM)}
    \label{eq: upper_limit}
,\end{equation}
using FWHM as the methanol value of 0.5 km~s$^{-1}$ for 869, 0.6 km~s$^{-1}$ for 1390, and 0.4 km~s$^{-1}$ for 4149.
The integrated intensities and their uncertainties as well as the upper limits are listed in Table~\ref{tab: lines_para}.

\subsection{Molecular complexity}
\label{subsec: Line}

\begin{figure*}
\center
\includegraphics[width=0.32\textwidth]{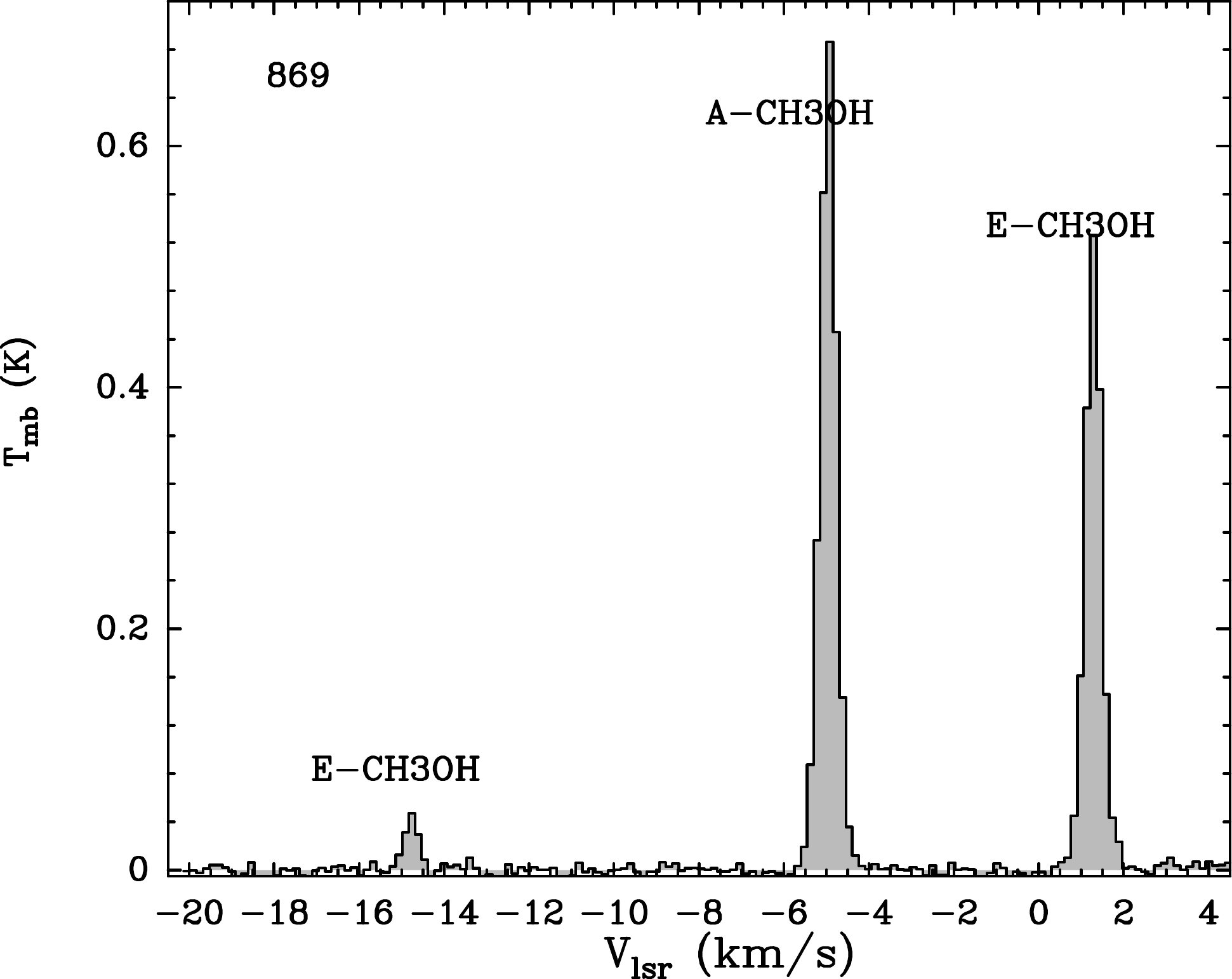}
\includegraphics[width=0.32\textwidth]{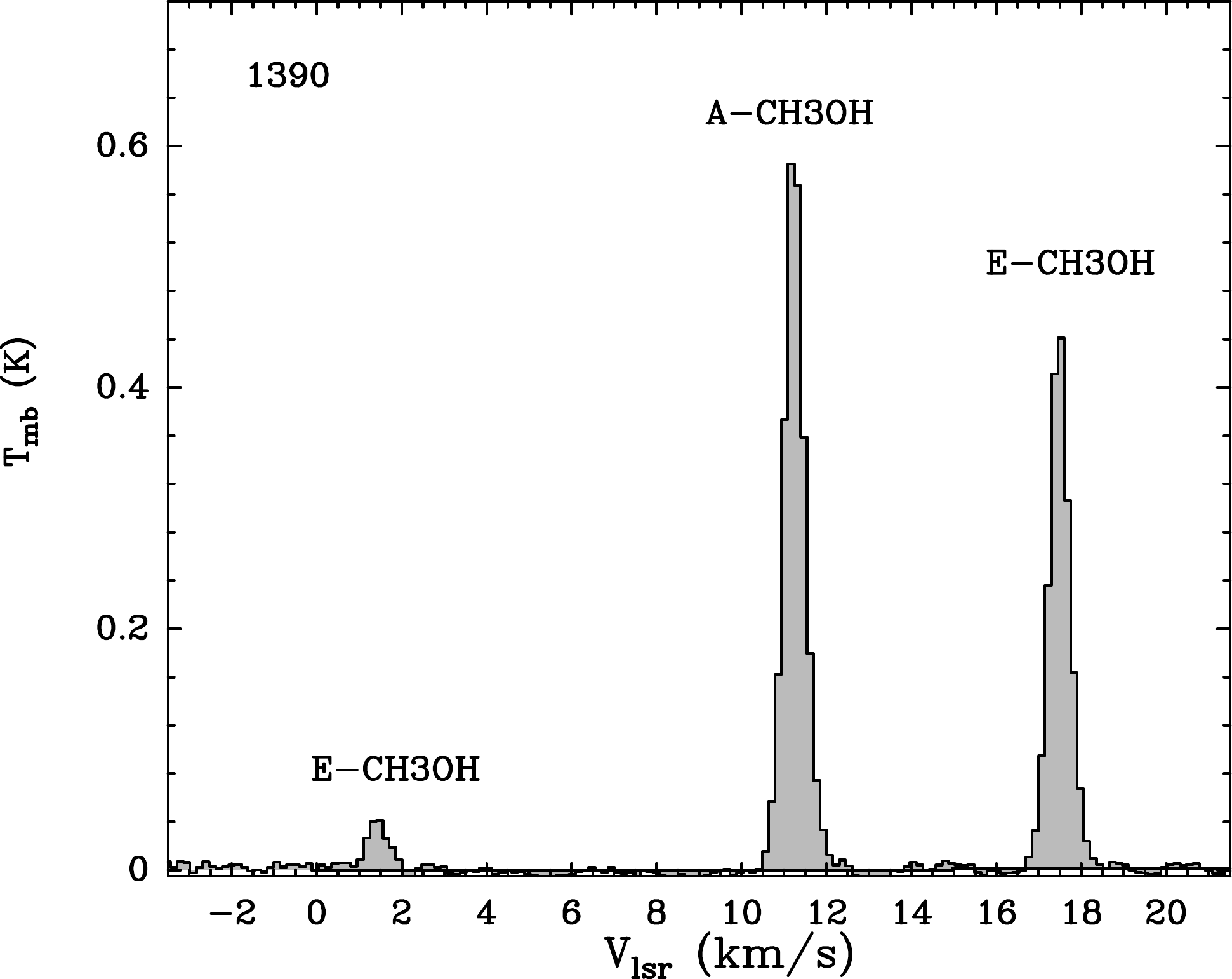}
\includegraphics[width=0.32\textwidth]{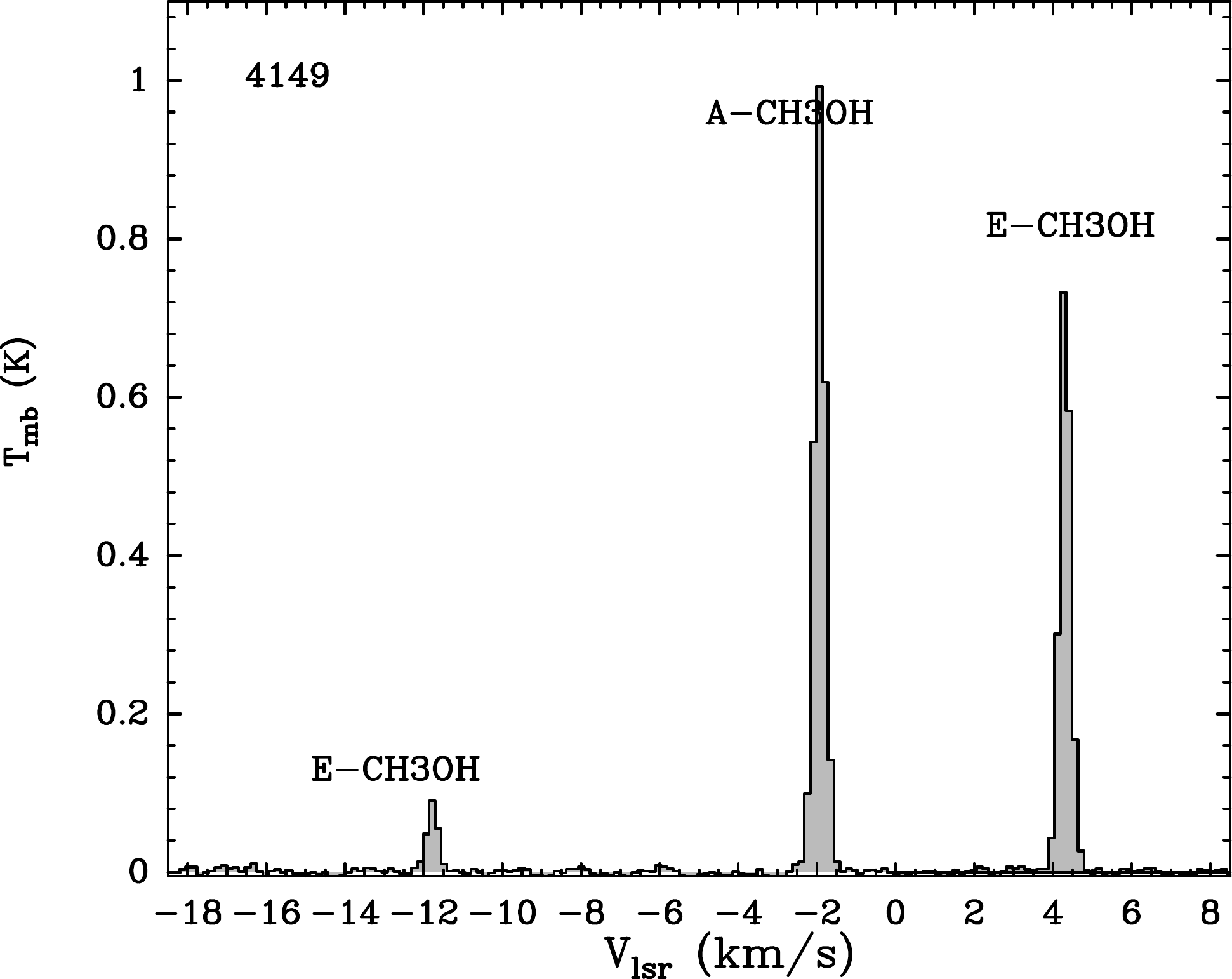}
\caption{Observations of the 2$_k$ $\rightarrow$ 1$_k$ methanol triplet toward 869 (left), 1390 (middle), and 4149 (right). The rest frequency for these plots is defined as the strongest transition, the A$^+$ line at 96.741 GHz. The E transition on the left side is 2$_0$$-$1$_0$ at 96.744 GHz, and the E transition on the right side is 2$_{-1}$$-$1$_{-1}$ at 96.739 GHz}
\label{fig: CH3OH}
\end{figure*}

We find large differences among the three sources: 21, 29, and 37 species are detected in 869, 1390, and 4149 with 29, 52, and 70 transitions, respectively.
In 869, we only detect four oxygen-bearing COMs, A/E-CH$_3$CHO and A/E-CH$_3$OH, and do not detect any deuterium-bearing molecules or nitrogen-bearing molecules, except for HNO. No carbon-chain species were detected, except for c-C$_3$H$_2$.
Conversely, we identified some deuterated molecules (C$_3$HD, DC$_3$N, D$_2$CS, NH$_2$D, etc.) in both 1390 and 4149. 
Carbon-chain molecules (HC$_3$N, CH$_{3}$CCH, C$_{3}$H, C$_4$H, C$_{3}$O, c-C$_{3}$H$_{2}$, and l-C$_{3}$H$_{2}$) are also detected towards these two cores.
Most of the sulphur molecules are detected in the three cores, although NS$^+$ is only detected in 869.
The greatest molecular diversity is observed in 4149, indicating that the three cores are in different evolutionary stages.
We also detect methyl mercaptan (CH$_3$SH) and methoxy (CH$_3$O) in 869 and 4149, but no detection in 1390 within $3\sigma$. The three detected CH$_3$OH lines are presented in Fig.~\ref{fig: CH3OH} for the three cores. The rest frequency is that of A-CH$_3$OH at 96741.37 MHz. 

\subsection{LTE and non-LTE analysis}
\label{subsec: analyses}

In this section we describe the method used for the LTE and non-LTE analysis of the detected species. We took into account the detected lines as well as the upper limits for the non-detected lines found in the frequency range of our spectral survey. Based on this analysis, we compute the column densities and give an estimate of the kinetic temperature and H$_{2}$ volume density.

\subsubsection{Methanol and ketene}

First, we performed the analysis for the species (H$_2$CCO, CH$_3$OH) where multiple transitions ($\geq 3$) were detected, covering at least 10 K in energy range.
CH$_3$OH is the most important species in our study because we detected at least three transitions with upper energy levels of $\rm E_{up}$ = 5, 7, 12, 22, and 32 K, with some upper limits at 84.5, 95.9, 96.8, and 97.6 GHz (where $\rm E_{up}$ is lower than 50 K) and for which the collision rates are available down to 10 K. Therefore, methanol is a very good probe to determine the density and the kinetic temperature of the medium explored. 
We used a non-LTE analysis based on the (at least) three detected transitions and the upper limits in our spectral survey assuming an A/E ratio of unity. 
We varied the kinetic temperature, column density, and H$_2$ volume density to make the models compatible with $T_\mathrm{mb}$ with the calibration uncertainty and statistical uncertainty of our detected transitions as well as with our undetected transitions. To do this, we used the collision rates for CH$_3$OH with para-H$_2$ computed by \citet{Rabli2010}, which are provided in the catalogue section of the CASSIS website. 
In addition, for the cold interstellar medium, we assumed the ortho-H$_2$ and para-H$_2$ ratio as 0.001 from \citet{Pagani2013}.
Unfortunately, the collision rate is computed with the range of T$_k$ from 10 to 200 K, so we cannot trace the temperatures lower than 10 K.
The results we obtained from this method are presented in Table~\ref{tab: analy869} (869), Table~\ref{tab: analy1390} (1390), and Table~\ref{tab: analy4149} (4149). The ranges correspond to the observed $T_\mathrm{mb}$ uncertainty including the calibration uncertainty and statistical uncertainty. The abundances were calculated using the H$_2$ column densities from the SED fitting results of \citet{Montillaud2015}.
However, it is important to emphasise that the H$_2$ column densities are averaged values within the area of the {\it Herschel} compact sources. We acknowledge the uncertainty of dust absorption cross sections, which could be up to a factor of two in these N(H$_{2}$) reference values in Table~\ref{tab: cores}. Because the SED fitting was performed using the Herschel fluxes of the whole elliptical area of the source (as presented in Fig.~\ref{fig: T&N} and Table~\ref{tab: cores}), which is larger than the IRAM-30m telescope beam size (29$^{\prime\prime}$), the obtained abundances are potentially biased towards higher values. We can also compare the H$_{2}$ volume density of the gas where methanol is emitted with the average value in the core, using N(H$_{2}$)/size: n(H$_{2}$) =  $9.1 \times 10^{4}$, $1.85 \times 10^{5}$ and $6.6 \times 10^{4}$ cm$^{-3}$ for 869, 1390, and 4149, respectively. Source 1390 is the densest of all cores.\\

\begin{figure*}
\center
\includegraphics[width=0.8\textwidth]{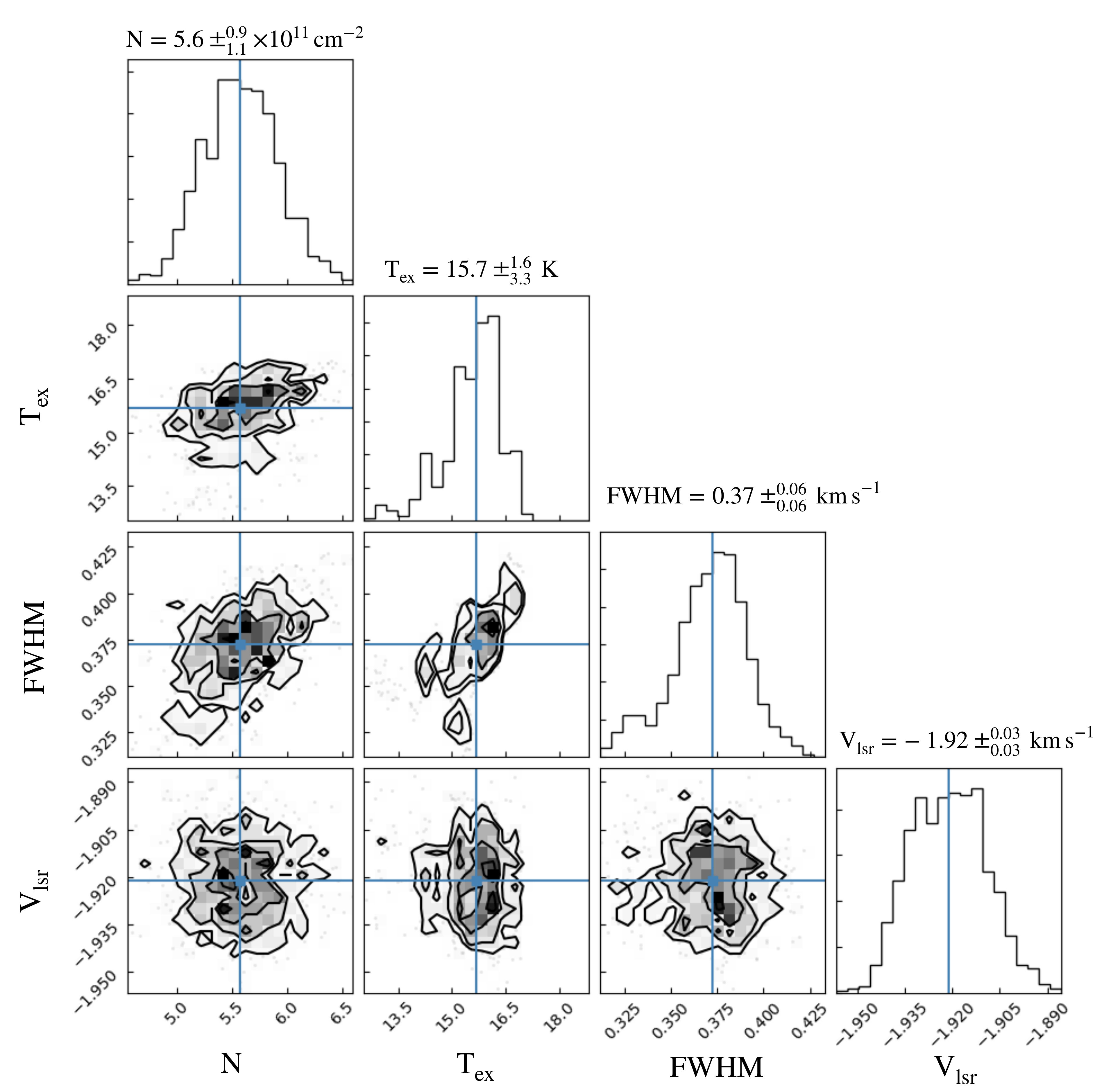} 

\includegraphics[width=0.8\textwidth]{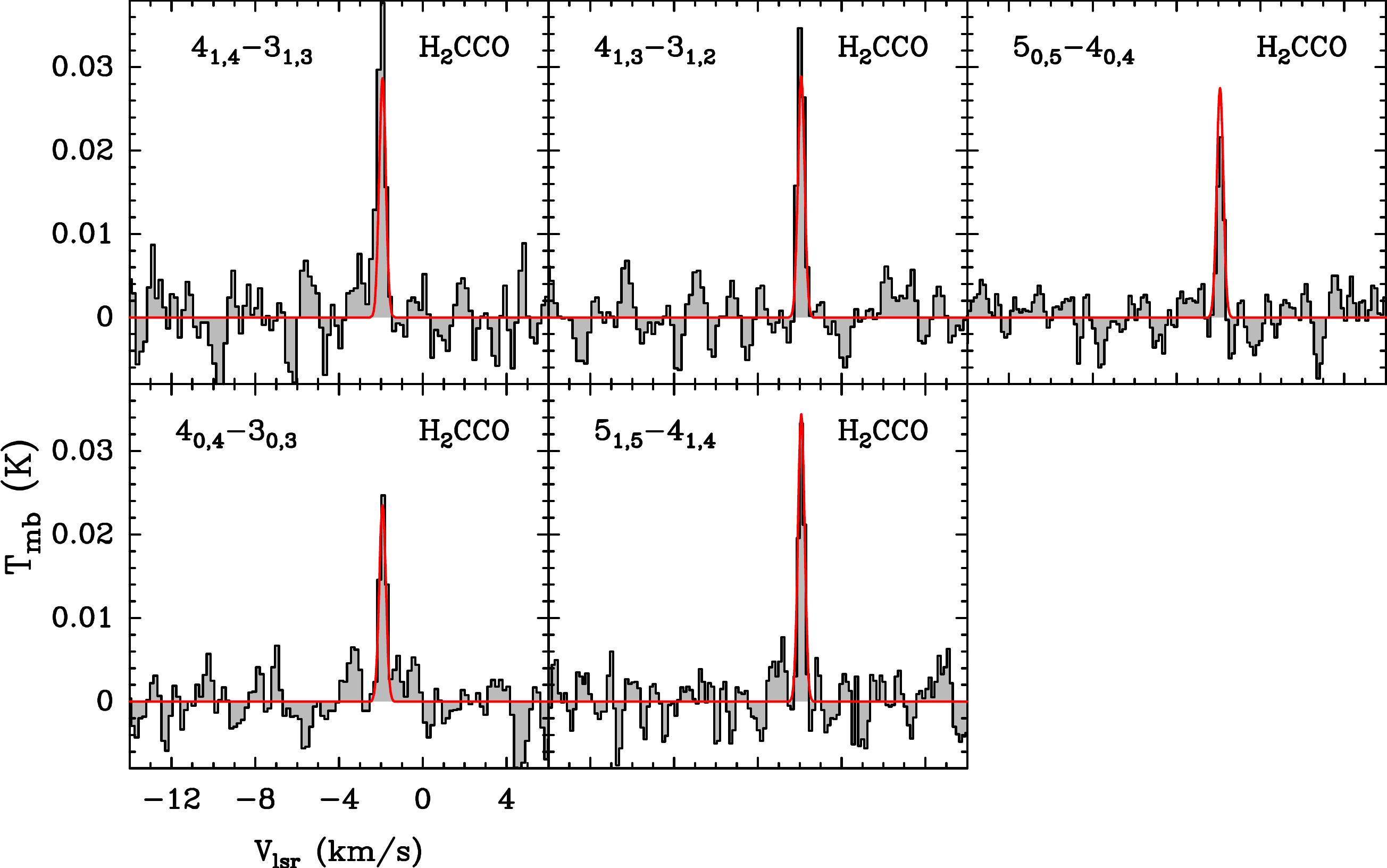}
\caption{MCMC corner plot produced by CASSIS of the fitting results of H$_2$CCO for source 4149 (fitting uncertainties of 3$\sigma;$ top). Modelled lines from the MCMC computation (red) for each transition (bottom).}
\label{fig: coner_plot}
\end{figure*}

At least four transitions of H$_{2}$CCO have been clearly detected with upper energy levels of $\rm E_{up}$ = 10, 14, 23, and 27 K. We used the Markov chain Monte Carlo (MCMC) method provided within CASSIS for the LTE analysis alone (no collision coefficients are available to our knowledge) and assumed that the sources fill the beam. 
The MCMC method is an iterative process that explores the parameter space with random walks in order to find the optimum parameters. 
A calibration uncertainty of 10\% is considered in the process.
All fitted parameters such as column density, excitation temperature (or kinetic temperature and H$_2$ volume density in the case of a non-LTE analysis), line width, and velocity can be varied to find the best-fit model. 
The computed results are also presented in Table~\ref{tab: analy869} (869), Table~\ref{tab: analy1390} (1390), and Table~\ref{tab: analy4149} (4149) with 3$\sigma$ uncertainty ranges.
Figure \ref{fig: coner_plot} shows the corner plot produced by CASSIS of the MCMC fitting results for H$_2$CCO in 4149. The plot shows the statistics of all the models, and the uncertainties in the plot are in 3$\sigma$. 
We show the modelled transitions for H$_2$CCO at the bottom of Fig.~\ref{fig: coner_plot} using the resulting MCMC parameters (N = 5.6 $\times$ 10$^{11}$ cm$^{-2}$, $\rm T_{ex}$ = 15.7 K, FWHM = 0.37 km~s$^{-1}$, and $\rm V_{LSR}$ = -1.92 km~s$^{-1}$). The excitation temperature is consistent within the error bars with the kinetic temperature found for methanol using the non-LTE analysis. The column densities and excitation temperatures are quoted in Table~\ref{tab: analy869} , Table~\ref{tab: analy1390}, and Table~\ref{tab: analy4149} for 869, 1390, and 4149, respectively. 

\subsubsection{Other species}

\begin{figure}
    \centering
    \includegraphics[width=0.45\textwidth]{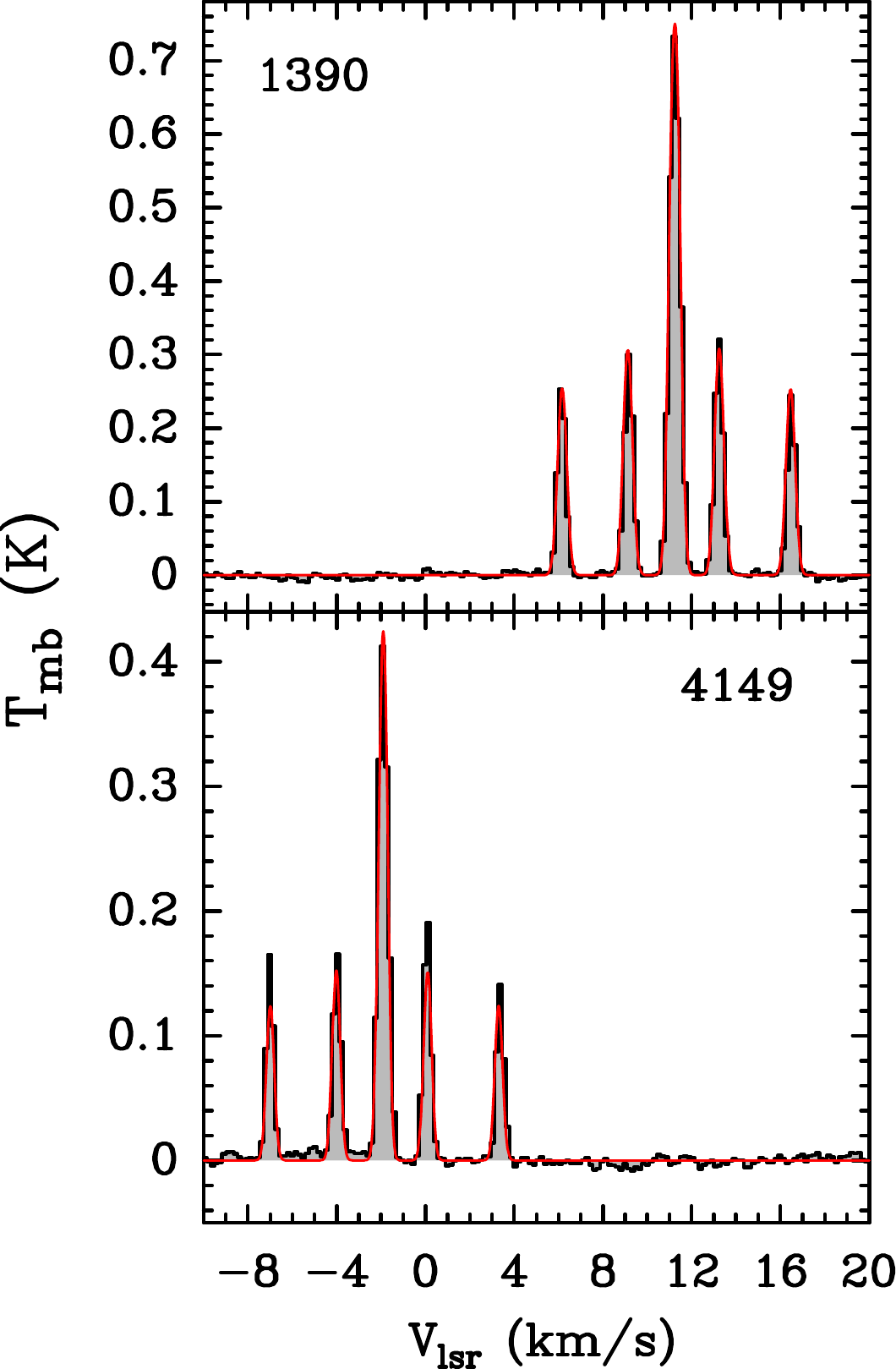} 
    \caption{Line profile of the ortho NH$_{2}$D transition at 86 GHz towards 1390 and 4149. The velocity is built relative to the 85926.2703 MHz transition. The best fit is plotted in red from the MCMC analysis. Core 4149: N(o-NH$_{2}$D) = 1.6 $\times$ 10$^{14}$ cm$^{-2}$, $\rm T_{ex}$ = 3.8 K, FWHM=0.41 km~s$^{-1}$, and $\rm V_{LSR}$ = -1.86 km~s$^{-1}$; core 1390: N(o-NH$_{2}$D) = 1.2 $\times$ 10$^{14}$ cm$^{-2}$, $\rm T_{ex}$ = 4.5 K, FWHM=0.47 km~s$^{-1}$, and $\rm V_{LSR}$ = 11.27 km~s$^{-1}$.}
    \label{fig: NH2D}
\end{figure}

For other species with fewer than three transitions detected or species with more than three transitions that cover a narrow energy range, we either conducted a non-LTE analysis when the collision coefficients were available, with T$_{k}$ and n(H$_2$) determined from the methanol transitions (see previous subsection), or an LTE analysis using the kinetic temperature range determined from methanol.  We varied the column densities to determine whether the models agree with the detected transitions and with undetected transitions, so that we could determine the best-fit range of column densities and abundances for these species. The uncertainty of the integrated intensity is taken into account as well.
We adopted the collision rates for HC$_3$N with ortho-/para-H$_2$ from  \citet{Faure2016}, HC$^{18}$O$^+$ with H$_2$ from \citet{Flower1999}, HCS$^+$ with H$_2$ from \citet{Flower1999}, CS with ortho-H$_2$ and para-H$_2$ from \citet{Denis-Alpizar2013}, SO with para-H$_2$ from \citet{Lique2007}, and OCS with H$_2$ from \citet{Green1978ApJS}. They are also provided on the CASSIS website, catalogue section, with temperatures down to at least 10 K. We also used the  hyperfine patterns of the ortho-NH$_{2}$D molecule (which has a high critical density of about $7 \times 10^{4}$ cm$^{-3}$) to estimate the excitation temperature and opacity (and therefore the column density). For this hyperfine structure (HFS) analysis, we used the CASSIS catalogue with the ortho and para separation using the CDMS catalogue. The results of Gaussian fits to the HFS give a V$_{LSR}$ of -1.86 km~s$^{-1}$ (resp. 11.27 km~s$^{-1}$) and an FWHM of 0.41 $\pm$ 0.01 km~s$^{-1}$ (0.47 $\pm$ 0.01 km~s$^{-1}$) for 4149 (1390). The hyperfine fit also gives an estimate for the $\rm T_{ex}$ of the transition (lower limit of the gas kinetic temperature) and column density quoted in Tables \ref{tab: analy1390} and \ref{tab: analy4149}. The derived total optical thicknesses of the detected lines are about unity, which means that the satellites are mainly optically thin and that we are in the regime where the integrated intensity is proportional to the column density of the molecule. The column densities were computed assuming that the source fills the telescope beam uniformly. However, NH$_{2}$D should be strongly concentrated in the centre of the core for chemical reasons (see Sec.~\ref{sec: deut}), and therefore is probably subject to potential dilution in the IRAM 30m beam. The column densities can therefore be considered upper limits. The ortho-NH$_{2}$D spectra and their best fit from the MCMC adjustment are presented in Fig.~\ref{fig: NH2D}. All the results are presented in Table~\ref{tab: analy869}, Table~\ref{tab: analy1390}, and Table~\ref{tab: analy4149}.\\

The presence of isotopologues is a useful probe of the validity of the column density with an estimate of the optical depth of the transitions. For example, CS and SO are usually optically thick in the observed transitions. In the ISM, \citet{Chin1996} found a relation between $^{32}$S/$^{34}$S isotope ratios and their galactocentric distance (DGC) of $^{32}$S/$^{34}$S=(3.3$\pm$0.5)(DGC/kpc)+(4.1$\pm$3.1) ($\sim$ 32$\pm$5 for our three cores), while no correlation was obtained between $^{34}$S/$^{33}$S ratios (6$\pm$1) and DGC. 
From Table~\ref{tab: analy869}, Table~\ref{tab: analy1390}, and Table~\ref{tab: analy4149}, the computed $^{32}$S/$^{34}$S ratio for SO is [9-60], [6-58], [7-29] for 869, 1390, and 4149, respectively, while for CS, we obtain [8-87], [7-125], and [2.4-12.7] for 869, 1390, and 4149, respectively. The C$^{32}$S/C$^{34}$S ratio for 4149 is lower than the value in the vicinity of the Sun \citep{Chin1996}, which prevents the determination of the CS column density. 
We therefore took the $^{34}$S isotopologues and $^{32}$S/$^{34}$S ratio for both CS and SO into account: N(CS$_{869}$)=[8.1--17.8] $\times$ 10$^{12}$ cm$^{-2}$, N(CS$_{1390}$)=[5.4--18.5] $\times$ 10$^{12}$ cm$^{-2}$, N(CS$_{4149}$)=[8.1--16.7]$\times$ 10$^{12}$ cm$^{-2}$, N(SO$_{869}$)=[13.5--33.3] $\times$ 10$^{12}$ cm$^{-2}$, N(SO$_{1390}$)=[16--55] $\times$ 10$^{12}$ cm$^{-2}$, N(SO$_{4149}$)= [12--24] $\times$ 10$^{12}$ cm$^{-2}$. These values are quoted in Table~\ref{tab: analy869}, Table~\ref{tab: analy1390}, and Table~\ref{tab: analy4149}. Additionally, in the case of 1390, we used the detection of one transition of $^{33}$SO, resulting in an observed $^{34}$S/$^{33}$S ratio of [1.2--6.0]. This is compatible with the value in the vicinity of the Sun. The total column density of HCO$^+$ was computed from the HC$^{18}$O isotopologue using the ISM $^{16}$O/$^{18}$O ratio of 557 taken from \citet{Wilson1999RPPh}.

\begin{table*}
\centering
\caption{LTE and non-LTE analysis results for core 869. }
\label{tab: analy869}
\begin{tabular}{*7c}
\hline \hline
Species & T$_\mathrm{ex}$ & T$_\mathrm{k}$ &  n$_\mathrm{H_2}$ & N & $X$\tablefootmark{a}\\
& (K) & (K) &(cm$^{-3}$) & (cm$^{-2}$) &  & \\
\hline
\multicolumn{7}{c}{LTE}\\
\hline
H$_2$CCO & $13 - 14$ & & &  $5 - 6\times10^{11}$ & $3.2 - 3.8\times10^{-11}$ \\
HNO & $10 - 14$\tablefootmark{b} & & & $8 - 16\times10^{11}$ & $5.1 - 10.2\times10^{-11}$ \\
CCS & $10 - 14$\tablefootmark{b} & & & $10 - 15\times10^{10}$ & $6.4 - 9.5\times10^{-12}$\\
o-c-C$_3$H$_2$  & $10 - 14$\tablefootmark{b} &  &  & $3 - 5\times10^{11}$ & $1.9 - 3.2\times10^{-11}$\\
p-c-C$_3$H$_2$ & $10 - 14$\tablefootmark{b} &  &  & $10 - 14\times10^{10}$ & $6.4 - 8.9\times10^{-12}$\\
CH$_3$O & $10 - 14$\tablefootmark{b} & & & $4 - 6\times10^{11}$ & $2.6 - 3.8\times10^{-11}$\\
OCS & $10 - 14$\tablefootmark{b} & & & $2.5 - 3.5\times10^{12}$ & $1.6 - 2.2\times10^{-10}$ \\
C$_4$H & $10 - 14 $\tablefootmark{b} & & & $9 - 13\times10^{11}$ & $5.7 - 8.3\times10^{-11}$ \\
E-CH$_3$CHO & $10 - 14$\tablefootmark{b}  & & & $6 - 11\times10^{10}$ & $3.8 - 7.0\times10^{-12}$ \\
A-CH$_3$CHO & $10 - 14$\tablefootmark{b} & & & $6 - 11\times10^{10}$ & $3.8 - 7.0\times10^{-12}$ \\
NS$^+$ & $10 - 14$\tablefootmark{b}  & & & $1.5 - 2.2\times10^{10}$ & $9.6 - 14.1\times10^{-13}$\\
CH$_3$SH & $10 - 14$\tablefootmark{b}  & & & $1.5 - 3.5\times10^{11}$ & $1.0 - 2.2\times10^{-11}$\\
o-H$_2$CS &$10 - 14$\tablefootmark{b} & & & $5 - 7\times10^{11}$ & $3.2 - 4.5\times10^{-11}$ \\
\hline
\multicolumn{7}{c}{non-LTE}\\
\hline
A/E-CH$_3$OH & & $10 - 14$ & $6 - 30\times10^3$ & $7 - 20\times10^{12}$ &  $5.1 - 12.7\times10^{-10}$\\
HC$^{18}$O$^+$ & & $10 - 14$\tablefootmark{b} & $6 - 30\times10^3$ & $1.2 - 6.5\times10^{10}$ &  $7.6 - 41.4\times10^{-13}$\\
HCS$^+$ & & $10 - 14$\tablefootmark{b} & $6 - 30\times10^3$ & $7 - 14\times10^{10}$ & $4.5 - 8.3\times10^{-12}$\\
CS & & $10 - 14$\tablefootmark{b} & $6 - 30\times10^3$ & $4 - 26\times10^{12}$ &  $2.5 - 16.6\times10^{-10}$ \\
C$^{34}$S & & $10 - 14$\tablefootmark{b} & $6 - 30\times10^3$ & $3 - 4.8\times10^{11}$ & $1.9 - 3.1\times10^{-11}$ \\ 
C$^{34}$S$\times ^{32}$S$/^{34}$S\tablefootmark{c} & &  &  & $8.1 - 17.8\times10^{12}$ & $5.2 - 11.3\times10^{-10}$ \\ 
SO & & $10 - 14$\tablefootmark{b} & $6 - 30\times10^3$ & $8 - 30\times10^{12}$ & $4.5 - 19.1\times10^{-10}$ \\
$^{34}$SO & & $10 - 14$\tablefootmark{b} & $6 - 30\times10^3$ & $5 - 9\times10^{11}$ & $3.2 - 5.7\times10^{-11}$\\ 
$^{34}$SO$\times ^{32}$S$/^{34}$S\tablefootmark{c} & &  &  & $13.5 - 33.3\times10^{12}$ & $8.6 - 21.2\times10^{-10}$ \\
OCS & & $10 - 14$\tablefootmark{b} & $6 - 30\times10^3$ & $2.5 - 4.0\times10^{12}$ &  $1.6 - 2.6\times10^{-10}$\\
\hline
\end{tabular}

\tablefoot{\\
The quoted error is the 3$\sigma$ (statistical and calibration) error from the fitting.\\
\tablefoottext{a}{Abundances derived using the H$_2$ column density N$(\mathrm{H}_2)=1.57\times10^{22}$ cm$^{-2}$, adopted from \citet{Montillaud2015}. The value is the average of the source ($r_\mathrm{eff} = 36\arcsec$).} \\
\tablefoottext{b}{The temperature (and the volume density) was fixed to results derived from methanol.} \\
\tablefoottext{c}{$^{32}$S$/^{34}$S$=(3.3 \pm 0.5)\,$(DGC/kpc)$+4.1 \pm 3.1  (\sim 32 \pm 5$ for our cores), adopted from \citet{Chin1996}. }
}
\end{table*}

\begin{table*}
\centering
\caption{LTE and non-LTE analysis results for core 1390. }
\label{tab: analy1390}
\begin{tabular}{*7c}
\hline \hline
Species & T$_\mathrm{ex}$ & T$_\mathrm{k}$ &  n$_\mathrm{H_2}$ & N & $X$\tablefootmark{a}\\
& (K) & (K) &(cm$^{-3}$) & (cm$^{-2}$) &  & \\
\hline
\multicolumn{7}{c}{LTE}\\
\hline
H$_2$CCO & $8 - 13$ & & &  $3.2 - 6.0\times10^{11}$ & $1.0 - 1.9\times10^{-11}$ \\
c-C$_3$HD & $10 - 15$\tablefootmark{b} & & & $8 - 13 \times10^{11}$ & $2.5 - 4.1\times10^{-11}$ \\
HNO & $10 - 15 $\tablefootmark{b}  & & & $2.5 - 3.5\times10^{11}$ & $7.8 - 10.9\times10^{-12}$ \\
CCS & $6 - 8 $ & & & $7 - 14\times10^{11}$ & $2.3 - 4.6\times10^{-11}$ \\
HC$_3$N & $10 - 15 $\tablefootmark{b} & & & $6 - 9\times10^{11}$ & $1.9 - 2.8\times10^{-11}$ \\
p-c-C$_3$H$_2$ & $10 - 15 $\tablefootmark{b} & & & $2 - 6\times10^{11}$ & $6.3 - 18.8\times10^{-12}$ \\
l-C$_3$H$_2$ & $10 - 15 $\tablefootmark{b} & & & $8 - 11\times10^{10}$ & $2.5 - 3.4\times10^{-12}$ \\
o-c-C$_3$H$_2$ & $10 - 15 $\tablefootmark{b} & & & $11 - 14\times10^{11}$ & $3.4 - 4.4\times10^{-11}$ \\
DC$_3$N & $10 - 15 $\tablefootmark{b} & & & $8 - 12\times10^{10}$ & $2.5 - 3.8\times10^{-12}$ \\
CH$_3$O & $10 - 15$\tablefootmark{b} & & & $2.5 - 3.8\times10^{11}$ & $7.8 - 11.9\times10^{-12}$\\
OCS & $10 - 15 $\tablefootmark{b} & & & $14 - 18\times10^{11}$ & $4.4 - 5.6\times10^{-11}$ \\
o-D$_2$CS & $10 - 15 $\tablefootmark{b} & & & $13 - 15\times10^{10}$ & $4.1 - 4.7\times10^{-11}$ \\
e-CH$_3$CCH & $10 - 15 $\tablefootmark{b} & & & $10 - 13\times10^{11}$ & $3.1 - 4.1\times10^{-11}$ \\
HOCO$^+$ & $10 - 15 $\tablefootmark{b} & & & $2 - 5.5\times10^{10}$ & $6.3 - 17.2\times10^{-13}$ \\
C$_4$H & $10 - 15 $\tablefootmark{b} & & & $7 - 10\times10^{12}$ & $2.2 - 3.1\times10^{-10}$ \\
NH$_2$D & $3.8 - 5.3$  & & & $0.6-3.4 10\times10^{14}$  & $1.7 - 10.6 \times 10^{-9}$ \\
E-CH$_3$CHO & $10 - 15\tablefootmark{b} $ & & & $15 - 21\times10^{10}$ & $4.7 - 6.6\times10^{-12}$ \\
A-CH$_3$CHO & $10 - 15\tablefootmark{b} $ & & & $14 - 19\times10^{10}$ & $4.4 - 5.9\times10^{-12}$ \\
c-C$_3$D$_2$ & $10 - 15 $\tablefootmark{b} & & & $7 - 9\times10^{10}$ & $2.2 - 2.8\times10^{-12}$ \\
l-C$_3$H & $10 - 15 $\tablefootmark{b} & & & $11 - 13\times10^{10}$ & $3.4 - 4.1\times10^{-12}$ \\
o-H$_2$CS& $10 - 15 $\tablefootmark{b} & & & $7 - 8\times10^{11}$ & $2.2 - 2.5\times10^{-11}$ \\
\hline
\multicolumn{7}{c}{non-LTE}\\
\hline
A/E-CH$_3$OH & & $10 - 15$ & $2 - 10\times10^4$ & $4 - 14\times10^{12}$ &  $1.3 - 4.4\times10^{-10}$ \\
HC$^{18}$O$^+$ & & $10 - 15$\tablefootmark{b} & $2 - 10\times10^4$ & $3 - 8\times10^{10}$ &  $9.4 - 25\times10^{-13}$ \\
HCS$^+$ & & $10 - 15$\tablefootmark{b} & $2 - 10\times10^4$ & $7 - 11\times10^{10}$ &  $2.2 - 3.4\times10^{-12}$ \\
CS & & $10 - 15$\tablefootmark{b} & $2 - 10\times10^4$ & $3.5 - 25\times10^{12}$ &  $1.1 - 7.8\times10^{-10}$ \\
C$^{34}$S & & $10 - 15 $\tablefootmark{b} & $2 - 10\times10^4$ & $2 - 5\times10^{11}$ & $6.3 - 15.6\times10^{-10}$ \\
C$^{34}$S$\times ^{32}$S$/^{34}$S\tablefootmark{c} & &  &  & $5.4 - 18.5\times10^{12}$ & $1.7 - 5.8\times10^{-10}$ \\
SO & & $10 - 15$\tablefootmark{b} & $2 - 10\times10^4$ & $9 - 35\times10^{12}$ &  $2.8 - 10.9\times10^{-10}$ \\
$^{33}$SO& & $10 - 15 $\tablefootmark{b} & $2 - 10\times10^4$ & $2.5 - 4.9\times10^{11}$ & $7.8 - 15.3\times10^{-12}$ \\
$^{34}$SO & & $10 - 15 $\tablefootmark{b} & $2 - 10\times10^4$ & $6 - 15\times10^{11}$ & $1.9 - 4.7\times10^{-11}$ \\
$^{34}$SO$\times ^{32}$S$/^{34}$S\tablefootmark{c} & & & & $16 - 55\times10^{12}$ & $5.1 - 17.3\times10^{-11}$ \\
OCS & & $10 - 15$\tablefootmark{b} & $2 - 10\times10^4$ & $9 - 18\times10^{11}$ &  $2.6 - 5.6\times10^{-11}$\\
\hline
\end{tabular}
\tablefoot{\\
The quoted error is the 3$\sigma$ (statistical and calibration) error from the fitting.\\
\tablefoottext{a}{Abundances derived using the H$_2$ column density N$(\mathrm{H}_2)=3.2\times10^{22}$ cm$^{-2}$, adopted from \citet{Montillaud2015}. The value is the average of the source ($r_\mathrm{eff} = 26\arcsec$).} \\
\tablefoottext{b}{The temperature was fixed to T$_\mathrm{k}$ derived from methanol.} \\
\tablefoottext{c}{$^{32}$S$/^{34}$S$=(3.3 \pm 0.5)\,$(DGC/kpc)$+4.1 \pm 3.1  (\sim 32 \pm 5$ for our cores), adopted from \citet{Chin1996}. }
}
\end{table*}

\begin{table*}
\centering
\caption{LTE and non-LTE analysis results for core 4149.}
\label{tab: analy4149}
\begin{tabular}{*7c}
\hline \hline
Species & T$_\mathrm{ex}$ & T$_\mathrm{k}$ &  n$_\mathrm{H_2}$ & N & $X$\tablefootmark{a}\\
& (K) & (K) &(cm$^{-3}$) & (cm$^{-2}$) &  & \\
\hline
\multicolumn{7}{c}{LTE}\\
\hline
H$_2$CCO & $12.4 - 17.3$ & & &  $4.5 - 6.5\times10^{11}$ & $4.4 - 6.4\times10^{-11}$ \\
c-C$_3$HD & $10 - 15$\tablefootmark{b} & & &  $10 - 15\times10^{11}$ & $9.8 - 12.7\times10^{-11}$ \\
C$_3$S & $10 - 15$\tablefootmark{b} & & &  $8 - 17\times10^{10}$ & $7.8 - 16.7\times10^{-12}$ \\
c-CC$^{13}$CH$_2$ & $10 - 15$\tablefootmark{b} & & &  $2.0 - 2.7\times10^{11}$ & $2.0 - 2.7\times10^{-11}$ \\
CH$_2$DCCH & $8 - 14$ & & &  $1.5 - 2.5\times10^{12}$ & $1.5 - 2.5\times10^{-10}$ \\
HNO & $10 - 15$\tablefootmark{b} & & &  $2.0 - 3.0\times10^{11}$ & $2.0 - 2.9\times10^{-10}$ \\
CCS & $5 - 6.5 $ & & & $2 - 5\times10^{12}$ & $1.9 - 4.9\times10^{-11}$ \\
HC$_3$N & $10 - 15 $\tablefootmark{b} & & & $8 - 15\times10^{11}$ & $7.8 - 12.7\times10^{-11}$ \\
p-c-C$_3$H$_2$ & $10 - 15 $\tablefootmark{b} & & & $4 - 11\times10^{11}$ & $3.9 - 10.8\times10^{-10}$ \\
l-C$_3$H$_2$ & $10 - 15 $\tablefootmark{b} & & & $4 - 8\times10^{10}$ & $3.9 - 7.8\times10^{-12}$ \\
o-c-C$_3$H$_2$ & $10 - 15 $\tablefootmark{b} & & & $7 - 15\times10^{11}$ & $6.9 - 14.7\times10^{-10}$ \\
CH$_3$O & $10 - 15 $\tablefootmark{b} & & & $2 - 4\times10^{11}$ & $1.9 - 3.9\times10^{-11}$ \\
HOCN & $10 - 15 $\tablefootmark{b} & & & $1.8 -2.2 \times10^{10}$ & $1.8 - 2.2\times10^{-12}$ \\
DC$_3$N & $10 - 15 $\tablefootmark{b} & & & $9 - 12\times10^{10}$ & $8.8 - 11.8\times10^{-12}$ \\
OCS & $10 - 15 $\tablefootmark{b} & & & $2.2 - 2.7\times10^{12}$ & $2.2 - 2.7\times10^{-10}$ \\
o-D$_2$CS & $10 - 15 $\tablefootmark{b} & & & $1.5 - 2.5\times10^{11}$ & $1.5 - 2.5\times10^{-11}$ \\
e-CH$_3$CCH & $10 - 15 $\tablefootmark{b} & & & $4 - 8\times10^{12}$ & $3.9 - 7.8\times10^{-11}$ \\
HOCO$^+$ & $10 - 15 $\tablefootmark{b} & & & $3.0 - 5.5\times10^{10}$ & $2.9 - 5.4\times10^{-12}$ \\
C$_3$O & $10 - 15$\tablefootmark{b} & & &  $4.0 - 8.5\times10^{10}$ & $3.9 - 8.3\times10^{-12}$ \\
C$_4$H & $10 - 15 $\tablefootmark{b} & & & $5.0 - 8.0\times10^{12}$ & $4.9 - 7.8\times10^{-10}$ \\
NH$_2$D & $3.4 - 4.2$ & & & $0.84 - 5 \times 10^{14}$ & $0.8 - 4.9 \times 10^{-8}$ \\
E-CH$_3$CHO & $10 - 15 $\tablefootmark{b} & & & $1.7 - 2.5\times10^{11}$ & $1.6 - 2.5\times10^{-11}$ \\
A-CH$_3$CHO & $10 - 15 $\tablefootmark{b} & & & $2.0 - 2.8\times10^{11}$ & $1.9 - 2.7\times10^{-11}$ \\
c-C$_3$D$_2$ & $10 - 15$\tablefootmark{b} & & &  $7.5 - 9.0\times10^{10}$ & $7.4 - 8.8\times10^{-12}$ \\
l-C$_3$H & $10 - 15$\tablefootmark{b} & & &  $9 - 11\times10^{10}$ & $8.9 - 10.8\times10^{-10}$ \\
CH$_3$SH & $10 - 15$\tablefootmark{b} & & &  $4 - 6\times10^{11}$ & $3.9 - 5.9\times10^{-11}$ \\
o-H$_2$C$^{34}$S & $10 - 15$\tablefootmark{b} & & &  $3.3 - 4.7\times10^{10}$ & $3.2 - 4.6\times10^{-10}$ \\
o-H$_2$CS & $10 - 15$\tablefootmark{b} & & &  $16.5 - 18.8\times10^{11}$ & $1.6 - 1.8\times10^{-10}$ \\
\hline
\multicolumn{7}{c}{non-LTE}\\
\hline
A/E-CH$_3$OH & & $10 - 15$ & $1.5 - 3.7\times10^4$ & $6 - 12\times10^{12}$ &  $5.9 - 11.8\times10^{-10}$ \\
HC$_3$N & & $10 - 15$\tablefootmark{b} & $1.5 - 3.7\times10^4$ & $1.8 - 6.0\times10^{12}$ &  $1.8 - 5.9\times10^{-10}$ \\
HC$^{18}$O$^+$ & & $10 - 15$\tablefootmark{b} & $1.5 - 3.7\times10^4$ & $4 - 10\times10^{10}$ &  $3.9 - 9.8\times10^{-12}$ \\
HCS$^+$& & $10 - 15$\tablefootmark{b} & $1.5 - 3.7\times10^4$ & $1.5 - 2.7\times10^{11}$ &  $1.5 - 2.7\times10^{-11}$ \\
CS & & $10 - 15$\tablefootmark{b} & $1.5 - 3.7\times10^4$ & $1.1 - 3.8\times10^{12}$ &  $1.1 - 3.7\times10^{-10}$ \\
C$^{34}$S & & $10 - 15 $\tablefootmark{b} & $1.5 - 3.7\times10^4$ & $3.0 - 4.5\times10^{11}$ & $2.9 - 4.4\times10^{-11}$ \\
C$^{34}$S$\times ^{32}$S$/^{34}$S\tablefootmark{c} & & & & $8.1 - 16.7\times10^{12}$ & $8 - 16\times10^{-10}$ \\
SO & & $10 - 15$\tablefootmark{b} & $1.5 - 3.7\times10^4$ & $4.5 - 13.1\times10^{12}$ &  $4.4 - 13.0\times10^{-10}$ \\
$^{34}$SO & & $10 - 15 $\tablefootmark{b} & $1.5 - 3.7\times10^4$ & $4.5 - 6.5\times10^{11}$ & $4.4 - 6.4\times10^{-11}$ \\
$^{34}$SO$\times ^{32}$S$/^{34}$S\tablefootmark{c} & & & & $12 - 24\times10^{12}$ & $12 - 24\times10^{-11}$ \\
OCS & & $10 - 15$\tablefootmark{b} & $1.5 - 3.7\times10^4$ & $2.0 - 4.0\times10^{12}$ &  $2.0 - 3.9\times10^{-10}$\\
\hline
\end{tabular}

\tablefoot{\\
The quoted error is the 3$\sigma$ (statistical and calibration) error from the fitting.\\
\tablefoottext{a}{Abundances derived using the H$_2$ column density N$(\mathrm{H}_2)=1.02\times10^{22}$ cm$^{-2}$, adopted from \citet{Montillaud2015}. The value is the average of the source ($r_\mathrm{eff} = 23\arcsec$).} \\
\tablefoottext{b}{The temperature was fixed to T$_\mathrm{k}$ derived from methanol.} \\
\tablefoottext{c}{$^{32}$S$/^{34}$S$=(3.3 \pm 0.5)\,$(DGC/kpc)$+4.1 \pm 3.1  (\sim 32 \pm 5$ for our cores), adopted from \citet{Chin1996}. }
}
\end{table*}

\section{Discussion}
\label{sec:discuss}

\subsection{Starless and prestellar cores}

Cold dense cores not associated with stars are called starless cores, and they represent the initial conditions in the process of star formation \citep{Caselli2012}.
Some of them reach configurations close to hydrostatic equilibrium and display kinematic features consistent with oscillations displaying a relatively flat density distribution, with central densities $n_{\mathrm{H}_2} < 10^5$ cm$^{-3}$. This is the critical density for gas cooling by gas-dust collisions, and it represents the dividing line for dynamical stability \citep{Goldsmith2001}. Starless cores with central densities below this critical density are thermally subcritical \citep{Keto2008ApJ}, and they may disperse back into the interstellar medium. When the central densities of H$_2$ molecules exceed $10^5$ cm$^{-3}$, starless cores become thermally supercritical, and gravitational forces take over. These are the so-called prestellar cores.

According to the non-LTE fitting results of methanol (see Sec.~\ref{subsec: analyses}), the gas emitting the lines in 869, 1390, and 4149 has a kinetic temperature range of 10--14 K, 10--15 K, and 10--15 K and an H$_2$ density $\rm n_{H_2}$ range of (6--30) $\times 10^3$ cm$^{-3}$, (2--10) $\times 10^4$ cm$^{-3}$ , and (1.5--3.7) $\times 10^3$ cm$^{-3}$. 
These values are quite consistent with the average density derived from the continuum (see Table~\ref{tab: comparison}), which, considering the large beam, is dominated by the lower end of the density range in each source.
Therefore, both the dust continuum and non-LTE analysis of the methanol lines suggest that the gas in the three sources is relatively tenuous ($\leq 1\times 10^5$ cm$^{-3}$). However, we recall that methanol, as any other species containing heavy elements is frozen onto the grain icy mantles at densities higher than about $10^5$ cm$^{-3}$ (e.g. \citealt{Caselli1999,Bacmann2002AA,Bergin2006ApJ}), so that by definition, these species generally probe the outer, relatively tenuous skin of the prestellar cores.
In the specific case of methanol, \citet{Vastel2014} have shown that the methanol lines in the prototype prestellar core L1544 originate in the outer layer of the core, where the density is $\sim 3\times 10^4$ cm$^{-3}$, from material that is (weakly) illuminated by interstellar UV photons. In this region, methanol is thought to be photo-desorbed from the icy mantles and injected into the gas phase.
Therefore, 869, 1390, and 4149 cannot be dismissed as prestellar core candidates based on the low-density estimates computed using methanol: the dust average and methanol-derived densities refer to the less dense part of the core.
In the following, we use the molecular composition derived in the previous section to better characterise the three sources and possibly confirm whether our assumption is correct. 

\subsection{Ions}
\label{sec: ions}

\begin{figure}
    \centering
    \includegraphics[width=1\columnwidth]{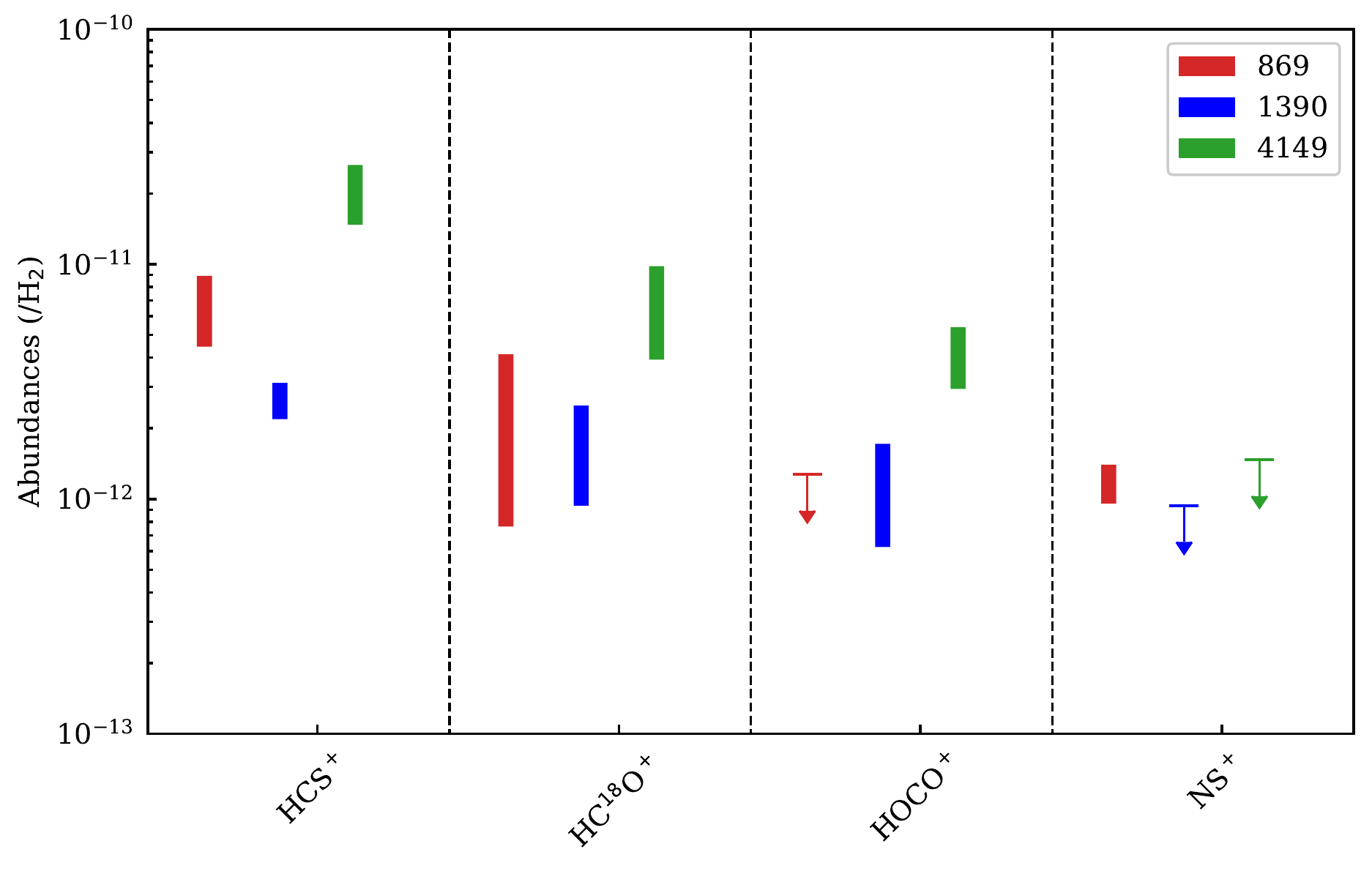}
    \caption{Abundance of ionised species. We estimated the upper limit of the column densities based on the T$_\mathrm{mb}$ upper limit for undetected species.}
    \label{fig: Ions}
\end{figure}

Several ions have been detected, such as HC$^{18}$O$^+$ and HCS$^+$ in all the three cores, HOCO$^+$ in  1390 and 4149, and NS$^+$ in 869. Figure \ref{fig: Ions} shows the abundances relative to H$_2$ for each species, calculated from the column densities derived from the ions (see Sec.~\ref{subsec: analyses}) and the H$_2$ column density taken from \citet{Montillaud2015}.
The figure shows that 4149 has the largest ion abundance, with the exception of NS$^+$, pointing to the highest ionisation degree of the three sources.
In the following, we separately discuss the information that can be extracted from HCS$^+$ and HCO$^+$,  and from HOCO$^+$ . We do not discuss the case of NS$^+$ further, as we only have one detection.

\paragraph{HCO$^+$ and HCS$^+$:} 
Because HCO$^+$ is the most abundant ion along with H$_3^+$ in cold gas, it provides an almost direct measurement of the gas ionisation degree.
In dense cores, the gas ionisation is mostly due to the cosmic rays, and the crucial parameter that measures it is the so-called cosmic-ray ionisation rate $\zeta_{CR}$. 
Although HCS$^+$ is a minor ion in terms of abundance, it is the most abundant sulphur-bearing ion, and it provides constraints on the sulphur abundance S/H.
With our new estimates of the ion abundances in the three cores, we can therefore provide constraints on $\zeta_{CR}$ and S/H and quantify the first differences or similarities in cores 869, 4149, and 1390.

To this end, we used the code MyNahoon, a modified version of Nahoon \citep{Wakelam2012}, which is a time-dependant code for calculating the gas-phase chemistry. The computations are based on the chemical network (GRETOBAPE.2020), which is an updated version of the KIDA.2014 network\footnote{\citealt{Wakelam2012}: \url{https://kida.astrochem-tools.org/}} and takes several recently revised reactions into account \citep[for details, see][]{Codella2020}.

The code has four main input parameters that characterise the source that is to be modelled: the initial elemental gaseous abundances, $\zeta_\mathrm{CR}$, the gas temperature, and the H-nucleus density. 
The initial element abundances are O/H=$2.6\times 10^{-5}$, C/H=$1.7\times 10^{-5}$, N/H=$6.2\times 10^{-6}$, Si/H=$8.0\times 10^{-9}$, Fe/H=$3.0\times 10^{-9}$ , and Mg/H=$7.0\times 10^{-9}$ from \citet{Jenkins2009ApJ}.
We left the sulphur abundance S/H a free parameter because it is the least constrained elemental abundance \citep[see][]{Jenkins2009ApJ} and has an important impact on the gas ionisation (see below).
The gas temperature and H-nucleus density were taken equal to the results of the non-LTE fitting of the methanol lines for each source. 
We ran a grid of models in which we varied both $\zeta_\mathrm{CR}$ and S/H, and followed the evolution of the gas until 10$^{6}$ years. We then compared the predicted results with our measured abundance of HCO$^+$ and HCS$^+$.

\begin{figure}
    \centering
    \includegraphics[width=0.8\columnwidth,trim=60 0 20 0 clip]{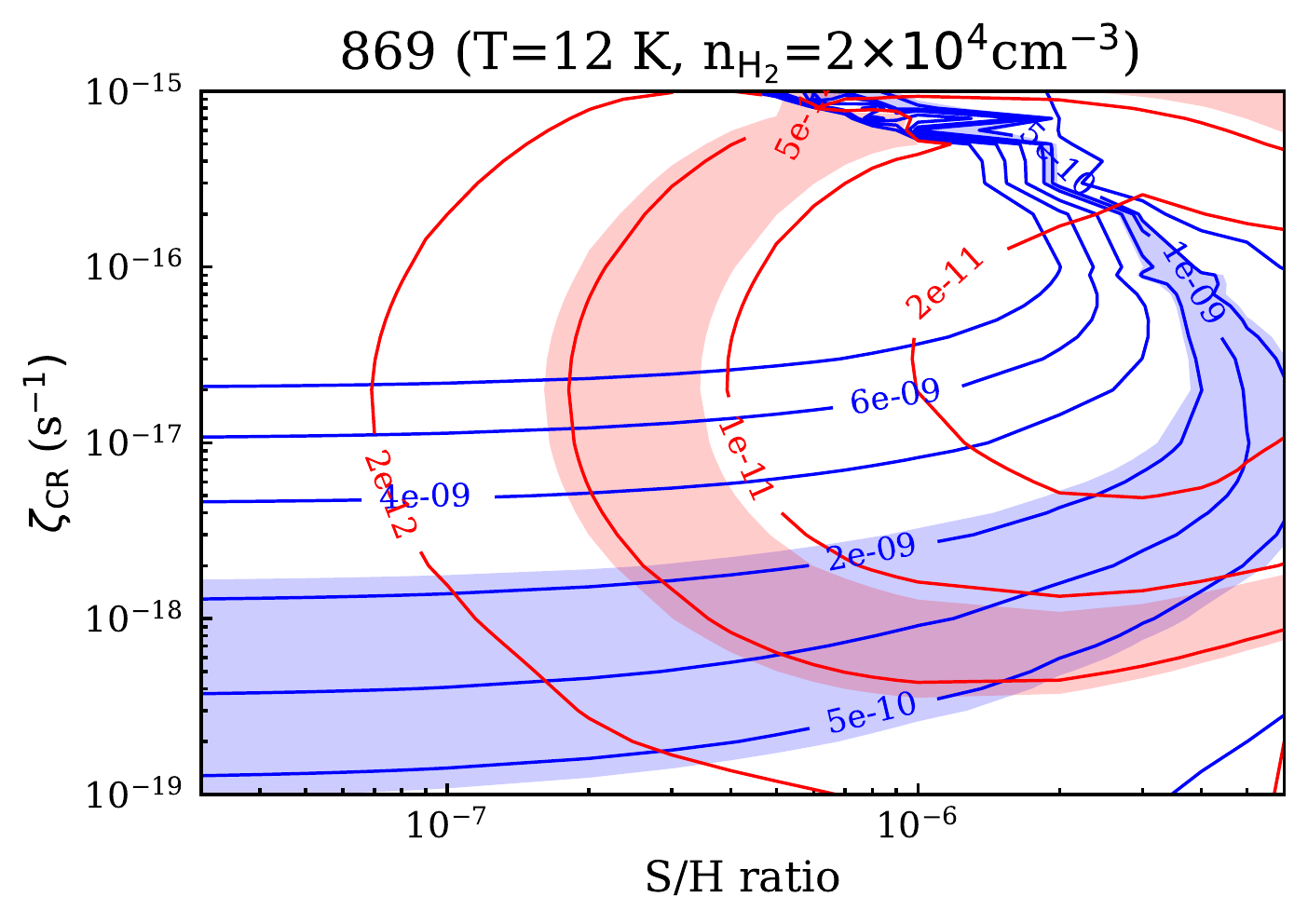} \ 
    \includegraphics[width=0.8\columnwidth,trim=60 0 20 0 clip]{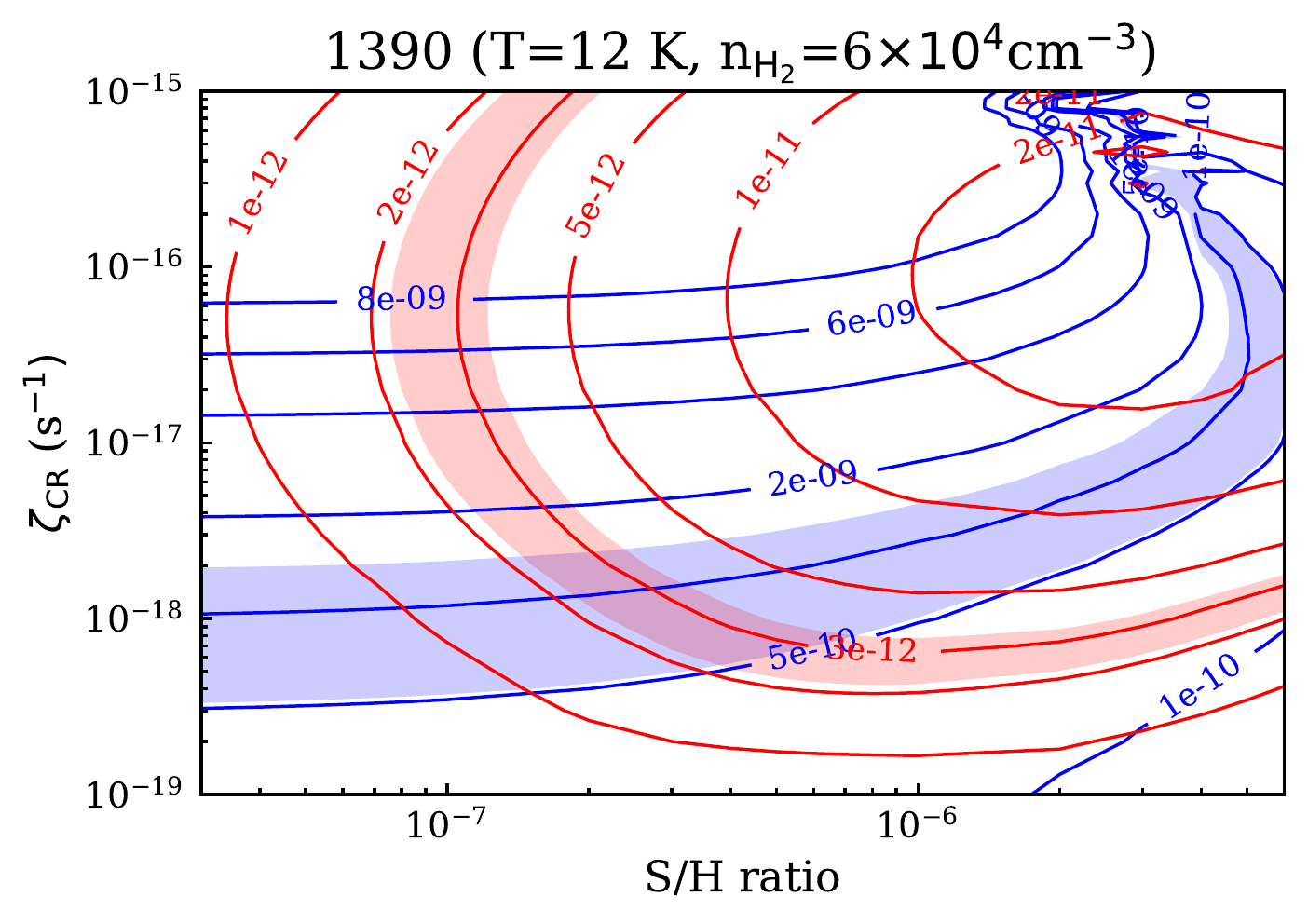} \ 
    \includegraphics[width=0.8\columnwidth,trim=60 0 20 0 clip]{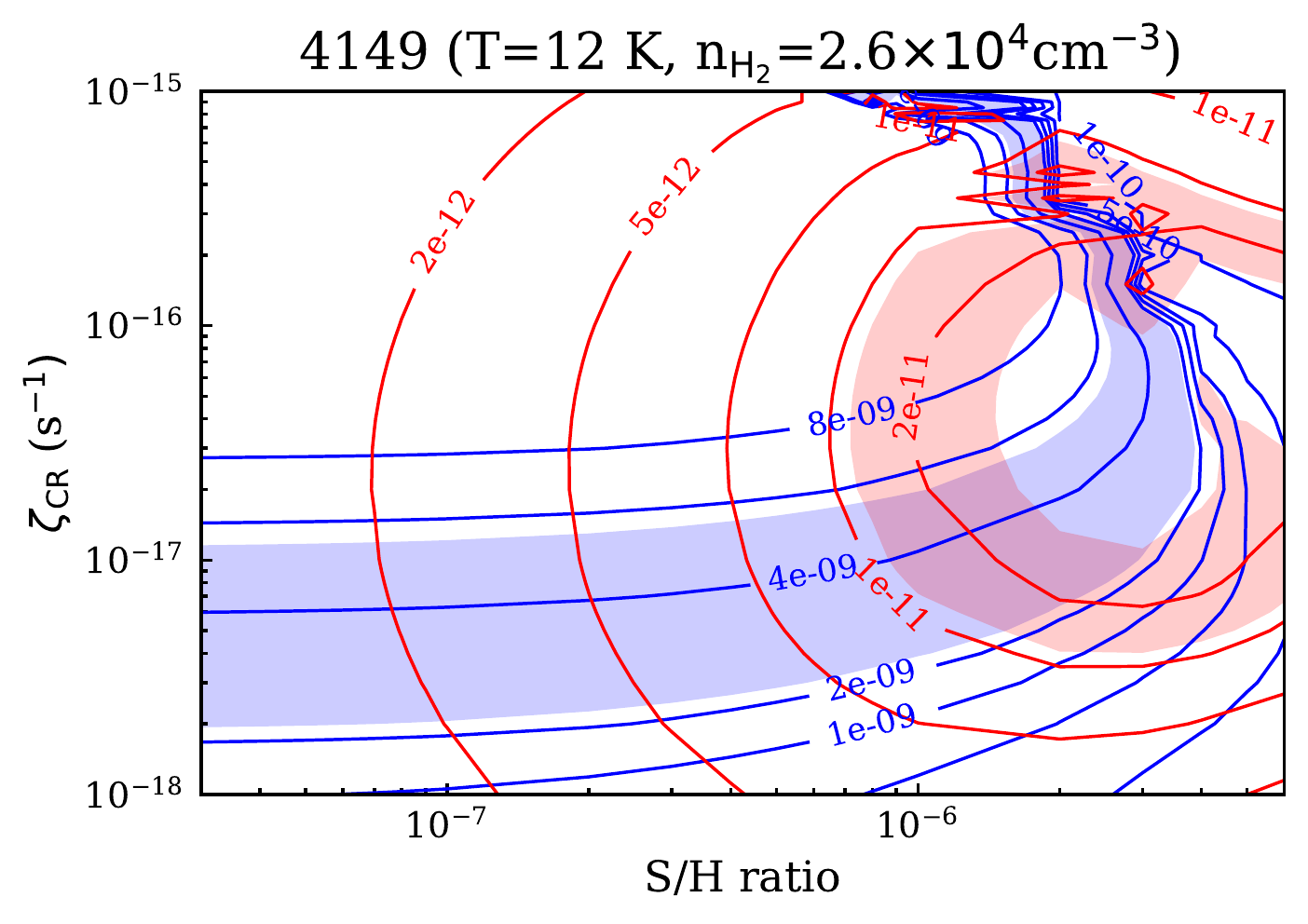}
    \caption{Contour plots of HCO$^+$/H$_2$ (blue line) and HCS$^+$/H$_2$ (red line) abundance ratios as a function of cosmic-ray ionisation rate and S/H gaseous elemental abundance. The blue and red zones mark the observational abundances. N(HCO$^+$)=N(HC$^{18}$O$^+$)$\times ^{16}$O/$^{18}$, where $^{16}$O/$^{18}$O=557. The models were obtained for a gas temperature of 12 K, an evolutionary time of 10 Myr, and an H$_2$ density of $2\times10^4$ cm$^{-3}$ (869), $6\times10^4$ cm$^{-3}$ (1390), and $2.6\times10^4$ cm$^{-3}$ (4149).}
    \label{fig: contour_plot}
\end{figure}

Figure \ref{fig: contour_plot} shows the contour plots of the predicted HCO$^+$ and HCS$^+$ abundances as a function of $\zeta_\mathrm{CR}$.
The figure shows that the predicted HCO$^+$ abundance mostly depends on  $\zeta_\mathrm{CR}$ for S/H lower than about $2\times 10^{-6}$.
Conversely, as expected, the predicted HCS$^+$ abundance strongly depends on both $\zeta_\mathrm{CR}$ and S/H.
Therefore, the comparison of the theoretical predictions with the measured abundances of HCO$^+$ and HCS$^+$, also shown in Fig. \ref{fig: contour_plot}, allows us to simultaneously constrain the two parameters $\zeta_\mathrm{CR}$ and S/H.
From the intersections of the two species, we could derive $\zeta_\mathrm{CR}$ equal to (3--15)$\times 10^{-19}$ s$^{-1}$, (4--20)$\times 10^{-19}$ s$^{-1}$ and (7--20)$\times 10^{-18}$ s$^{-1}$, and S/H equal to (3--30)$\times 10^{-7}$, (1.5--6)$\times 10^{-7}$ and (7--20)$\times 10^{-7}$ for 869, 1390, and 4149, respectively.
For 4149, the solution with $\zeta_\mathrm{CR}$ $\sim 10^{-16}$ s$^{-1}$ is in principle possible as well (Fig. \ref{fig: contour_plot}). 
However, we failed to detect possible cosmic rays (i.e. supernova remnant) or X-ray sources in the vicinity of 4149, therefore we discarded this high value.

The $\zeta_\mathrm{CR}$ value we derive for 4149 is compatible with the values reported in other starless or prestellar cores, $\sim 10^{-17}$ s$^{-1}$ \citep[see e.g.][]{Redaelli2021}. Conversely, the $\zeta_\mathrm{CR}$ values in 869 and 1390 are about one order of magnitude lower. 
The reason for these differences is not obvious. Core 869 is located in MBM12, an isolated high-latitude cloud, far from any known cosmic-ray or X-ray source, which could explain the lower cosmic-ray ionisation rate. Core 1390 is located in the $\lambda$ Ori region, where a supernova explosion is thought to have occurred about a million years ago \citep{Dolan2002}. Therefore, the origin of the differences in $\zeta_\mathrm{CR}$ must be searched for elsewhere, possibly in the local and surrounding magnetic field strength and geometry \citep{Owen2021}. Among the three cores, 1390 seems to have the lowest S/H elemental abundance compatible with a higher ionisation  $\zeta_\mathrm{CR}$ value, a higher gas density, propitious to a stronger sulphur depletion, namely sulphur-bearing species frozen onto the grain mantles.


The ionisation in dense cores, which is mainly provided by cosmic rays, is thought to be playing a crucial role in astrochemistry. Cosmic rays promote the formation of H$_3^+$ ions, which can easily donate a proton to elements such as C and O, and thus eventually form more complex molecules. Cosmic rays are an important heating source in cold and dense environment \citep{Goldsmith2001}. The higher cosmic-ray ionisation found in 4149 is therefore compatible with the higher number of COMs detected in this core.
They can directly heat the dust grain and the icy mantle, activating the solid-phase chemistry and releasing molecules from the icy mantle \citep{Ivlev2015ApJ,Shingledecker2017PCCP}. The higher cosmic-ray ionisation found in 4149 is therefore compatible with the higher number of COMs detected in this core. Calculating the cosmic-ray ionisation rate requires both an accurate knowledge of the spectrum of MeV to GeV protons at the edge of the cloud and an accurate model for the propagation of cosmic rays into molecular clouds: they can stream freely along magnetic field lines, and/or they can propagate diffusively due to turbulence.  We need to understand the mechanisms controlling cosmic-ray penetration into the clouds better in order to explain our observations.

Previous studies suggested that the cosmic-ray ionisation rate of molecular clouds in the Galaxy is at the magnitude of $10^{-16}$ s$^{-1}$ \citep{Indriolo2012ApJ, Indriolo2015ApJ, Bacalla2019}. $\zeta_\mathrm{CR}$ can reach a higher value ($\sim 10^{-14}$ s$^{-1}$) in regions close to supernovae remnants \citep[e.g.][]{Ceccarelli2011,Vaupre2014} and some massive star formation regions  (OMC-2: \citealt{Ceccarelli2014, Fontani2017, Favre2018ApJ}), while it decreases in denser regions such as starless core ($\sim 10^{-17}$ s$^{-1}$ or lower: \citealt{Caselli1998ApJ, Maret2007ApJ, Fuente2016, Bovino2020MNRAS, Redaelli2021}). 
\citet{Neufeld2017} also found the similar dependence of $\zeta_\mathrm{CR}$ on the gas column density ($\zeta_\mathrm{CR}$ $\propto$ N$^{-1}$). 
For our three cores, the $\zeta_\mathrm{CR}$ value of 4149 is compatible with the value estimated towards L1544 \citep[$\rm 2-3 \times10^{-17} s^{-1}$:][]{Bovino2020MNRAS, Redaelli2021}. The possible explanation for the seemingly lower $\zeta_\mathrm{CR}$ value towards 869 and 1390 could be that the core is located in an isolated environment and far from supernovae (869), or the core is embedded in a dense region of the molecular cloud and is less exposed to the background cosmic rays (1390).  It should be noted that there are large uncertainties on the determination of the cosmic-ray ionisation rate using different methods. Two low values have been reported by \citet{padovani2013} (their Fig. 6), compatible with our low estimates towards 869. However, a better constraint on the cosmic-ray ionisation rate throughout each region is necessary to conclude, and observations with  a higher spatial resolution that trace lower density envelopes (using low critical density tracers) are needed to properly measure the cosmic-ray ionisation rate throughout the cores.

\paragraph{HOCO$^+$:}
CO$_2$ is highly abundant in the interstellar ices \citep{Boogert2015}, in comets, and in planetary atmospheres \citep{Hoang2017AA}. 
The amount of frozen CO$_2$ is as large as that of frozen CO or CH$_3$OH, which implies that this species is a major carbon reservoir on the ices.
Unfortunately, because gaseous CO$_2$ lacks a permanent dipole moment, it cannot be detected in cold environments, so that we ignored its abundance in the gas phase. However, the CO$_2$ protonated ion, HOCO$^+$, can be and has been used to derive constraints on the abundance of gaseous CO$_2$ \citep[e.g.][]{Podio2014AA,Vastel2016AA}. In the prestellar core L1544, HOCO$^+$ has been detected in emission in the external layer of the core where non-thermal desorption of other species has previously been detected. Modelling of the chemistry involving the formation and destruction of HOCO$^+$ provided a gaseous CO$_{2}$/H$_{2}$ ratio of 2 $\times$ 10$^{-7}$, with an upper limit of  2 $\times$ 10$^{-6}$ \citep{Vastel2016AA}. This study was the first indirect estimate of the CO$_{2}$ presence in a cold prestellar core, with obviously no confusion between hot, warm, and cold gas. The inferred abundance seems to be lower than the solid-state abundances of $\sim (1-3) \times 10^{-6}$ \citep{Gerakines1999} measured through the interstellar CO$_{2}$ ice absorption features detected by the ISO telescope. In the past, HOCO$^+$ has been detected towards the Galactic centre \citep{Thaddeus1981, Minh1991}, diffuse and translucent clouds \citep{Turner1999ApJ}, low-mass prestellar cores \citep{Vastel2016AA}, low-mass protostars and their associated molecular shocks \citep{Sakai2008ApJ, Podio2014AA, Majumdar2018MNRAS}, and massive star-forming clumps \citep{Fontani2018MNRAS}.

From the LTE analysis (Sect. \ref{subsec: analyses}), we derived a HOCO$^+$ abundance of (6--17)$\times 10^{-13}$ and (3--5)$\times 10^{-12}$ for 1390 and 4149, respectively. These values can be compared to the high value found toward L1544, $\sim 5 \times 10^{-11}$ \citep{Vastel2016AA}, suggesting a lower CO$_2$ gaseous abundance in 1390 and 4149 than in L1544.
Likewise, the HOCO$^+$ abundance in the cold cores TMC-1 and L183 is $\sim3\times 10^{-10}$ and $\sim5\times 10^{-11}$ , respectively \citep{Turner1999ApJ}.
In contrast, the HOCO$^+$ abundance is larger than that in the cold envelope of the Class 0 protostar IRAS 16293-2422 ($\sim 10^{-13}$: \citealt{Majumdar2018MNRAS}).
In the cold and dense gas, two pathways of HOCO$^+$ formation are suggested by the literature: (1) the protonation of CO$_2$ by H$_3^+$, and (2) the oxidation of HCO$^+$ by OH \citep{Podio2014AA,Vastel2016AA, Bizzocchi2017AA}.
Therefore, the lower HOCO$^+$ abundance in 1390 and 4149 with respect to other cold core in which it has also been detected could be due to the lower cosmic-ray ionisation rate that we measured in these sources and not necessarily due to the fact that 1390 and 4149 are deprived of gaseous CO$_2$.

\subsection{Deuterated species and D/H ratio}
\label{sec: deut}

When the density increases above 10$^5$ cm$^{-3}$ and the temperature drops below $20$\,K, species heavier than helium tend to disappear from the gas phase due to the process of freeze-out (the adsorption of species onto dust grain surfaces). In the central positions of the cold starless cores, CO and its isotopologues are heavily depleted via freeze-out onto dust grains \citep{Caselli1999, Bergin2002}. These are perfect conditions for an extreme molecular deuteration. Although the deuterium abundance is 1.5 $\times$ 10$^{-5}$ relative to hydrogen \citet{Linsky2007}, singly, doubly, and even triply deuterated molecules have been detected to be up to 13 orders of magnitude more abundant than the elemental D/H ratio \citep{Parise2004}.
In the cold environments of prestellar cores, deuterium fractionation takes place, beginning from H$^+_3$ + HD $\rightarrow$ H$_2$D$^+$ + H$_2$ + 230K \citep{Watson1974ApJ} and continuing even further to D$_2$H$^+$ and D$_3^+$ \citep{Vastel2004ApJ, Vastel2006RSPTA}. This reaction cannot proceed from right to left when the kinetic temperature is below 20\,K. Because CO is the main destruction partner of H$_2$D$^+$, deuterium fractionation is further enhanced when the freeze-out of neutral species becomes strong. The deuterated H$^+_3$ isotopologues transfer a deuteron to other species (e.g. CO or N$_2$), forming a diversity of deuterated species such as ions. The dissociative recombination of deuterated ions may also release atomic deuterium into the gas, which can then be adsorbed onto grain surfaces and thus contributes to deuteration on the surface by addition or abstraction reactions. Desorption through transient heating of the grains by cosmic rays or photons then releases the deuterated species into the gas phase \citep{Aikawa2012ApJ,Sipila2013,Sipila2016}. 
Therefore CO depletion and a high D/H ratio in the gas phase are significant characteristics of prestellar cores.

\begin{figure}
    \centering
    \includegraphics[width = 0.95\columnwidth]{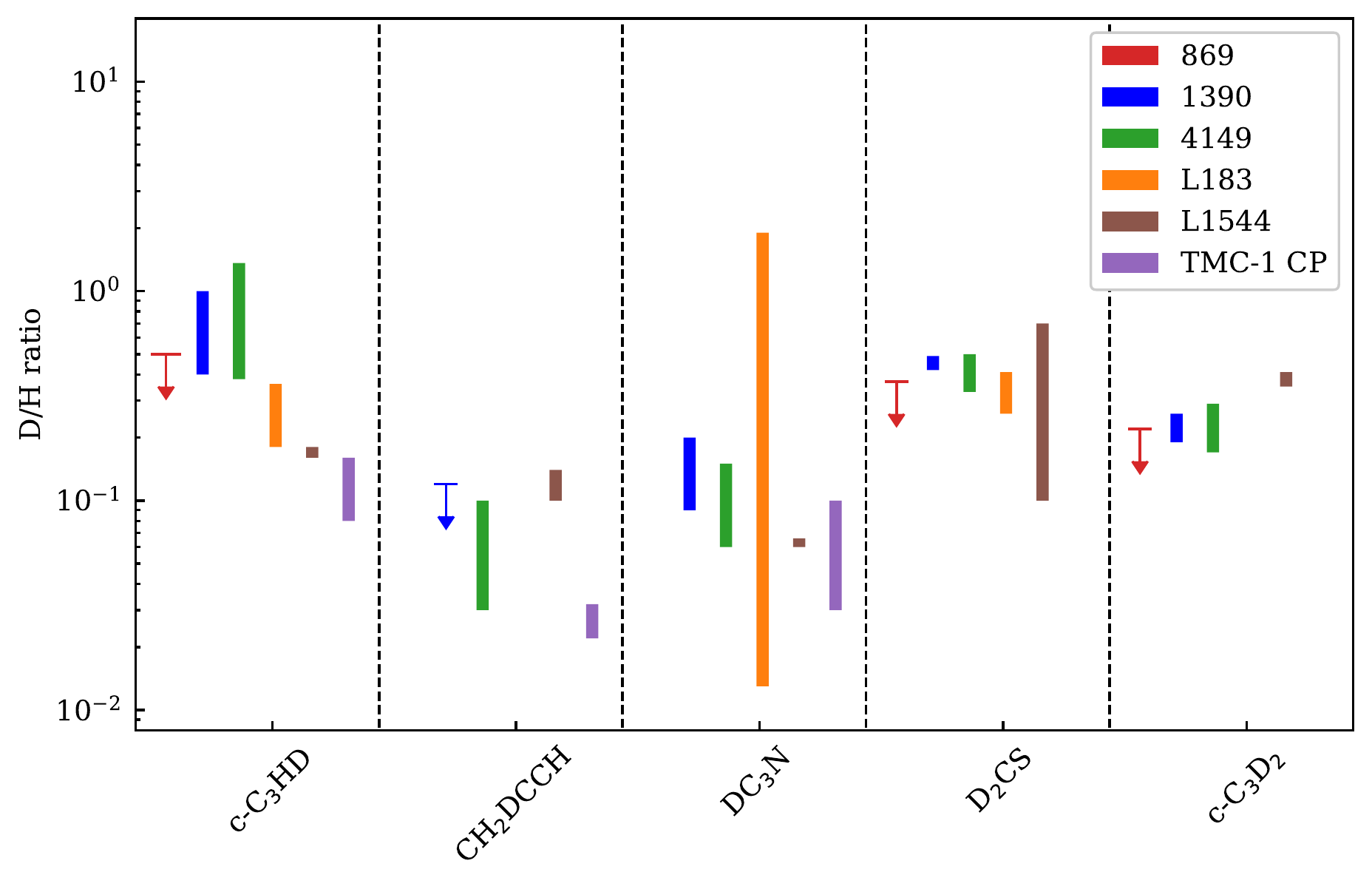}
    \caption{D/H ratio derived from each deuterium-bearing species and their counterparts. We also plot the D/H ratio of L183, L1544, and TMC-CP (see Sec.~\ref{sec: deut} ).}
    \label{fig:D-species}
\end{figure}

From the spectral survey of our three sources, we detected deuterium species only in 1390 and 4149, which either reflects an inefficient deuterium fractionation in 869 despite the low temperature or the fact that 869 in general presents  lower column densities, apparently in disagreement with the high values found in the prototypical prestellar core L1544: c-C$_{3}$H$_{2}$ $\sim$ 4 $\times$ 10$^{13}$ cm$^{-2}$ and c-C$_{3}$HD $\sim$ 6 $\times$ 10$^{12}$ cm$^{-2}$ \citep{Spezzano2013}, and HC$_{3}$N=[2-8] $\times$ 10$^{13}$ cm$^{-2}$ \citep{Quenard2017,Lattanzi2020}. This suggests that the three cores are in different evolutionary stages, 869 being younger while the other two are more evolved, with a higher CO depletion evolution. No deuterated species (not even NH$_2$D) have been detected in 869. For the other sources, we calculated the D/H ratio via the abundance ratio for the deuterium species with a single deuterium atom and their hydrogen counterpart, and the square root of the abundance ratio for the deuterium species with two deuterium atoms such as D$_2$CS and c-C$_3$D$_2$. For CH$_2$DCCH, we computed the molecular fractionation and divided by a factor 3 corresponding to the possibilities for deuteration on the three hydrogen atoms. We used an A/E ratio of unity for CH$_{3}$CCH \citep{Markwick2005}. We used a statistical value of 2 for o/p D$_2$CS and 3 for H$_2$CS and used o-H$_2$C$^{34}$S when available. The values of the D/H ratio are represented in Fig.~\ref{fig:D-species} for each detected D-species.

No obvious trend can be highlighted in this figure, with the D/H ratio in the 0.1--1 range for both 1390 and 4149. The calculated upper limits are within the error bars of the 1390 and 4149 cores, and we cannot conclude on a possible trend. We then show a comparison of the D/H ratio for our cores with  two prestellar cores, L183 and L1544, and the cyanoployyne peak of the Taurus molecular cloud-1 (TMC-1 CP), towards which spectral surveys have been performed \citep[e.g.][]{Vastel2014,Lattanzi2020}. Towards TMC-1 CP, the CH$_{2}$DCCH/CH$_3$CCH ratio was adopted from \citep{Markwick2005}, the c-C$_3$HD/c-C$_3$H$_2$ from \citet{Bell1988}, and DC$_3$N/HC$_3$N from \citet{Howe1994}. The TMC-1 CP presents the lowest deuterium fractionation, as expected from a cold dark cloud, before the prestellar core stage. Towards L1544, we used the c-C$_3$D$_{2}$/c-C$_3$H$_2$ from \citet{Spezzano2013}, D$_2$CS/H$_2$CS, DC$_3$N/HC$_3$N and c-C$_3$HD/c-C$_3$H$_2$ from \citet{Lattanzi2020}, and the ASAI spectral survey for CH$_2$DCCH/CH$_3$CCH \citep{Vastel2014}. From a rotational diagram analysis of 11 detected transitions of CH$_2$DCCH, the resulting column density is (1.4 $\pm$ 0.2) $\times$ 10$^{13}$ cm$^{-2}$ for an excitation temperature of (9.1 $\pm$ 0.5) K (Vastel, unpublished results). For L183, we used the values quoted in \citet{Lattanzi2020}. The very high uncertainty on the D/H ratio for DC$_{3}$N results from the uncertainty in the excitation temperature between 5 and 10 K, as discussed by \citet{Lattanzi2020}. Assuming the same excitation temperature in L183 as was found for L1544 (i.e. 7.2 K), they obtained a column density in L183 $\sim$ 50 times lower than in L1544, therefore 50 times higher for the D/H.
Except for c-C$_3$HD, for which 1390 and 4149 are higher that the others, they both have D/H values similar to the prototypical prestellar core L1544. 

\subsection{Sulphur species}
\label{sec: sulphur}

Sulphur is the tenth most abundant element in the Milky Way. The observations indicated only a small fraction of the cosmic sulphur abundance in the cold dense regions \citep{Tieftrunk1994AA, Palumbo1997ApJ}. \citet{Jenkins2009ApJ} showed the elemental depletions for sulphur as well as for other elements with increasing density.
More observational works report the sulphur elemental gas-phase abundances range from $\sim 10^{-8}$ (dense core, \citealt{Agundez2013ChRv, Vastel2018b}) to $\sim 10^{-5}$ (diffuse clouds, \citealt{Jenkins2009ApJ}).
\citet{Spezzano2017AA} reported that organic sulphur-bearing molecules (e.g. CS, CCS, and HCS$^+$) follow the c-C$_3$H$_2$ peak in L1544, suggesting that torwards L1544, sulphur-bearing molecules trace chemically young dense gas.
Recent observational and modelling works suggested that significant sulphur depletion takes place during the dynamical evolution of dense cores from the starless to pre-stellar stage, as also found by modelling work of \citet{Laas2019AA, Nagy2019}.

\begin{figure*}[h!]
    \centering
    \includegraphics[width=0.9\textwidth]{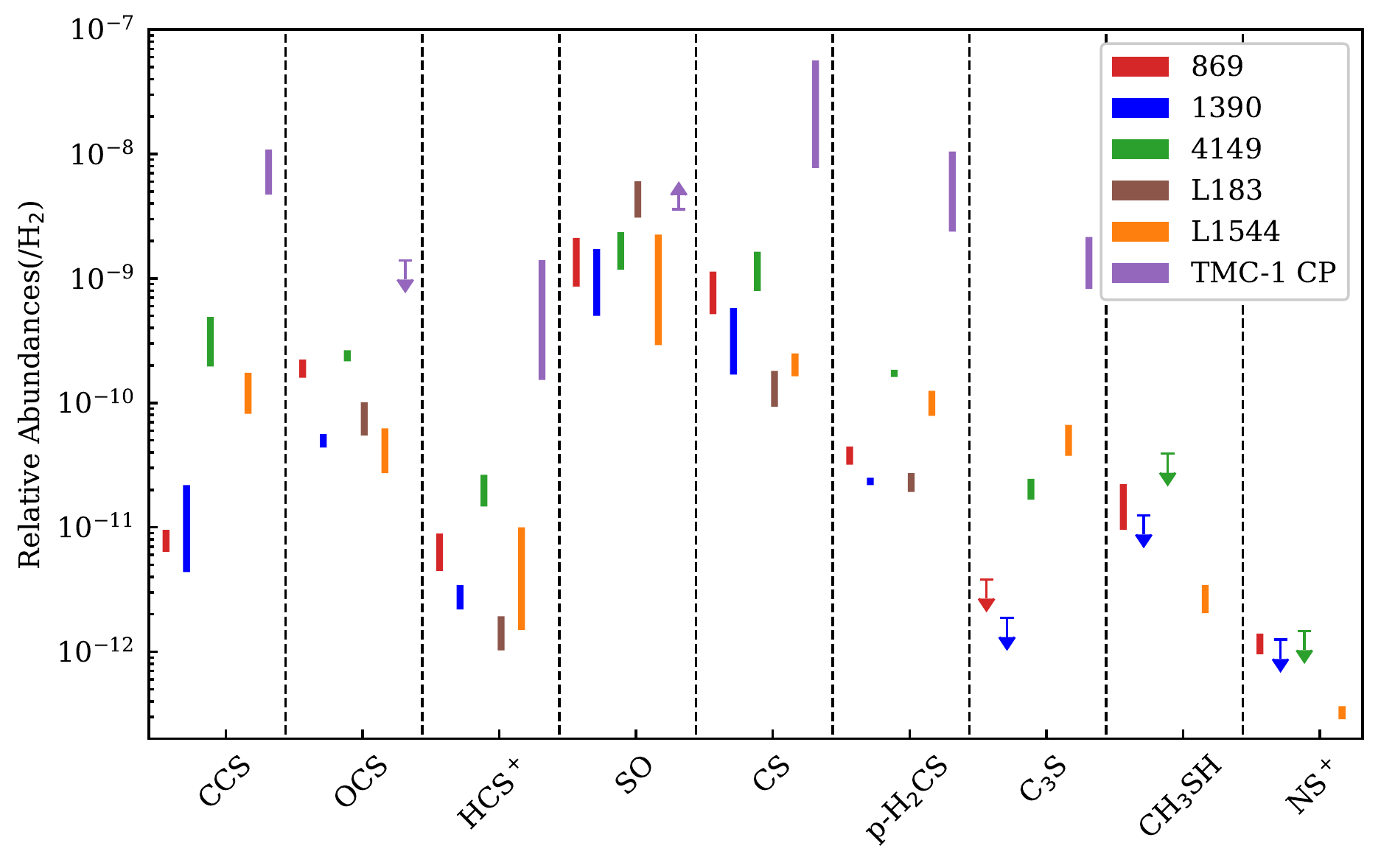} 
    \includegraphics[width=0.9\textwidth]{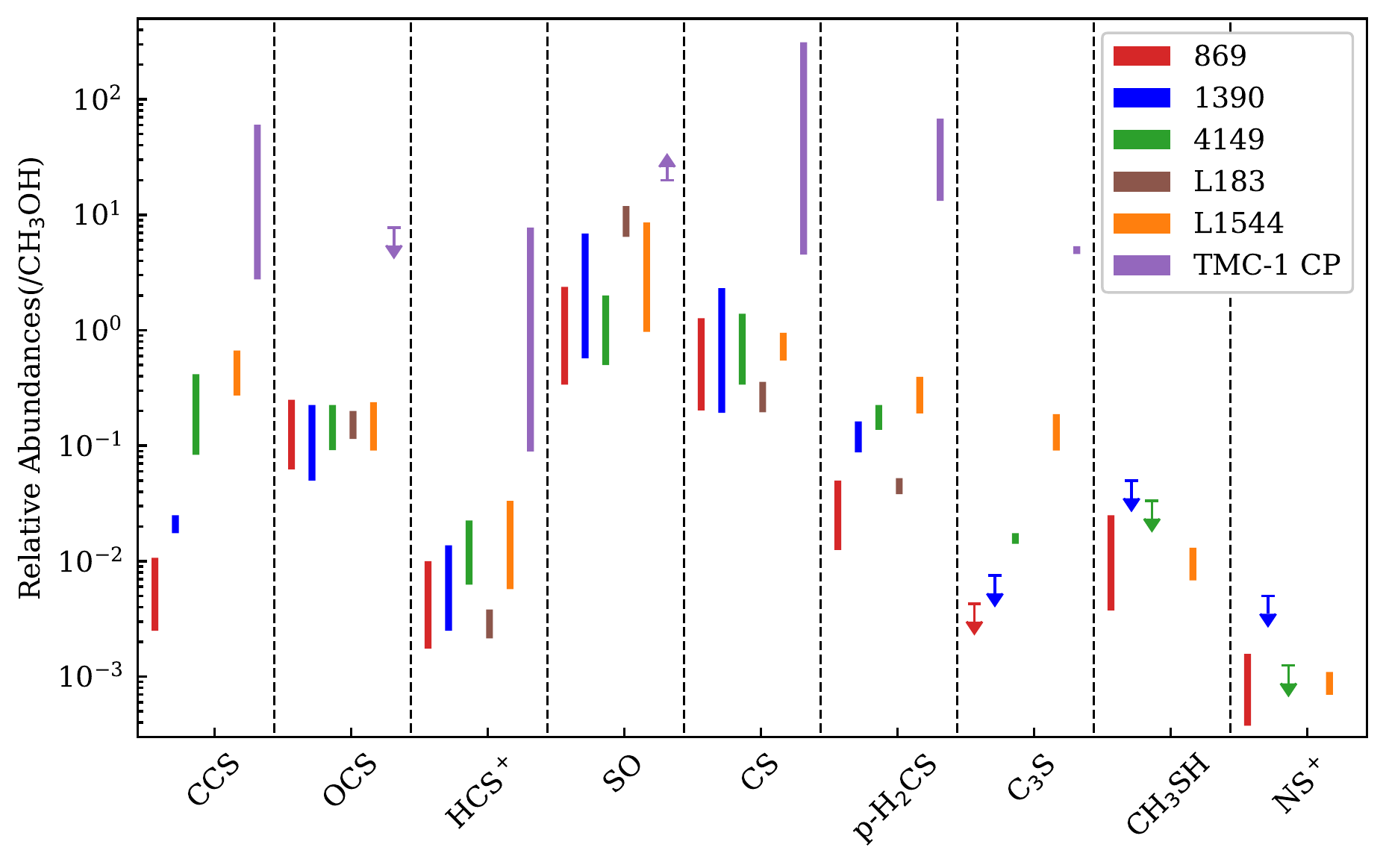}
    \caption{Relative abundances of sulphur-bearing species (main isotopologue) with H$_2$ (top). The $N_\mathrm{CS}$ and $N_\mathrm{SO}$ are calculated from $N_\mathrm{C^{34}S}$, $N_\mathrm{^{34}SO}$ and $^{32}$S$/^{34}$S$=(3.3 \pm 0.5)\,$(DGC/kpc)$+4.1 \pm 3.1  (\sim 32 \pm 5$ for our cores), adopted from \citet{Chin1996}. For the undetected species, we estimated the upper limit of their column densities based on the integrated area upper limit given in Table~\ref{tab: lines_para}. The sulphur-bearing species abundances are adopted from \citet{Lattanzi2020}, \citet{Vastel2018b}, and \citet{Gratier2016} for L183, L1544, and TMC-1 CP, respectively. The average N(H$_2$) of L183 and L1544 are $(10.0 \pm 1.7) \times 10^{22}$ cm$^{-2}$ and $(9.4 \pm 1.6) \times 10^{22}$ cm$^{-2}$ \citep{Crapsi2005ApJ} , and the average N(H$_2$) of TMC-1 CP is $(1.4 \pm 0.1) \times 10^{22}$ cm$^{-2}$ \citep{Kauffmann2008AA}. Relative abundances of sulphur-bearing molecules (main isotopologue) relative to CH$_3$OH (bottom).}
    \label{fig: S-bearing}
\end{figure*}

Here, we show the relative abundance compared to H$_2$ of the detected sulphur-bearing molecule (main isotopologue) towards our three cores along with L183, L1544, and TMC-1 in the upper panel of Fig.~\ref{fig: S-bearing}.  
L183 and L1544 are considered to be more evolved cores with a higher mass and column density, along with CO/CS depletion and deuterium-bearing species detections (see \citealt{Spezzano2017AA} and their references). In contrast, TMC-1 with a lower column density and D/H ratio is considered to be at an earlier evolutionary stage.
The main isotopologue abundances are computed by their $^{34}$S isotopologue. For L183 and L1544, the abundances of sulphur-bearing species are adopted from \citet{Lattanzi2020} and \citet{Vastel2018b}, respectively, and the abundances of sulphur-bearing species  of TMC-1 are adopted from \citet{Gratier2016}. The average N(H$_2$) are $(10.0 \pm 1.7) \times 10^{22}$ cm$^{-2}$ and $(9.4 \pm 1.6) \times 10^{22}$ cm$^{-2}$ for L183 and L1544, adopted from \citet{Crapsi2005ApJ}. The average N(H$_2$) of TMC-1 finally is $(1.4 \pm 0.1) \times 10^{22}$ cm$^{-2}$, derived by \citet{Kauffmann2008AA}. 

The highest abundances of sulphur-bearing species in our three cores, except for SO, are found toward 4149, while the lowest are found toward 1390. This is consistent with the result that the estimated S/H ratio of 1390 is lowest in the three cores. This strong sulphur depletion in the densest core 1390 supports the idea that sulphur depletion takes place during the evolution and contraction of dense cores.

Compared to L183, L1544, and TMC-1, the abundances of all sulphur-bearing molecules, except for SO, are highest in TMC-1. 
4149 has the second highest abundances of OCS, HCS$^+$, CS and H$_2$CS. The abundances of OCS, HCS$^+$ , and CS towards 1390, L183, and L1544 are comparable. In general, the relatively higher abundances of sulphur-bearing molecules towards 4149 indicate that this core is chemically younger than 1390 (more dense), not taking into account the effects of the environment, for which we still lack many constraints. All of our sources along with L183 and L1544 show a stronger sulphur depletion, which indicates that they are more evolved than TMC-1.

As sulphur-bearing molecules may deplete in the central densest region of the core, they mainly emit in the external layer, as shown in \citet{Vastel2018b}. The molecular abundances for the prestellar core candidates would therefore be underestimated as they are compared to the average H$_2$ column density of the whole core. Thus, we also show sulphur-bearing molecular abundances relative to methanol (Fig.~\ref{fig: S-bearing}, lower panel) because their emissions are co-existent \citep{Spezzano2017AA,Vastel2018b}.
The methanol column densities of L183, L1544 and TMC-1 (with similar spatial resolution to the current observations) are adopted from the same references providing sulphur-bearing molecule abundances. The relative abundances with methanol of CCS, OCS, HCS$^+$, CS, o-H$_2$CS toward 1390 and 4149 are comparable with the values toward L183 and L1544, while toward TMC-1, the relative abundances are still the highest. Thus the sulphur depletion of 1390 and 4149 still suggest that  they are more evolved than TMC-1.

\subsection{Carbon chains}

Carbon-chain chemistry has long been detected in dark clouds \citep[e.g.][]{Suzuki1992ApJ}, notably involving polyynes C$_n$H \citep{Tucker1974ApJ}, cyanopolyynes HC$_n$N \citep{Turner1970IAUC}, or cyclopropenylidene c-C$_3$H$_2$ \citep{Thaddeus1985ApJ}. Higher abundances of carbon chains were traditionally associated with younger cores, in the sense of less chemically evolved cores, making their comparison with oxygen-rich and highly-reduced species a convenient tool for chemical evolution diagnostics \citep[e.g.][]{Dickens2000ApJ}. Although the interest in carbon chains has somewhat shifted to protostellar objects following the discovery of warm carbon-chain chemistry (WCCC) in the envelope of a young stellar object \citep{Sakai2007ApJ}, the understanding of carbon-chain chemistry in starless and prestellar cores remains of high interest for several reasons. The presence of carbon chains near protostars may partly originate in the prestellar chemistry \citep{Aikawa2020ApJ, Kalvans2021ApJ}, but more importantly for our study, the current understanding is that the chemical differentiation observed from the comparison of carbon chains and oxygen-rich species such as methanol is caused by different physical (e.g. gas temperature) and environmental conditions. Recent observational \citep{Higuchi2018ApJS,Lattanzi2020} and modelling \citep{Aikawa2020ApJ,Kalvans2021ApJ} studies showed that the irradiation by interstellar photons can play a major role because in their absence, carbon atoms are locked in CO, which leads to oxygen-bearing species.

\begin{figure}[h!]
    \centering
    \includegraphics[width=0.95\columnwidth]{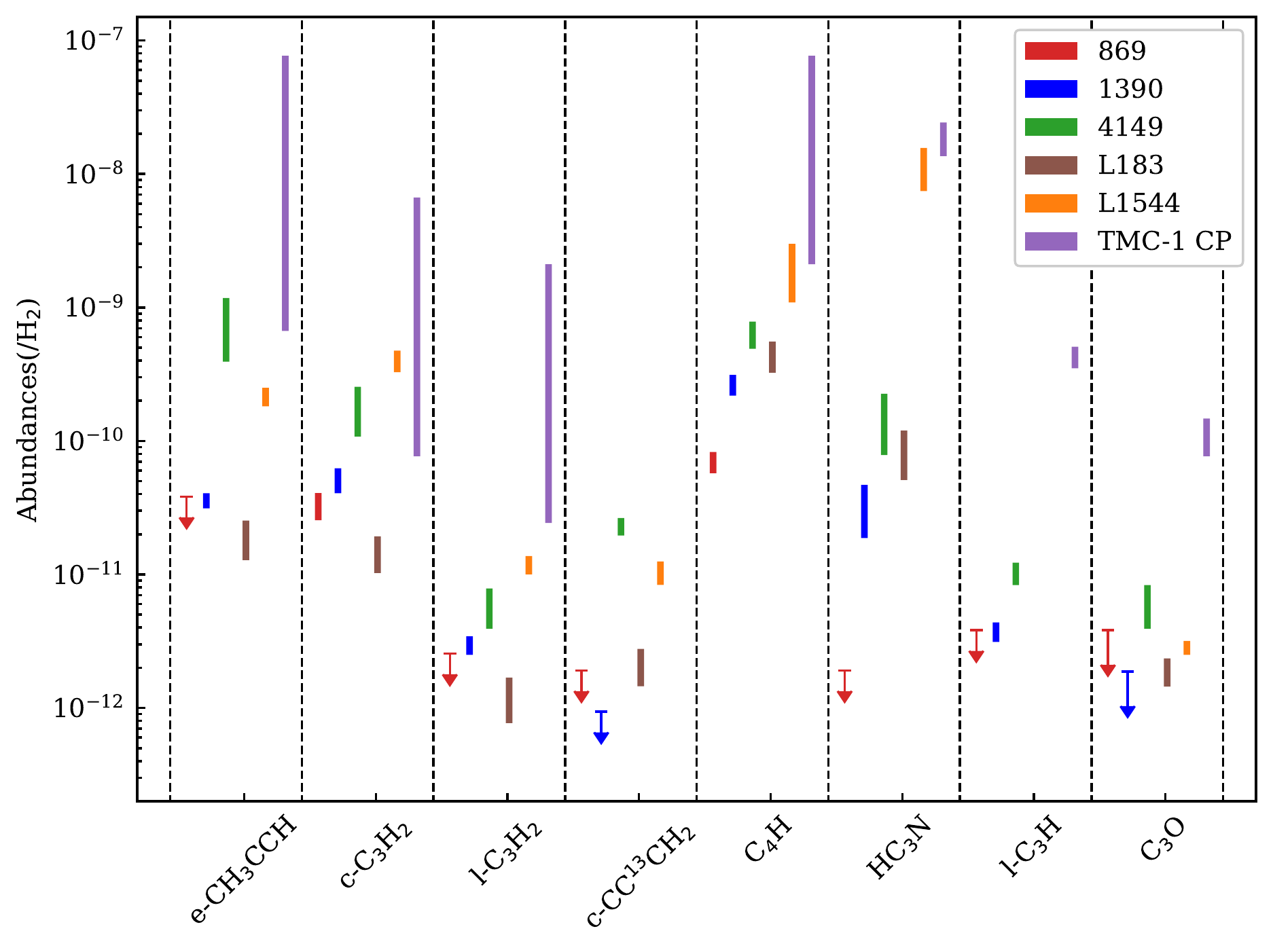} 
     
    \includegraphics[width=0.95\columnwidth]{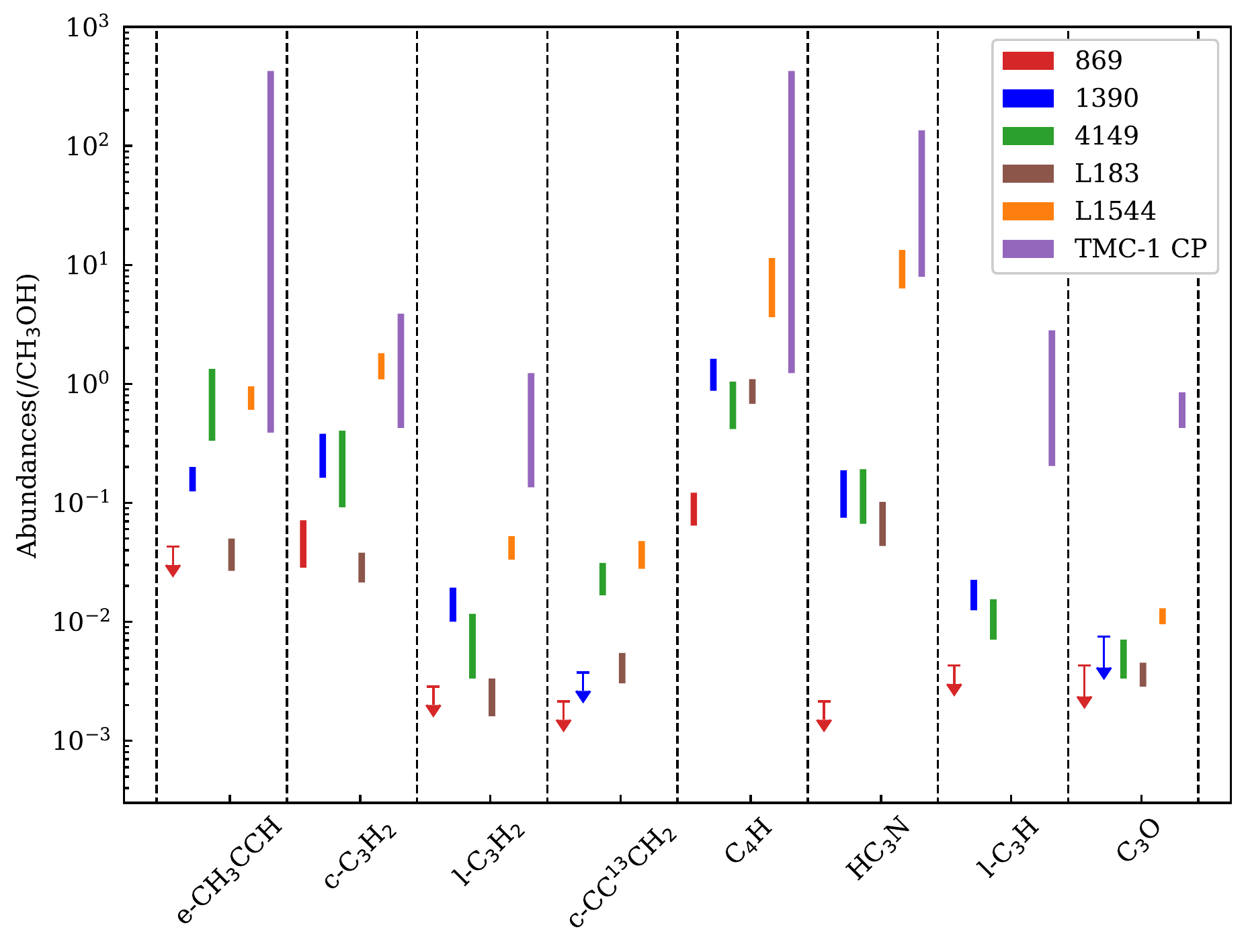}
    \caption{Abundances of carbon-chain species relative to molecular hydrogen (upper) and to methanol (lower). The rectangles show the lower and upper limits of the ratios. For the undetected species, we estimated the upper limit of their column densities based on the upper limit of the integrated intensity. The carbon-chain species and methanol abundances of L183, L1544, and TMC-1 are adopted from \citet{Lattanzi2020} and \citet{Gratier2016}, respectively. The l-C$_{3}$H column density in L183 ([2.5--3] $\times$ 10$^{11}$ cm$^{-2}$) is adopted from \citet{Turner2000} using a [10--15] K excitation temperature range. }
    \label{fig: CChain}
\end{figure}

Figure~\ref{fig: CChain} presents the abundances of the carbon chains detected in this study relative to H$_2$ as derived from dust continuum observations \citep{Montillaud2015}. The abundances of C$_4$H in our three prestellar targets are approximately within the $10^{-10}-10^{-9}$ range. As a comparison, \citet{Sakai2009ApJ} reported C$_4$H abundances $\gtrsim 10^{-9}$ in the two prototypical WCCC protostellar sources L1527 and IRAS15398, and $\sim 10^{-11}$ in the cold envelope surrounding the hot corino IRAS 16293. Although the large C$_4$H abundances might be thought to indicate a WCCC future for our cores, we recall that the abundance of C-chains varies rapidly with time in forming molecular cold cores, while they are the result of the sublimation of CH$_4$-rich ices in WCCC sources \citep[][and references therein]{Aikawa2020ApJ}. In addition, the photon-dominated region in which the cores are embedded can contribute to the observed abundance \citep[see e.g.][]{Bouvier2020}.

In c-C$_3$H$_2$ and C$_4$H, the abundances seem to increase from source 869 to 1390 and to 4149, in positive correlation with the overall chemical activity of each source as evaluated by the number of detected transitions (Table~\ref{tab: lines}). This trend is possibly even more significant for c-CC$^{13}$CH$_2$, although its transition was detected only in 4149, because the upper limits in the two other sources are lower than in this source by more than one order of magnitude. Similarly, for e-CH$_3$CCH, the abundance ratio of 4149 and 1390 is greater than ten and the abundance in 869 is probably lower than that in 1390. 

The interpretation of the differences among our three targets is difficult because the various possible causes are partly degenerate. To our knowledge, no comprehensive chemo-dynamical models dedicated to prestellar core evolution have been published so far. A few studies \citep{Aikawa2020ApJ, Kalvans2021ApJ} included the prestellar phase in their calculations, but only with the explicit goal of providing realistic conditions for the protostellar phase that was the focus of the studies. We attempt a comparison of these recent modelling works with our data below, however, while keeping this limitation in mind. 

\citet{Aikawa2020ApJ} modelled the chemical evolution of a Lagrangian fluid particle inside a $\sim 4$\,M$_\odot$ dense core during a static prestellar phase and then during its collapse. For C$_4$H and C$_3$H$_2$, they obtain abundances of about $10^{-10}$, in good agreement with our values, for $\sim$ 10$^3$ and 10$^4$ years, respectively (their Fig. 3). In their model, the abundance of both species increases during the first few $10^3$ to $10^4$ yr, and then is rather constant until a few $10^5$ yr. Assuming that the three sources are young, the increase in the abundance of C$_3$H$_2$ and C$_4$H from 869 to 4149 may reflect an increase in age. However, it would remain possible, for example, that source 1390 is the oldest of the three sources and has already reached the abundance decrease due to freeze-out after a few $10^5$ yr. 

\citet{Kalvans2021ApJ} computed the time evolution of chemical abundances during the collapse of a 1 M$_\odot$ dense core for most of the species detected here. They provide the time evolution for cores with strong or weak shielding from the external interstellar radiation field (ISRF) and cosmic rays. The order of magnitude of the observational abundances of the three sources is consistent with the models with weak irradiation for c-C$_3$H$_2$ and C$_4$H (their Fig.~5). Interestingly, the models predict larger abundances for more irradiated sources, while in our sample, the source we expect to be exposed to the most intense irradiation (1390) has only intermediate abundances for these carbon chains. This suggests that this source is strongly shielded from the $\lambda$ Ori flux.

These authors also computed the time evolution of propyne, also called methyl acetylene (CH$_3$CCH), for cases where either the ISRF or the  cosmic rays were not shielded (their Fig.~9). In their reference model, they find values of about $10^{-11}$ for most of the prestellar phase, which is broadly compatible with our estimates for 869 and 1390. However, the large abundance of $\sim 10^{-9}$ observed in 4149 only occurs in the reference model very soon before the appearance of the protostar. Alternatively, the model with a less shielded exposition to cosmic rays can also produce high propyne abundances. It is particularly interesting to put this point in perspective with the constraints on the cosmic ionisation rate $\zeta$ derived from HCO$^+$ and HCS$^+$ lines (Fig.~\ref{fig: contour_plot}), where 4149 was found to have a $\zeta$ value about ten times higher than the two other sources. 

The lower panel of Fig.~\ref{fig: CChain} shows the abundance of carbon chains with respect to methanol. Interestingly, because the relative abundance of methanol with respect to H$_2$ in 1390 is approximately three times lower than in the two other sources, this source presents overall higher ratios of carbon chain to methanol for l-C$_3$H$_2$, C$_4$H, and l-C$_3$H, although within the uncertainties, the values remain compatible with those in 4149. If this trend is genuine, following the interpretation of the ratio of carbon chain to methanol by \citet{Lattanzi2020}, 1390 would be the most highly irradiated of our three targets. This is in line with the fact that it lies at the border of the $\lambda$ Ori HII region, but contradicts the conclusion derived from the comparison of c-C$_3$H$_2$ and C$_4$H abundances in the model by \citet{Kalvans2021ApJ} and in our data. This discrepancy may be a sign of the model limitations in the prestellar phase. Alternatively, the source 1390 could be a source of similar evolutionary stage as 869, but with boosted c-C$_3$H$_2$ and C$_4$H abundances compared to 869 due to irradiation. It contradicts the fact that we detected deuterated species toward 1390, however, abundant in the most evolved cores, in the prestellar core phase. \\

Fig.~\ref{fig: CChain} shows carbon-chain abundances for the L183, TMC-1 CP, and L1544 dense cores. The L183 dense core is deeply embedded within the molecular cloud, protecting the molecular content from the interstellar radiation field. It is composed of an outer cloud region where $A_V < 15$ , a central part where $15 < A_V < 90$, and a central core for the densest region with $A_V > 90$ \citep{Pagani2004} with a peak extinction of $A_{V} \sim 150$ and a dust temperature close to 7 K. As a consequence, most of carbon is likely to be converted into CO in the central core region, leaving only a few C atoms to build up carbon chains. L1544, in contrast, lies close to the sharp edge of the cloud, making it more freely exposed to the interstellar radiation field. Chemical differentiation is therefore likely to be affected more by environmental conditions. For most species in Fig.~\ref{fig: CChain}, the abundances of our targets are framed by those observed in L183 and L1544, suggesting that our targets are systems with conditions intermediate between the well-shielded case of L183 and the well-irradiated case of L1544. A notable exception is C$_{4}$H, for which 869 and 1390 have abundances below L183.
The TMC-1 carbon-rich cyanopolyyne peak is embedded in the translucent filament TMC 1 with abundant carbon chains, as shown in Fig.~\ref{fig: CChain}, and it presents greater abundances in virtually all species than all other cores. The formation chemistry of cyanopolyynes and other unsaturated carbon-chain species is dominated by gas-phase reactions. Contribution of grain-surface reactions is negligible, as unsaturated carbon chains rapidly react with hydrogen on grain surfaces, producing saturated hydrocarbons. Unsaturated carbon-chain molecules are likely formed from reactions between smaller carbon chains through the addition of one or more carbon atoms to the backbone. It is still unclear why carbon chains are heavily produced in this particular core with no clear evidence of any large-scale influence of star formation.

A significant exception to the trend of the ratio of carbon chain to methanol is propyne, which presents increasing CH$_3$CCH/CH$_3$OH ratios from 869 to 1390 and to 4149: the value found for 4149 is indeed similar to that in the prototypical prestellar core L1544 \citep{Vastel2014}, while it is not detected in the earliest core of our sample, 869. This is likely to originate in the formation process of CH$_3$CCH on the grain surface, which, similarly to methanol, occurs through successive hydrogenation of physisorbed C$_3$ \citep{Hickson2016MolAs} by
\begin{eqnarray*}
\text{C$_3$} \xrightarrow {\text{H}} \text{c-C$_3$H} \xrightarrow {\text{H}} \text{c-C$_3$H$_2$} \xrightarrow {\text{H}} \text{CH$_2$CCH} \xrightarrow {\text{H}} \text{CH$_3$CCH.}
\end{eqnarray*}
However, in spite of the similarities in their formation process, propyne and methanol have very different time evolutions. The methanol abundance increases slowly, while propyne quickly reaches a plateau and suddenly increases after $\sim 10^5$ years (compare Fig.~2 in \citealt{Ruaud2015MNRAS} and Fig.~3 in \citealt{Hickson2016MolAs}). Based on these models, the propyne-to-methanol ratios suggest that 4149 is older than the two other sources. This contradicts the higher density found in 1390, but does not take environmental effects into account.\\

Finally, C$_3$O is the longest oxygen-bearing carbon chain observed in the interstellar medium. Very few detections have been reported, with one in the carbon-chain-rich cold core, TMC-1 with an abundance of $\sim$ 10$^{-11}$ \citep{Matthews1984Natur,Brown1985ApJ,Kaifu2004PASJ,Agundez2013ChRv}. It has also been detected in the L1544 prestellar core  by \citet{Vastel2014} with a higher abundance of 5 $\times$ 10$^{-11}$ and in L1498 and Elias 18 \citep{Urso2019}. \citet{Loison2014MNRAS} showed that the reactions between carbon-chain molecules and radicals (C$_{n}$ , C$_n$H, C$_n$H$_2$, C$_{2n+1}$O, etc.) with O atoms will produce large enough abundances of C$_3$O, whereas \citet{Urso2019} found the opposite and claimed that energetic processing of the ices would be a better explanation for the observed C$_3$O abundances. We report here the detection of C$_3$O in core 4149 with a lower abundance ([4-8] $\times$ 10$^{-12}$) compared to L1544 and TMC-1.

\subsection{iCOMs and precursors}

\begin{figure}[!h]
    \centering
    \includegraphics[width=0.95\columnwidth]{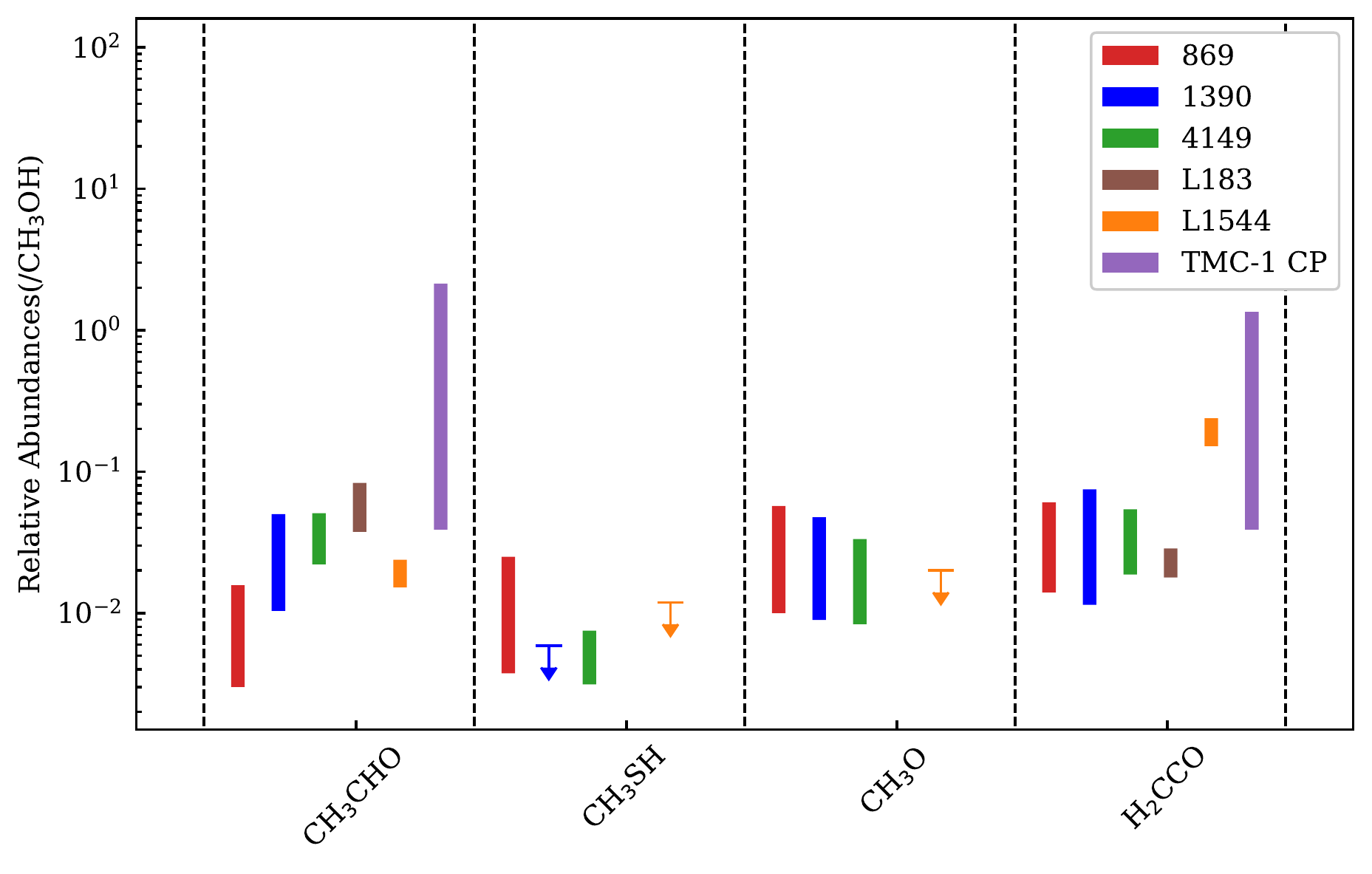}
    \caption{Abundances of iCOMs and precursors relative to methanol. For the undetected species, we estimated the upper limit  of their column densities based on the integrated intensity upper limit. }
    \label{fig: iCOMs}
\end{figure}

Figure \ref{fig: iCOMs} shows the variation of the detected iCOMs (and the methoxy precursor) in all three cores relative to methanol, which is abundant, detected in all three cores and supposedly a precursor to the major oxygen-bearing iCOMS. Only three iCOMS have been detected in all three cores: CH$_3$OH, CH$_3$CHO, and CH$_3$CCH (although the latter has been explained in Sec. \ref{sec: deut}).\\
Methanol is the most representative saturated organic molecule formed on dust grains, and it has been detected in all three sources, with a high signal-to-noise ratio. More complex species, such as methyl formate (HCOOCH$_3$) and dimethyl ether (CH$_3$OCH$_3$), are detected in the gas phase towards only a few prestellar cores. It is not clear yet how these are synthesised since, at the low temperatures, no grain surface reactions other than hydrogenation can occur and these large molecules cannot simply be released from the grain surfaces into the gas phase. Thus, it is proposed that they form via gas-phase reactions from simple species such as methanol (e.g. \citealt{Vasyunin2013, Vastel2014, Balucani2015, Jimenez-Serra2016}), injected from the grain mantles by non-thermal desorption processes such as chemical desorption and  cosmic rays \citep{Dartois2019}. 

In grain-surface chemistry, it has been proposed that CH$_3$CHO is formed as an intermediate species during sequential hydrogenation of H$_2$CCO (e.g., \citealt{Herbst2009}), leading to H$_3$CCO, CH$_3$CHO, C$_2$H$_5$O, and C$_2$H$_5$OH. H$_2$CCO should therefore exist on dust grains as a parent molecule, forming CH$_3$CHO. The detection of ketene and CH$_3$CHO in cold dense cores \citep{Ohishi1991ASPC, Vastel2014, Bacmann2012} may indicate that non-thermal desorption mechanisms are at work, in the same way as for methanol. From Fig. \ref{fig: iCOMs}, H$_2$CCO does not present any variation (compared to methanol) from one core to another. The values are a factor of ten lower than in L1544, which has a ratio of $\sim$ 0.2 \citep{Vastel2014}, although this value should be taken with caution, considering the high excitation temperature (27 K) compared to an average kinetic temperature of 10 K in this core. This might be explained by a variation in abundance from the densest and depleted regions of L1544 to the external less dense layer combined with opacity and non-LTE effects. Moreover, the CH$_3$CHO/CH$_3$OH ratios in 1390 and 4149 are comparable to the values found in the literature for prestellar cores (0.02 in L1544 \citep{Vastel2014} and 0.05 in L1689B \citep{Bacmann2012}), with a lower value for the less evolved core 869. Therefore, the suggested hydrogenation path for the formation of acetaldehyde is problematic. However, efficient paths of acetaldehyde in the gas phase exist, and they easily reproduce the observed abundances \citep[see the review by][]{Vazart2020}.

Methoxy (CH$_3$O) is an important species in this context as a simple precursor of iCOMs in general \citep[e.g.][]{Balucani2015}. It has first been identified through stacking measurements in the cold core B1-b \citep{Cernicharo2012ApJ}. It has not been detected in L1544 and has an abundance lower than 1.5 $\times$ 10$^{-10}$ (CH$_3$O/CH$_3$OH $<$ 2$\%$). It is clearly detected in core 869, but with a lower signal-to-noise ratio for 1390 and 4149, with CH$_3$O/CH$_3$OH similar to the limit in L1544. It is formed in the gas phase from the reaction between methanol and OH and possibly leads to the formation of methyl formate and dimethyl ether, which have not been detected in our cores, possibly due to the sensitivity limit of our observations. From Fig. \ref{fig: iCOMs}, methoxy seems to vary very little compared to methanol in the three cores, with a constant value for the CH$_3$CHO/CH$_3$O ratio. 

We report here the 5$\sigma$ detections of two transitions and 5$\sigma$ detection of one transition of methyl mercaptan in core 869 and core 4149. The detection has already been claimed in the cold environment of a cold core by \citet{Cernicharo2012ApJ} and in a prestellar core by \citet{Vastel2018b}. The CH$_3$SH/CH$_3$OH ratio is lower than the cosmic S/O ratio (0.024), but we cannot see a clear variation in the three cores. Methyl mercaptan has been proposed to be formed in the solid state via subsequent hydrogenation of CS (in the same way as methanol) and released in the gas phase through non-thermal desorption, so that no variation is expected between the sources.

\subsection{Physical properties compared with the prototypical prestellar core L1544}

In this section we review and compare the physical properties of our three cores and L1544 in Table~\ref{tab: comparison}. 
Dust temperatures, hydrogen column densities, and equivalent radius of our cores were adopted from \citet{Montillaud2015}.  The visual extinction was estimated assuming a standard N(H)/A$_{v}$ ratio of 1.9 $\times$ 10$^{21}$ atoms/cm$^{2}$/magnitude. The H$_{2}$ column densities (and therefore visual extinctions) are lower limits, as explained in Sec.~\ref{sec: Select}. 
The dust temperature of L1544 is 6 --13 K, adopted from \citet{Lattanzi2020}, and the gas temperatures toward our cores are determined by non-LTE fitting of methanol, while towards L1544, it is determined by NH$_3$ \citep{Crapsi2007}. Dust temperatures are similar in all four cores, but the gas temperature is much lower in L1544. This can be explained to the fact that NH$_{3}$, used to measure $\rm T_{gas}$ in L1544, is supposed to deplete at a later time, and therefore is a good tracer of the central depleted regions of the prestellar cores compared to methanol, which is mainly emitted in the external layers due to non-thermal desorption mechanisms \citep{Vastel2014}. The visual extinction in L1544 can be explained by the fact that the column densities measured from the Herschel observations in our three cores are probably underestimated. The sizes of the cores are similar and smaller than 0.1 pc. However, for cores 869, 1390, and 4149, the size values come from Herschel observations and are barely resolved: the sizes are close to the data resolution. The volume density computed from the emission lines of methanol are similar in all four cores, indicating a similar non-thermal desorption mechanism in the external layers of the cores. The cosmic-ray ionisation rate $\zeta_\mathrm{CR}$ of L1544 was obtained by \citet{Bovino2020MNRAS}, simulating the chemical evolution with a collapsing core, and comparing their results with the observations of H$_{2}$D$^{+}$ \citep{Vastel2006ApJ}. 4149 is comparable to L1544 in terms of the cosmic-ray ionisation rate and molecular complexity. The sulphur elemental abundance of L1544 was determined by \citet{Vastel2018b} using 21 sulphur-bearing species (ions, isotopomers, and deuteration) that have been detected in this ASAI spectral-survey through 51 transitions: CH$_3$SH, CS, CCS, C$_3$S, SO, SO$_2$, H$_2$CS, OCS, HSCN, NS, HCS$^+$, NS$^+$, and H$_2$S. LTE and non-LTE radiative transfer modelling was performed, and they used the NAUTILUS chemical code updated with the most recent chemical network for sulphur to explain their observations and therefore constrain the sulphur elemental abundance to 0.5$\%$ of the cosmic value of 1.5 $\times$ 10$^{-5}$. The sulphur elemental abundance in the gas phase is therefore much higher in our three cores, with values in the [0.2--7] $\times$ 10$^{-6}$ range, although constrained on HCS$^{+}$ only (see Sec.~\ref{sec: ions}). We also have to keep in mind that the analysis performed for L1544 took the density profile into account to compare with the abundances computed by the chemical model. Since we still lack a constrained density profile for our three cores, we only assumed a constant density and it might not make a difference in the case of the 869 cold core, but it certainly will make a difference in the case of the more advanced cores 1390 and 4149. L1544 seems to be in the most advanced stage of our three cores, with much higher visual extinction, a high CO depletion (for which we still lack information for our three cores) and a much higher density in its centre, which can explain the high sulphur depletion. \citet{Caselli2010} detected water vapour in the L1544 core and concluded that the emission and absorption signatures of the water spectrum indicate that the cloud is undergoing gravitational contraction and is therefore collapsing to form a new star. This prestellar core is likely in a much more advanced stage than 869, and we need more data on a larger scale for both 1390 and 4149 to compare them with L1544.

\begin{table*}
{\small
    \centering
    \caption{Comparison of the physical properties of our three cores and the prototypical prestellar core L1544.}
    \label{tab: comparison}
    \begin{tabular}{ccccc}
    \hline \hline
         &  869 & 1390 & 4149 & L1544 \\
         \hline
         Mass (M$_{\sun}$)   &  2.9\tablefootmark{b}  &  5.9\tablefootmark{b}  & 2.4\tablefootmark{b}  &  2.7 $\pm$ 0.7\tablefootmark{a}\\
        $T_\mathrm{dust}$ (K) & 12.0\tablefootmark{b} & 8.8\tablefootmark{b} & 7.5\tablefootmark{b} & 6 -- 13\tablefootmark{c} \\
        $T_\mathrm{gas}$ (K)  & [10 -- 14]\tablefootmark{b} & [10 -- 15]\tablefootmark{b} & [10 -- 15]\tablefootmark{b} & [5.5 -- 12.5]\tablefootmark{d} \\
        (Dust) $N_\mathrm{H_2} \,\times 10^{22}$ (cm$^{-2}$) & > 1.57\tablefootmark{b} & > 3.2\tablefootmark{b} & > 1.02\tablefootmark{b} & 6 -- 13\tablefootmark{a} \\
        $A_v$ (mag) & $\ge$ 16.5 & $\ge$ 33.7 & $\ge$ 10.7 & 63 -- 137 \\
        Size (pc) & 0.086\tablefootmark{b} & 0.102\tablefootmark{b} & 0.104\tablefootmark{b} & 0.086\tablefootmark{a} \\
        (Dust) $n_\mathrm{H_2} = \frac{N_\mathrm{H_2}}{Size}$ ($\times 10^3\,\mathrm{cm}^{-3}$) & > 59 & > 102 & > 32 & 226 -- 490 \\
        (non-LTE) $n_\mathrm{H_2}$ ($\times 10^3\,\mathrm{cm}^{-3}$) where CH$_{3}$OH emits  & 6 -- 30 & 20--100 & 15--37 & 30\tablefootmark{e} \\
        $\zeta_\mathrm{CR}$ ($\times 10^{-18\,}$s$^{-1}$) & 0.3 -- 1.5 & 0.4 -- 20 & 7 -- 10 & 20 -- 30\tablefootmark{f} \\
        S/H ($\times$ 10$^{-6}$) & 0.3 -- 3 & 0.2 --0.6 & 0.8 -- 7 & $\sim$0.08\tablefootmark{g} \\
        \hline
    \end{tabular}
    \tablefoot{
    \tablefoottext{a}{\citet{Doty2005}. }
    \tablefoottext{b}{The data of our cores are adopted from \citet{Montillaud2015}.}
    \tablefoottext{c}{\citet{Lattanzi2020}. }
    \tablefoottext{d}{\citet{Crapsi2007}. }
    \tablefoottext{e}{\citet{Vastel2014}.}
    \tablefoottext{f}{\citet{Bovino2020MNRAS}.}
    \tablefoottext{g}{\citet{Vastel2018b}.}
    
    }
    }
\end{table*} 

\section{Conclusions}
Three cold-core candidates ($\rm T_{dust}$ lower than 12 K) have been identified in the Galactic Cold Core catalogue. They represent diverse galactic environments: source 869 in the high-latitude MBM 12 molecular cloud (distance $\sim$ 325 pc), source 1390 in the compressed shell of the $\lambda$ Ori star (distance $\sim$ 445 pc), therefore affected by stellar feedback, and source 4149 in the California molecular cloud (distance $\sim$ 450 pc). 
\\
We obtained a spectral survey with the IRAM 30m telescope to explore the molecular complexity of the cores, and found large differences:  21, 29, and 37 species are detected in 869, 1390, and 4149, with 29, 52, and 70 transitions, respectively. We performed LTE and non-LTE radiative transfer modelling for all the detected species and compared their chemical evolution with the TMC-1 CP cold core and the L183 and L1544 prestellar cores. All three cores are cold (10-15 K) and dense ($\sim$ 10$^5$ cm$^{-3}$ for 869 and 4149, and a factor of two higher for 1390). Non-LTE modelling of methanol, which is likely emitted in the external layers of the densest cores, which are less strongly affected by molecular depletion, is consistent with a lower densities of a few 10$^{4}$ cm$^{-3}$.\\ 
Deuterated species have been detected in 1390 and 4149, leading to the conclusion that they are both prestellar cores. iCOMS and precursors have been also detected (CH$_3$OH, CH$_3$CCH, CH$_3$CHO, and C$_3$O, CH$_3$O), indicating  that non-thermal desorption mechanisms and gas-phase reactions are at work. Sulphur-bearing species are also detected in all three cores (with CH$_3$SH and C$_3$S). The higher abundances of sulphur-bearing molecules towards 4149 indicate that this core is chemically less evolved than 1390. We used the detected ionised species in our survey to estimate the cosmic-ray ionisation rate: [0.7--1] $\times$ 10$^{-17}$ s$^{-1}$ for 4149 (towards which CH$_3$CHO has the largest abundance), with lower values for 869 and 1390 ([0.3--1.5] $\times$ 10$^{-18}$ s$^{-1}$ and [0.4--20] $\times$ 10$^{-18}$ s$^{-1}$ , respectively). The abundance of sulphur is also a parameter in the chemical modelling, and we found that of the three cores, 1390 seems to have the lowest S/H elemental abundance, while 4149 has the highest. This is compatible with the fact that 1390 has the highest H$_2$ density and H$_2$ column density, which are propitious to a larger sulphur depletion, namely sulphur-bearing species frozen onto the grain mantles. Many carbon chains have also been detected, such as CH$_3$CCH, and the large abundance found in 4149 could be compatible with a later evolution, just before the formation of a protostar, or more likely in a case of a less shielded exposition to cosmic rays. 
\\
The cosmic-ray ionisation rate should be measured with more accuracy using mapping observations of HCO$^+$/DCO$^+$ and N$_2$H$^+$/N$_2$D$^+$. Moreover, to compare the results from our chemical exploration, the structure of the cores should be measured, for example with measurements of the NH$_3$ transitions, as depletion is an important factor for the chemistry of cold cores and prestellar cores.

\begin{acknowledgements}
This project has received funding from the European Union's Horizon 2020 research and innovation programme under the Marie Sklodowska-Curie grant agreement No 811312 for the Project "Astro-Chemical Origins'' (ACO).  This work is based on observations carried out under projects numbers 008-18 and 119-18 with the IRAM 30m telescope. IRAM is supported by INSU/CNRS (France), MPG (Germany) and IGN (Spain). Charlotte Vastel would like to acknowledge the IRAM staff for their help before and during the observations.
\end{acknowledgements}

\bibliographystyle{aa}
\bibliography{ColdCore_CV_Final.bbl}

\clearpage
\onecolumn

\begin{appendix}
\section{Details on the observed transitions}

\begin{longtable}{ccccccc}
\caption{\label{tab: lines} Identified molecular transitions in the three cores. } \\
\hline\hline
Transitions  & Frequency & E$_\mathrm{up}$ & A$_\mathrm{ij}$ & \multicolumn{3}{c}{S/N}\\
 & (MHz) & (K) & (s$^{-1}$) & 869  &  1390 & 4149 \\
\hline
\endfirsthead
\caption{continued.}\\
\hline\hline
Transitions  & Frequency & E$_\mathrm{up}$ & A$_\mathrm{ij}$ & \multicolumn{3}{c}{S/N}\\
 & (MHz) & (K) & (s$^{-1}$) & 869  &  1390 & 4149\\
\hline
\endhead
\hline
\endfoot

c-C$_3$HD  3$_{2, 2}$ -- 3$_{1, 3}$ & 79643.09 & 14.70 & 8.54$\times 10^{-6}$  & & 4 & \\
c-C$_3$HD  2$_{1, 2}$ -- 1$_{0, 1}$     & 79812.33 & 5.85        & 1.64$\times 10^{-5}$ &  & 60 & 147 \\
c-C$_3$HD  2$_{1, 1}$ -- 1$_{1, 0}$ & 95994.08 & 7.56 & 4.52$\times 10^{-6}$ & & 12 & 27\\
c-CC$^{13}$CH$_2$        2$_{0, 2}$ -- 1$_{1, 1}$ &     80047.54        & 6.30 & 1.66$\times 10^{-5}$ & & & 16\\
c-CC$^{13}$CH$_2$  3$_{1, 2}$ -- 3$_{0, 3}$ & 80775.35 & 15.72 & 9.20$\times 10^{-6}$ &  &  & 5\\
c-CC$^{13}$CH$_2$ 2$_{1, 2}$ -- 1$_{0, 1}$ & 84185.63 & 6.33 & 2.17$\times 10^{-5}$ & & & 16\\
H$_2$CCO 4$_{1,4}$ -- 3$_{1,3}$ & 80076.65 & 22.66 & 5.04$\times 10^{-6}$ & 16 & 15 & 15\\ 
H$_2$CCO 4$_{0,4}$ -- 3$_{0,3}$ & 80832.12 & 9.70 & 5.52$\times 10^{-6}$ & 11 & 7 & 10\\
H$_2$CCO 4$_{1,3}$ -- 3$_{1,2}$ & 81586.23 & 22.84 & 5.33$\times 10^{-6}$ & 13 & 6 & 13\\
H$_2$CCO 5$_{1,5}$ -- 4$_{1,4}$ & 100094.51 & 27.46 & 1.03$\times 10^{-5}$ & 12 & 8 & 10\\
H$_2$CCO 5$_{0,5}$ -- 4$_{0,4}$ & 101036.63 & 14.55 & 1.10$\times 10^{-5}$ & 7 &  & 7\\
C$_3$S 14 -- 13 & 80928.18 & 29.13 & 4.09$\times 10^{-5}$ & & & 9\\
CH$_2$DCCH 5$_{1, 5}$ -- 4$_{1, 4}$ & 80577.16 & 17.06 & 1.63$\times 10^{-6}$ & & & 6\\
CH$_2$DCCH 5$_{0, 5}$ -- 4$_{0, 4}$ &   80902.23 &      11.65 &         1.72$\times 10^{-6}$ & & & 10\\
CH$_2$DCCH 5$_{1, 4}$ -- 4$_{1, 3}$     & 81228.14 & 17.15 &    1.67$\times 10^{-5}$ & & & 6\\
CH$_2$DCCH 6$_{1, 6}$ -- 5$_{1, 5}$ & 96691.59 & 21.70 & 2.90$\times 10^{-6}$ & & & 5\\
CH$_2$DCCH 6$_{0, 6}$ -- 5$_{0, 5}$ &   97080.73 & 16.31 &      3.02$\times 10^{-6}$ & & & 10\\
CH$_2$DCCH 6$_{1, 5}$ -- 5$_{1, 4}$ & 97472.74 & 21.83 & 2.97$\times 10^{-6}$ & & & 7\\
HNO  1$_{0,1}$ -- 0$_{0,0}$ & 81477.49 & 3.91   & 2.23$\times 10^{-6}$ & 39 & 8 & 6\\
CCS  J$_N$=7$_6$ -- 6$_5$ & 81505.17 & 15.39 & 2.43$\times 10^{-5}$ & 9 & 53 & 129\\
CCS J$_N$=7$_8$ -- 6$_7$        & 99866.52 & 28.13 & 4.40$\times 10^{-5}$ & & 10 & 13 \\
HC$_3$N 9 -- 8 & 81881.47 & 19.65 & 4.21$\times 10^{-5}$ & & 198 & 226 \\
HC$_3$N 11 --10 & 100076.39 & 28.82 & 7.77$\times 10^{-5}$ & & 54 & 67\\
p-c-C$_3$H$_2$ 2$_{0, 2}$ -- 1$_{1, 1}$ & 82093.54 & 6.42 & 1.89$\times 10^{-5}$ & 13& 121 & 122\\
p-c-C$_3$H$_2$ 3$_{2, 2}$ -- 3$_{1, 3}$ & 84727.69 & 16.14 & 1.04$\times 10^{-5}$ &  & 16 & 12\\
l-C$_3$H$_2$ 4$_{1, 4}$ -- 3$_{1, 3}$ & 82395.09 & 23.24 & 4.56$\times 10^{-5}$ & & 12 & 8\\
l-C$_3$H$_2$ 4$_{0, 4}$ -- 3$_{0, 3}$ & 83165.34 & 9.98 & 5.00$\times 10^{-5}$ & & 7 & 5\\
l-C$_3$H$_2$ 4$_{1, 3}$ -- 3$_{1, 2}$ & 83933.70 &      23.43 &  4.82$\times 10^{-5}$ & & 15 & 10 \\
o-c-C$_3$H$_2$ 3$_{1, 2}$ -- 3$_{0, 3}$ & 82966.20 &    13.70 & 9.92$\times 10^{-6}$ & & 52 & 24\\
o-c-C$_3$H$_2$ 2$_{1, 2}$ -- 1$_{0, 1}$ & 85338.89 & 4.10 & 2.32$\times 10^{-5}$ & 43 & 264 & 261\\
CH$_3$O $N=1 - 0, K=0, J=3/2 - 1/2, F= 1 - 1, \Lambda =+1$  & 82455.98 & 3.96 & 6.52$\times 10^{-6}$ & 5 &  &  \\
CH$_3$O $N=1 - 0, K=0, J=3/2 - 1/2, F= 2 -1, \Lambda =-1$  & 82458.25 & 3.96 & 9.78$\times 10^{-6}$ & 6 &  & 4 \\
CH$_3$O $N=1 - 0, K=0, J=3/2 - 1/2, F=2 - 1, \Lambda =+1$  & 82471.83 & 3.97 & 9.78$\times 10^{-6}$ & 7 & 6 &  \\
HOCN  4$_{0,4}$ -- 3$_{0,3}$ & 83900.57 & 10.07 & 4.22$\times 10^{-5}$ & & & 7 \\
DC$_3$N 10 -- 9 & 84429.81      & 22.29 & 4.67$\times 10^{-5}$ & & 14 & 13 \\
OCS 7 -- 6 & 85139.10 & 16.34 & 1.71$\times 10^{-6}$ & 28 & 19 & 22\\
OCS 8 -- 7 & 97301.21 & 21.01 & 2.58$\times 10^{-6}$ & 21 & 12 & 14\\
o-D$_2$CS 3$_{0,3}$ -- 2$_{0,2}$ & 85153.92     & 8.18 & 8.47$\times 10^{-6}$ & & 17 & 22 \\
HC$^{18}$O$^+$ 1 -- 0 & 85162.22 & 4.09 & 3.64$\times 10^{-5}$ & 7 & 27 & 27\\
HCS$^+$  2 -- 1 & 85347.89 & 6.14 & 1.11$\times 10^{-5}$ & 17 & 13 & 35\\
e-CH$_3$CCH  5$_2$ -- 4$_2$ &   85450.77        & 41.21 & 1.70$\times 10^{-6}$ & & & 4\\
e-CH$_3$CCH 5$_1$ -- 4$_1$ &    85455.67        & 19.53 & 1.95$\times 10^{-6}$ & & 11 & 35\\
e-CH$_3$CCH 5$_0$ -- 4$_0$      & 85457.30 & 12.30 & 2.03$\times 10^{-6}$ & & 9 & 36\\
HOCO$^+$  4$_{0,4}$ -- 3$_{0,3}$ & 85531.50     & 10.26 & 2.36$\times 10^{-5}$ & & 9 & 9\\
C$_4$H N = 9--8 J = 9.5--8.5\tablefootmark{a}& 85634.00 & 20.55 & 2.60$\times 10^{-6}$ & 5 & 44 & 46\\
C$_4$H N = 9--8 J = 8.5--7.5\tablefootmark{b} & 85672.58 & 20.56 & 2.59$\times 10^{-6}$ & 4 & 40 & 51\\
C$_4$H N = 10--0 J= 10.5--9.5\tablefootmark{c} & 95150.39 & 25.11 & 3.60$\times 10^{-6}$ &  & 28 & 18\\
C$_4$H N = 10--0 J= 9.5--8.5\tablefootmark{d} & 95188.95        & 25.13 & 3.58$\times 10^{-6}$ &  & 26 & 16\\
NH$_2$D 1$_{1,1}$ -- 1$_{0,1}$ F= 0--1  & 85924.78 & 20.09 & 7.81$\times 10^{-6}$ &  & 87 & 41\\
NH$_2$D 1$_{1,1}$ -- 1$_{0,1}$ F= 2--1  &       85925.70 & 20.09 & 1.95$\times 10^{-6}$ &  & 107 & 58\\
NH$_2$D 1$_{1,1}$ -- 1$_{0,1}$ F= 2--2\tablefootmark{e}  & 85926.27     & 20.09   & 5.86$\times 10^{-6}$ &  & 112 & 117\\
NH$_2$D 1$_{1,1}$ -- 1$_{0,1}$ F= 1--2  & 85926.88      & 20.09 & 3.25$\times 10^{-6}$ &  & 107 & 47\\
NH$_2$D 1$_{1,1}$ -- 1$_{0,1}$ F= 1--0  &       85927.73        & 20.09 & 2.60$\times 10^{-6}$ &  & 94 & 48\\
E-CH$_3$CHO 5$_{0, 5}$ -- 4$_{0, 4}$ & 95947.44 & 13.84 & 2.95$\times 10^{-5}$ & 4 & 10 & 7\\
E-CH$_3$CHO 3$_{1, 3}$ -- 2$_{0, 2}$ & 101343.44 & 7.64 & 3.90$\times 10^{-6}$ &  &  & 4\\
A-CH$_3$CHO 5$_{0, 5}$ -- 4$_{0, 4}$ & 95963.46 & 13.84 & 2.95$\times 10^{-5}$ & 5 & 5 & 9\\
A-CH$_3$CHO 5$_{1, 4}$ -- 4$_{1, 3}$ & 98900.94 & 16.51 & 3.10$\times 10^{-5}$ &  &  & 5\\
C$_3$O 10 -- 9 & 96214.62 & 25.40 & 2.82$\times 10^{-5}$ & & & 5\\
C$^{34}$S 2 -- 1 & 96412.95 & 6.94 & 1.60$\times 10^{-5}$ & 57 & 64 & 65\\
E-CH$_3$OH 5$_{-1}$ -- 4$_{0}$ & 84521.17 & 32.49       & 1.97$\times 10^{-6}$ &  & 4 & 6\\
E-CH$_3$OH 2$_{-1}$ -- 1$_{-1}$ & 96739.36 & 4.64 & 2.55$\times 10^{-6}$ & 204 & 229 & 251\\
E-CH$_3$OH 2$_{0}$ -- 1$_{0}$ & 96744.55 & 12.19 & 3.40$\times 10^{-6}$ & 18 & 16 & 29\\
A-CH$_3$OH 2$_{0}$ -- 1$_{0}$ & 96741.37 & 6.97 & 3.40$\times 10^{-6}$  & 267 & 169 & 321\\
A-CH$_3$OH 2$_{1}$ -- 1$_{1}$ & 97582.80 & 21.56 & 2.62$\times 10^{-6}$ &  &  & 4\\
$^{34}$SO  2$_3$ -- 1$_2$ & 97715.32     & 9.09 & 1.07$\times 10^{-5}$  & 55 & 69 & 33\\
c-C$_3$D$_2$ 3$_{1,3}$ -- 2$_{0,2}$ & 97761.98  & 9.88 & 3.88$\times 10^{-5}$ &  & 6 & 6\\
CS 2 -- 1 & 97980.953 & 7.05 & 1.68$\times 10^{-5}$ & 143 & 407 & 38\\
l-C$_3$H $^{2}\Pi_{\frac{1}{2}}$ J=9/2-7/2 b F=4-3 & 97995.91 & 12.54 & 5.95$\times 10^{-5}$ & & & 7\\
l-C$_3$H $^{2}\Pi_{\frac{1}{2}}$ J=9/2-7/2 a F=5-4\tablefootmark{a} & 98011.61  & 12.54 & 6.13$\times 10^{-5}$ & & 7 & 7\\
l-C$_3$H $^{2}\Pi_{\frac{1}{2}}$ J=9/2-7/2 a F=4-3 & 98012.52 & 12.54 & 5.96$\times 10^{-5}$ & & 7 & 7\\
$^{33}$SO 2$_{3,4.5}$ -- 1$_{2,3.5}$ & 98493.64 & 9.16 & 1.10$\times 10^{-5}$ & & 6 & \\
SO N$_{J}$ = 2$_3$ -- 1$_2$ & 99299.87 & 9.23   & 1.13$\times 10^{-5}$ & 428 & 819 & 223\\
NS$^+$ 2--1 & 100198.47 & 7.21 & 2.21$\times 10^{-5}$ & 4 & & \\
CH$_3$SH 4$_{0, 4, 0}$ -- 3$_{0, 3, 0}$ & 101139.11 & 12.14 & 9.23$\times 10^{-6}$ & 5 & & 5\\
CH$_3$SH 4$_{1, 3, 1}$ -- 3$_{1, 2, 1}$ & 101284.35 & 18.33 & 8.69$\times 10^{-6}$ & 5 & & \\
o-H$_2$C$^{34}$S 3$_{0,3}$ -- 2$_{0,2}$ & 101284.31 & 9.72 & 1.41$\times 10^{-5}$ & & & 5\\
o-H$_2$CS 3$_{1, 3}$ -- 2$_{1, 2}$ &    101477.80       & 8.12 & 1.26$\times 10^{-5}$ & 22 & 36 & 82 \\

\end{longtable}

\tablefoot{\\
S/N represents the signal-to-noise ratio, and we only quote values higher than or equal to 4$\sigma$.\\
\tablefoottext{a}{The nearby l-C$_3$H 97995.1660 MHz transition $^{2}\Pi_{\frac{1}{2}}$ J=9/2-7/2 b F=5-4 falls in one of the frequency-switching absorption lines. We adopted the quantum classification by \citet{Loison2017}.}
\tablefoottext{a}{The line is blended with C$_4$H N,$_{J, K}$=9$_{9.5, 10}$ -- 8$_{8.5, 9}$.} \\
\tablefoottext{b}{The line is blended with C$_4$H N,$_{J, K}$=9$_{8.5, 9}$ -- 8$_{7.5, 8}$.} \\
\tablefoottext{c}{The line is blended with C$_4$H N,$_{J, K}$=10$_{10.5, 11}$ -- 9$_{9.5, 10}$.} \\
\tablefoottext{d}{The line is blended with C$_4$H N,$_{J, K}$=10$_{9.5, 10}$ -- 9$_{8.5, 9}$.} \\
\tablefoottext{e}{The line is blended with 1$_{1,1}$ -- 1$_{0,1}$ F= 1--1.}
}

\begin{landscape}
\begin{longtable}{*{10}c}
\caption{\label{tab: lines_para} Measured parameters of identified molecular transitions in the three cores. } \\
\hline\hline
Transitions  & \multicolumn{3}{c}{869} & \multicolumn{3}{c}{1390} & \multicolumn{3}{c}{4149}\\
  & W & FWHM & V$_\mathrm{LSR}$  & W & FWHM & V$_\mathrm{LSR}$  & W & FWHM & V$_\mathrm{LSR}$\\
 & (mK km\,s$^{-1}$) & (km\,s$^{-1}$) & (km\,s$^{-1}$) & (mK km\,s$^{-1}$) & (km\,s$^{-1}$) & (km\,s$^{-1}$) & (mK km\,s$^{-1}$) & (km\,s$^{-1}$) & (km\,s$^{-1}$) \\
\hline
\endfirsthead
\caption{continued.}\\
\hline\hline
Transitions  & \multicolumn{3}{c}{869} & \multicolumn{3}{c}{1390} & \multicolumn{3}{c}{4149} \\
  &W & FWHM & V$_\mathrm{LSR}$  & W & FWHM & V$_\mathrm{LSR}$  & W & FWHM & V$_\mathrm{LSR}$\\
 & (mK km\,s$^{-1}$) & (km\,s$^{-1}$) & (km\,s$^{-1}$) & (mK km\,s$^{-1}$) & (km\,s$^{-1}$) & (km\,s$^{-1}$) & (mK km\,s$^{-1}$) & (km\,s$^{-1}$) & (km\,s$^{-1}$) \\
\hline
\endhead
\hline
\endfoot

c-C$_3$HD  3$_{2, 2}$ -- 3$_{1, 3}$  & $\leq$4.0 & &  & 4.5(1.2) & 0.35(0.10) & 11.2 & $\leq$3.6 & &\\
c-C$_3$HD  2$_{1, 2}$ -- 1$_{0, 1}$ & $\leq$4.0 & & & 86.9(8.8) & 0.62(0.01) & 11.2  & 193(19.3) & 0.51(0.01) & -2.0\\
c-C$_3$HD  2$_{1, 1}$ -- 1$_{1, 0}$& $\leq$4.9 & &  & 19.7(2.6) & 0.58(0.05) & 11.2 & 39.3(4.2) & 0.42(0.03) & -2.0\\
c-CC$^{13}$CH$_2$        2$_{0, 2}$ -- 1$_{1, 1}$ & $\leq$4.0 & & & $\leq$4.4 & & & 17.9(2.1) & 0.37(0.03) & -2.1\\
c-CC$^{13}$CH$_2$ 2$_{1, 2}$ -- 1$_{0, 1}$ & $\leq$4.0 & & & $\leq$4.4 & & & 18.1(2.2) & 0.39(0.03) & -1.8\\
c-CC$^{13}$CH$_2$  3$_{1, 2}$ -- 3$_{0, 3}$ & $\leq$3.9 & & & & $\leq$4.3 & & 5.8(1.4) & 0.48(0.14) & -2.0\\
H$_2$CCO 4$_{1,4}$ -- 3$_{1,3}$ & 23.6(2.7) & 0.61(0.05) & -5.0 & 23.1(2.7) & 0.75(0.06) & 11.4 & 19.1(2.2)& 0.47(0.04) & -2.0\\ 
H$_2$CCO 4$_{0,4}$ -- 3$_{0,3}$  & 15.8(2.0) & 0.58(0.07) & -5.0 & 12.7(1.9)& 0.87(0.15) & 11.4 & 13.0(1.8) & 0.46(0.06) & -1.9\\
H$_2$CCO 4$_{1,3}$ -- 3$_{1,2}$ & 19.4(2.4) & 0.63(0.06) & -5.0  & 11.2(1.8) & 0.89(0.13) & 11.4 & 15.3(1.9) & 0.41(0.03) & -2.0\\
H$_2$CCO 5$_{1,5}$ -- 4$_{1,4}$ & 18.0(2.3) & 0.47(0.04) & -5.0 & 14.1(2.3) & 0.7(0.3) & 11.4 & 13.0(1.9) & 0.38(0.04) & -1.9\\
H$_2$CCO 5$_{0,5}$ -- 4$_{0,4}$ & 11.7(2.0) & 0.53(0.07) & -4.9 & $\leq$5.3 &&& 10.2(1.7) & 0.41(0.05) & -1.9\\
C$_3$S 14 -- 13 & $\leq$ 1.6 &  &  & $\leq$4.0 & & & 10.3(1.6) & 0.41(0.05) & -1.9\\
CH$_2$DCCH 5$_{1, 5}$ -- 4$_{1, 4}$ & $\leq$4.0 & & & $\leq$4.4 & & & 6.7(1.2) & 0.32(0.07) & -1.9\\
CH$_2$DCCH 5$_{0, 5}$ -- 4$_{0, 4}$ & $\leq$4.0 & & & $\leq$4.4 & & & 11.9(1.6) & 0.44(0.06) & -1.9\\
CH$_2$DCCH 5$_{1, 4}$ -- 4$_{1, 3}$ & $\leq$4.0 & & & $\leq$4.4 & & & 7.7(1.4) & 0.43(0.07) & -1.9\\
CH$_2$DCCH 6$_{1, 6}$ -- 5$_{1, 5}$ & $\leq$4.9 & & & $\leq$5.4 & & & 7.2(1.6) & 0.40(0.04) & -2.0\\
CH$_2$DCCH 6$_{0, 6}$ -- 5$_{0, 5}$ & $\leq$4.9 & & & $\leq$5.4 & & & 12.8(1.9) & 0.34(0.03) & -2.0\\
CH$_2$DCCH 6$_{1, 5}$ -- 5$_{1, 4}$ & $\leq$4.9 & & & $\leq$5.4  & & & 8.9(1.5) & 0.32(0.04) & -2.0\\
HNO  1$_{0,1}$ -- 0$_{0,0}$ & 57.8(5.9) & 0.67(0.02) & -5.3 & 12.1(1.8) & 0.62(0.07) & 11.1 & 7.3(1.4) & 0.46(0.08) & -2.2\\
CCS  N$_J$=6$_7$ -- 5$_6$  & 11.7(1.7) & 0.49(0.02) &-5.1 & 72.4(7.3) & 0.57(0.01) & 11.1 & 157.1(15.8) & 0.45(0.01) & -2.1\\
CCS N$_J$=8$_7$ -- 7$_6$ & $\leq$4.8 & & & 13.5(1.9) & 0.39(0.05) & 11.3 & 15.7(2.0) & 0.29(0.03) & -1.8 \\
HC$_3$N 9 -- 8  & $\leq$4.0 & & & 273.6(27.4) & 0.58(0.01) & 11.3 & 265.8(26.7) & 0.42(0.01) & -1.9 \\
HC$_3$N 11 --10 & $\leq$4.8 & & & 88.9(9.0) & 0.57(0.01) & 11.3 & 88.6(8.9) & 0.36(0.01) & -1.9\\
p-c-C$_3$H$_2$ 2$_{0, 2}$ -- 1$_{1, 1}$ & 17.9(2.2) & 0.61(0.05) & -5.0 & 168.1(16.9) & 0.59(0.01) & 11.3 & 151.7(15.2) & 0.47(0.01) & -1.9\\
p-c-C$_3$H$_2$ 3$_{2, 2}$ -- 3$_{1, 3}$ & $\leq$3.9 & & & 20.4(2.5) & 0.53(0.05) & 11.3 & 11.5(1.6) & 0.31(0.04) & -1.9\\
l-C$_3$H$_2$ 4$_{1, 4}$ -- 3$_{1, 3}$ & $\leq$4.0 & & & 18.6(2.3) & 0.69(0.07) & 11.2 & 10.9(11.3) & 0.51(0.05) & -2.0\\
l-C$_3$H$_2$ 4$_{0, 4}$ -- 3$_{0, 3}$ & $\leq$4.0 & & & 8.9(1.6) & 0.51(0.08) & 11.4 & 7.7(1.3) & 0.62(0.08) & -2.1\\
l-C$_3$H$_2$ 4$_{1, 3}$ -- 3$_{1, 2}$ & $\leq$3.9 & & & 19.2(2.3) & 0.53(0.04) & 11.3 & 11.1(1.6) & 0.38(0.05) & -1.8\\
o-c-C$_3$H$_2$ 3$_{1, 2}$ -- 3$_{0, 3}$  & $\leq$3.9 & & & 67.1(6.8) & 0.52(0.02) & 11.2 & 28.1(3.1) & 0.41(0.02) & -2.0\\
o-c-C$_3$H$_2$ 2$_{1, 2}$ -- 1$_{0, 1}$  & 62.7(6.4) & 0.66(0.02) & -4.9 & 366.8(36.7) & 0.61(0.01) & 11.3 & 302(30.4) & 0.42(0.01) & -1.9\\
CH$_3$O {\small $N=1 - 0, K=0, J=3/2 - 1/2, F=1 - 1, \Lambda =-1$}  & 5.9(1.5) & 0.46(0.04) & -4.9 & $\leq$4.3 & & & $\leq$3.5 &  &  \\
CH$_3$O {\small $N=1 - 0, K=0, J=3/2 - 1/2, F=2 - 1, \Lambda =-1$}  & 7.6(1.5) & 0.49(0.05) & -4.9 & $\leq$4.3 & & & 4.8(1.3) & 0.47(0.11) & -1.9 \\
CH$_3$O {\small $N=1 - 0, K=0, J=3/2 - 1/2, F=2 - 1, \Lambda =+1$}  & 9.9(1.6) & 0.63(0.10) & -4.9 & 7.7(1.6) & 0.46(0.10) & 11.4 & $\leq$3.5 &  &  \\
HOCN  4$_{0,4}$ -- 3$_{0,3}$  & $\leq$3.9 & & & $\leq$4.3 & & & 8.5(1.4) & 0.52(0.07) & -1.9 \\
DC$_3$N 10 -- 9 & $\leq$3.9 & & & 19.1(2.3) & 0.62(0.05) & 11.2 & 14.5(1.8) & 0.41(0.03) & -1.9 \\
OCS 7 -- 6 & 38.1(4.0) & 0.57(0.03) & -4.9 & 29.5(3.4) & 0.84(0.06) & 11.2 & 24.2(2.7) & 0.38(0.02) & -1.9\\
OCS 8 -- 7 & 34.4(3.8) & 0.53(0.03) & -4.9 & 24.1(3.1) & 0.80(0.08) & 11.3 &  18.7(2.3) & 0.37(0.02) & -1.9\\
o-D$_2$CS  & $\leq$3.9 & & & 18.9(2.3) & 0.39(0.05) & 11.3 & 23.1(2.5) & 0.36(0.02) & -1.9 \\
HC$^{18}$O$^+$ 1 -- 0  & 8.8(1.5) & 0.49(0.08) & -4.9 & 37.7(4.0) & 0.59(0.03) & 11.2 & 28.5(3.0) & 0.36(0.02) & -2.0\\
HCS$^+$  2 -- 1  & 21.7(2.5) & 0.52(0.04) & -4.8 & 18.5(2.3) & 0.68(0.06) & 11.5 & 40.0(4.2) & 0.42(0.01) & -1.8\\
e-CH$_3$CCH  5$_2$ -- 4$_2$  & $\leq$3.9   & & & $\leq$4.3   & & & 4.7(1.2) & 0.43(0.11) & -2.0\\
e-CH$_3$CCH 5$_1$ -- 4$_1$  & $\leq$3.9   & & & 18.6(2.5) & 0.92(0.09) & 11.4 & 40.0(4.1) & 0.43(0.01) & -1.9\\
e-CH$_3$CCH 5$_0$ -- 4$_0$  & $\leq$3.9        & & & 13.4(1.9) & 0.78(0.10) & 11.3 & 42.1(4.4) & 0.42(0.01) & -1.9\\
HOCO$^+$  4$_{0,4}$ -- 3$_{0,3}$ & $\leq$3.9   & & & 10.2(1.7) & 0.45(0.08) & 11.3 & 10.2(1.5)  & 0.42(0.05) & -1.9\\
C$_4$H N = 9--8 J = 9.5--8.5   & 9.6(2.0) & 0.95(0.19) & -5.0 & 58.8(6.0) & 0.58(0.02) & 11.4 & 48.6(5.0) & 0.35(0.01) & -1.9\\
C$_4$H N = 9--8 J = 8.5--7.5   & 4.3(1.2) & 0.43(0.13) & -5.0 & 53.5(5.7) & 0.58(0.02) & 11.3 & 49.5(5.1) & 0.30(0.01) & -1.9\\
C$_4$H N = 10--0 J= 10.5--9.5 & $\leq$4.9   & & & 49.8(5.2) & 0.63(0.02) & 11.4 & 23.9(2.7) & 0.36(0.02) & -1.9\\
C$_4$H N = 10--0 J= 9.5--8.5   & $\leq$4.9   & & & 47.8(5.1) & 0.66(0.02) & 11.4 & 22.1(2.6) & 0.37(0.02) & -1.9\\
NH$_2$D 1$_{1,1}$ -- 1$_{0,1}$ F = 0--1   & $\leq$3.9   & & &  111.1(11.2) & 0.52(0.01) & 11.3 & 49.8(5.1) & 0.48(0.01) & -1.8\\
NH$_2$D 1$_{1,1}$ -- 1$_{0,1}$ F = 2--1   & $\leq$3.9   & & & 141.2(14.2) & 0.55(0.01) & 11.2 & 70.7(7.2) & 0.47(0.01) & -1.9\\
NH$_2$D 1$_{1,1}$ -- 1$_{0,1}$ F = 2--2  & $\leq$3.9   & & & 145.3(14.6) & 0.54(0.01) & 11.3 & 159.8(16.0) & 0.59(0.01) & -1.9\\
NH$_2$D 1$_{1,1}$ -- 1$_{0,1}$ F = 1--2   & $\leq$3.9   & & & 139.4(14.1) & 0.54(0.01) & 11.3 & 58.8(6.0) & 0.50(0.01) & -1.8\\
NH$_2$D 1$_{1,1}$ -- 1$_{0,1}$ F = 1--0 &  $\leq$3.9 & & & 133.6(13.9) & 0.65(0.01) & 11.3 & 58.1(5.9) & 0.46(0.01) & -1.9\\
E-CH$_3$CHO 5$_{0, 5}$ -- 4$_{0, 4}$ & 5.8(1.6) & 0.45(0.11) & -4.9 & 14.5(2.0) & 0.40(0.05) & 11.4 & 10.6(1.8) & 0.43(0.04) & -2.0\\
E-CH$_3$CHO 3$_{1, 3}$ -- 2$_{0, 2}$  & $\leq$4.8   & & & $\leq$5.2  & & & 5.1(1.4) & 0.34(0.08) & -2.0\\
A-CH$_3$CHO 5$_{0, 5}$ -- 4$_{0, 4}$ & 5.7(1.4) & 0.31(0.06) & -4.9 & $\leq$5.3 &  &  & 12.3(1.8) & 0.38(0.04) & -1.9\\
A-CH$_3$CHO 5$_{1, 4}$ -- 4$_{1, 3}$  & $\leq$4.8   & & & 11.3(1.1)   & 0.42(0.11)& 11.3 & 5.9(1.3) & 0.27(0.08) & -2.0\\
C$_3$O 10 -- 9 & $\leq$3.9   & & & $\leq$4.9   & & & 5.7(1.4) & 0.31(0.07) & -1.9\\
C$^{34}$S 2 -- 1 & 97.8(9.9) & 0.59(0.01) & -5.0 & 114.4(11.5) & 0.64(0.01) & 11.3 & 89.5(9.0) & 0.38(0.01) & -2.0\\
E-CH$_3$OH 5$_{-1}$ -- 4$_{0}$  & $\leq$3.9   & &  &  7.0(1.9) & 0.93(0.19) & 11.6 & 5.6(1.3) & 0.32(0.09) & -1.9\\
E-CH$_3$OH 2$_{-1}$ -- 1$_{-1}$ & 308.1(30.8) & 0.46(0.01) & -5.0 &  381.5(38.2) & 0.56(0.01) & 11.2 & 325.5(32.5) & 0.34(0.01) & -1.9\\
E-CH$_3$OH 2$_{0}$ -- 1$_{0}$ & 27.5(3.1) & 0.49(0.03) & -4.9 & 26.8(3.2) & 0.60(0.04) & 11.3 & 38.3(4.1) & 0.35(0.01) & -1.9\\
A-CH$_3$OH 2$_{0}$ -- 1$_{0}$  & 402.4(40.3) & 0.46(0.01) & -5.0 & 283.7(28.4) & 0.57(0.01) & 11.3 & 423.0(42.3) & 0.35(0.01) & -1.9\\
A-CH$_3$OH 2$_{1}$ -- 1$_{1}$  & $\leq$4.9   & & & $\leq$5.4   & & & 5.0(1.3) & 0.30(0.09) & -1.9\\
$^{34}$SO  2$_3$ -- 1$_2$  & 86.7(8.7) & 0.51(0.01) & -5.2 & 114.4(11.5) & 0.56(0.01) & 11.1 & 46.9(4.9) & 0.42(0.01) & -2.2\\
c-C$_3$D$_2$ 3$_{1,3}$ -- 2$_{0,2}$  & $\leq$4.9   & & & 8.9(1.6) & 0.39(0.08) & 11.2 & 8.5(1.7) & 0.45(0.05) & -1.9\\
CS 2 -- 1 & 307.5(30.8) & 0.94(0.01) & -4.8 & 794.8(79.5) & 0.78(0.01) & 11.3 & 78.3(7.9) & 0.87(0.02) & -1.9\\
l-C$_3$H  $^{2}\Pi_{\frac{1}{2}}$ J = 9/2--7/2 b F = 4--3 & $\leq$4.8  & & & $\leq$5.3   & & & 8.5(1.5) & 0.31(0.05) & -2.0\\
l-C$_3$H   $^{2}\Pi_{\frac{1}{2}}$ J = 9/2--7/2 a F = 5--4  & $\leq$4.8   & & & 15.0(2.5) & 0.83(0.13) & 11.1 & 9.8(1.7) & 0.38(0.05) & -2.0\\
l-C$_3$H  $^{2}\Pi_{\frac{1}{2}}$ J = 9/2--7/2 a F = 4--3  & $\leq$4.8  & & & 10.7(1.9) & 0.51(0.09) & 11.1 & 9.8(1.8) & 0.45(0.05) & -2.1\\
$^{33}$SO 2$_{3}$ -- 1$_{2}$ F=$\frac{9}{2}$--$\frac{7}{2}$  & $\leq$4.8   & & &  10.9(2.1) & 0.62(0.08) & 11.2 & $\leq$4.3   & & \\
SO N$_{J}$ 2$_3$ -- 1$_2$ & 780.9 (78.1) & 0.69(0.02) & -5.0 & 1429.4(143.0) & 0.63(0.01) & 11.2 & 328.4(32.9) & 0.45(0.01) & -2.0\\
NS$^+$ 2--1 & 6.6(1.8) & 0.62(0.14) & -5.3 & $\leq$5.3   & & & $\leq$4.3   & & \\
CH$_3$SH 4$_{0, 4, 0}$ -- 3$_{0, 3, 0}$ & 6.9(1.5) & 0.38(0.11) & -5.1 &  $\leq$5.3   & & & 7.2(1.7) & 0.52(0.10) & -2.0\\
CH$_3$SH 4$_{1, 3, 1}$ -- 3$_{1, 2, 1}$ & 7.9(1.7) & 0.50(0.09) & -5.0 &  $\leq$5.3   & & & $\leq$4.3   & &\\
o-H$_2$C$^{34}$S & $\leq$4.8   & & & $\leq$4.3   & & & 6.8(1.6) & 0.41(0.07) & -1.9\\
o-H$_2$CS & 37.8(4.0) & 0.61(0.02) & -5.0 & 58.2(6.0) & 0.55(0.02) & 11.2 & 111.2(11.2) & 0.39(0.01) & -2.0\\

\end{longtable}
\tablefoot{\\
The noise is measured around each line within 20 km\,s$^{-1}$. The uncertainty of the integrated intensity W is given by Eq.~\ref{eq: error}.  For the undetected transitions, we estimated the upper limit of 3$\sigma$ based on Eq.~\ref{eq: upper_limit}.}
\end{landscape}

\end{appendix}

\end{document}